\documentclass[aps,prd,amsmath,nofootinbib,preprint,superscriptaddress]{revtex4-2}
\usepackage{graphicx}
\usepackage{color}
\usepackage{epsfig}
\usepackage{epsf}
\usepackage[flushleft]{threeparttable}
\usepackage{bm}
\usepackage{multirow}
\usepackage{dsfont}
\usepackage{braket}
\usepackage{slashed}
\usepackage{ulem}
\usepackage[CJKbookmarks=true, colorlinks=true, linkcolor=blue, urlcolor=blue,citecolor=blue]{hyperref}
\usepackage[utf8]{inputenc}
\usepackage{bbold}
\newcommand{\F}{h}

\newcommand{\dd}{\text{d}}
\newcommand{\eeLL}{e^+e^-\to J/\psi\to\Lambda\overline\Lambda}
\newcommand{\CP}{{\rm CP}}
\DeclareMathOperator{\sgn}{sgn}
\newcommand{\aaa}{g_S}
\newcommand{\bb}{g_P}
\newcommand\myeq{\mathrel{\overset{\makebox[0pt]{\mbox{\normalfont\tiny\sffamily exp}}}{=}}}
\graphicspath{{figures/}}
\newcommand\snowmass{\begin{center}\rule[-0.2in]{\hsize}{0.01in}\\\rule{\hsize}{0.01in}\\
\vskip 0.1in Submitted to the  Proceedings of the US Community Study\\ 
on the Future of Particle Physics (Snowmass 2021)\\ 
\rule{\hsize}{0.01in}\\\rule[+0.2in]{\hsize}{0.01in} \end{center}}

\begin{document}

\title{Study of CP violation in  hyperon decays at Super Charm--Tau Factories with a polarized electron beam}
\author{Nora Salone}
\affiliation{National Centre for Nuclear Research,  Pasteura 7, 02-093 Warsaw, Poland}
 \author{Patrik Adlarson}
\affiliation{Department of
  Physics and Astronomy, Uppsala University, Box 516, SE-75120
  Uppsala, Sweden}
 \author{Varvara Batozskaya}
\affiliation{Institute of High Energy Physics,  Beijing 100049, People's Republic of China}
\affiliation{National Centre for Nuclear Research,  Pasteura 7, 02-093 Warsaw, Poland}
\author{Andrzej Kupsc}
\email[]{Andrzej.Kupsc@physics.uu.se}
\affiliation{Department of
  Physics and Astronomy, Uppsala University, Box 516, SE-75120
  Uppsala, Sweden}
\affiliation{National Centre for Nuclear Research,  Pasteura 7, 02-093 Warsaw, Poland}
\author{Stefan Leupold}
\affiliation{Department of
  Physics and Astronomy, Uppsala University, Box 516, SE-75120
  Uppsala, Sweden}
\author{Jusak Tandean}
\affiliation{Surabaya, Indonesia}
\date{\today}
\begin{abstract}
Non-leptonic two-body weak decays of baryons are an important tool to probe  the combined charge-conjugation--parity symmetry (CP) violation.  We explain why the decays of strange baryons provide complementary information to the decays of kaons.  A model-independent parameterization of the non-leptonic decays of the $\Lambda$- and $\Xi$-baryons is reviewed, and the amplitudes are updated according to the 
latest experimental input. We demonstrate the  potential of performing precision tests in strange baryon decays at the next generation electron--positron $J/\psi$ factories with luminosity of $10^{35}$ cm$^{-2}$s$^{-1}$.
The copious production of spin-entangled hyperon--antihyperon pairs via the $J/\psi$ resonance allows for a direct comparison of the baryon and antibaryon decay properties. Using analytic approximations and numerical calculations, we study the  
quantitative impact of spin correlations and polarization in such CP tests. We show that by using a longitudinally-polarized electron beam, the statistical precision of the CP tests can be significantly improved compared to the experiments without polarized beams. Furthermore, we map out further directions for possible improvements, like analysis of incompletely reconstructed events or a combination of the isospin related processes.
Altogether, these methods are promising for the observation of a statistically significant CP-violation signal with a strength corresponding to the standard model predictions. Our conclusions should encourage more detailed feasibility studies, including optimisation of the measurement methods and studies of systematic effects. Finally, our results call for an update of the theory predictions with increased precision.
\end{abstract}
\maketitle
\snowmass{}
\section{Introduction and summary}
Although the standard model (SM) of elementary particle physics can describe the subatomic world accurately, there are several theoretical and experimental indications that it needs to be completed. In general, precision tests of symmetries and their violation patterns provide guidelines towards a deeper understanding of elementary particles and their interactions. Here we focus on charge-conjugation parity (CP) violation as a means of teasing out new physics. It is well known that the CP-violating mechanism in the SM is not sufficient to explain the observed imbalance between matter and antimatter in our Universe as a dynamic effect~\cite{Sakharov:1967dj}. On the other hand, the processes included in the SM are strong enough to wash out any initial imbalance before the electroweak phase transition \cite{Kuzmin:1985mm, Shaposhnikov:1991cu}. Thus, a CP violation beyond the SM is required. In the quark sector, the existence of CP violation in kaon and beauty meson decays is well established~\cite{Christenson:1964fg,Aubert:2001nu,Abe:2001xe} and so far most observations are consistent with the SM expectations. {There are tensions like the $B\to\pi K$ decay puzzle which require further exploration~\cite{Buras:2003yc}}. The first CP-violating signal for charmed mesons,  reported by the LHCb experiment \cite{Aaij:2019kcg}, is at the upper edge of the SM prediction. As CP-violating effects are subtle, a detailed understanding requires a systematical mapping of various hadronic systems studied with complementary approaches.

\begin{figure}
    \centering
   \includegraphics[width=0.95\textwidth]{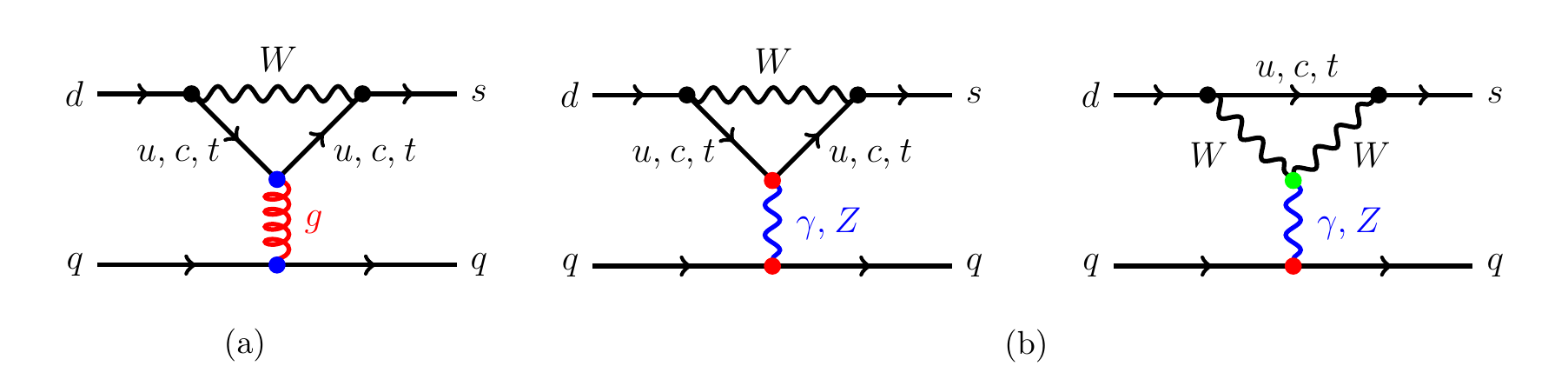}    
\caption{Quark diagrams relevant for kaon and hyperon decays. Direct CP-violation effects in kaon and hyperon decays in the SM are given by the (a)  QCD-penguin operators  and (b) electroweak penguin operators. 
This figure was created using a modified script from Ref.~\cite{Gisbert:2017vvj}.}
\label{fig:diag}
\end{figure}

In the strange-quark sector, one of the most sensitive probes of non-SM contributions is direct CP violation. The experimental result is given by the value $\mathrm{Re}(\epsilon'/\epsilon)=(16.6 \pm 2.3)\times 10^{-4}$ ~\cite{Batley:2002gn,AlaviHarati:2002ye,Abouzaid:2010ny} determined from the
decay amplitude ratios of $K_L$ and $K_S$ mesons into pion pairs,
\begin{equation}
    \frac{{\cal A}(K_L\to\pi^+\pi^-)}{{\cal A}(K_S\to\pi^+\pi^-)}=: \epsilon+\epsilon'
    ~\ {\rm and}~\ 
    \frac{{\cal A}(K_L\to\pi^0\pi^0)}{{\cal A}(K_S\to\pi^0\pi^0)}=: \epsilon-2\epsilon'\ .
\end{equation}
This direct CP-violating effect arises in the weak part of the transition amplitudes to pions due to the interference between isospin $I=0$ and $I=2$ final states ($|\Delta I|=1/2$ and $|\Delta I|=3/2$ transitions, respectively).
The CP-violation mechanism in the SM requires loop diagrams where all three quark families are involved, the so-called penguin diagrams, like those shown in Fig.~\ref{fig:diag}. Predictions for the kaon decays have been a challenge for many years since there are partially cancelling contributions from sub-leading types of the penguin diagrams, where the gluon line is replaced by $\gamma,Z^0$, see \textit{e.g.} Ref.~\cite{Buras_2021} and references therein. Recently, a satisfactory understanding was reached using Lattice~\cite{Bai:2015nea,Abbott:2020hxn} and effective field theory~\cite{Gisbert:2017vvj,Aebischer:2020jto} approaches to Quantum Chromodynamics (QCD). 
This progress ensures that the kaon decays continue to be an important precision test of the SM. 

The subject of our paper is a complementary approach to study CP violation (CPV) in two-body non-leptonic $\Delta S=1$ transitions of hyperons~\cite{Brown:1983wd,Chau:1983ei,Donoghue:1985ww}. For such weak two-body decays, one also needs an interference pattern: this time between parity-even and parity-odd decay amplitudes. These emerge from the spin degrees of freedom of the initial and final baryon. 
Since we will consider decays of a spin-$1/2$ baryon $B$ to a spin-$1/2$ baryon $b$ and a pion, the parity-even amplitude leads to a \textit{p}-wave final state while the parity-odd amplitude to an \textit{s}-wave final state. The two amplitudes are denoted $P$ and $S$, respectively. 
In the following, we will often write the decay generically as $D(B\!\to\! b\pi)$. When we need to be more specific, we use indices $\Lambda$ and $\Xi$ to denote $\Lambda\to p\pi^-$ and $\Xi^-\!\to\Lambda\pi^-$, respectively.
The decay amplitude is
\begin{equation}
   {\cal A}\sim S\sigma_0 +P{\boldsymbol{\sigma}}\cdot {\bf\hat{n}}\ , \label{eq:Ampli}
   \end{equation}
where $\sigma_0$ is the $2\times 2$ unit matrix, $\boldsymbol{\sigma}:=(\sigma_1,\sigma_2,\sigma_3)$ are the Pauli matrices and  ${\bf\hat{n}}={\bf q}/|{\bf q}|$ is the direction of the $b$-baryon momentum ${\bf q}$ in the $B$-baryon rest frame. 
It is important to note that these amplitudes depend on the initial (weak) decay, which produces the two final particles, but depend also on the (strong) final-state interaction.
These $S$ and $P$ amplitudes are Lorentz scalars, which can depend only on the invariant mass of the two-body system. Yet this quantity is fixed for a two-body decay: if we disregard the unmeasurable overall phase, the two complex amplitudes $S$ and $P$ can be fully specified by the overall normalisation $|S|^2+|P|^2$ and the size and relative phase of the interference term $S^* P$. These are directly related to the partial decay width and the following two parameters~\cite{Lee:1957qs}:
\begin{align}
    \alpha_D&:=\frac{2\ {\rm Re}(S^*P)}{|S|^2+|P|^2} \quad {\rm and} \quad
\beta_D:=\frac{2\ {\rm Im}(S^*P)}{|S|^2+|P|^2} \ .
\label{eq:decPar}
\end{align}
The relation of the parameters to the shape of the angular distribution, including the polarization, of the baryon $b$ will be shown in Sec.~\ref{sec:General}. 
In the CP-conserving limit, the amplitudes $\overline{S}$ and  $\overline{P}$ for the charge-conjugated (c.c.) decay mode of the antibaryon $\overline D(\overline{B}\to\overline{b}+\overline{\pi})$ 
are $\overline{S}=-S$ and  $\overline{P}=P$. Therefore, the decay parameters 
have the opposite values: $\overline{\alpha}_D=-\alpha_D$ and $\overline{\beta}_D=-\beta_D$. 

Two independent experimental CPV tests can be defined using these parameters,
\begin{equation}
    A_{\CP}^D:=\frac{\alpha_D+\overline\alpha_D}{\alpha_D-\overline\alpha_D}\quad {\rm and }\quad 
B_{\CP}^D:=\frac{\beta_D+\overline{\beta}_D}{\alpha_D-\overline\alpha_D}\ , \label{eq:CPV}
\end{equation}
where $A_{\CP}^D(B_{\CP}^D)\neq 0$ indicates CP violation in the $D$ decay. 
The $A_{\CP}^D$ test requires measurement of the angular $b(\overline{b})$ distribution from polarized $B(\overline{B})$-baryon decay. 
The $B_{\CP}^D$ test probes time reversal-odd transitions and can be potentially much more sensitive but it requires in addition a measurement of the $b(\overline{b})$-baryon polarization.
In the SM, CPV effects in the hyperon decays are dominated by the QCD-penguin contribution, Fig.~\ref{fig:diag}(a).

In the 1960s, hyperon decays were a tool for discrete symmetry tests on equal footing with the kaons. The last dedicated programme to observe CP violation in hyperons was performed by the Fermilab experiments E756~\cite{Ho:1991rz} and HyperCP~\cite{HyperCP:2004kbv} at the dawn of this century. In these experiments, 
the sum of the $A_{\CP}$ observables for $\Xi^-\to \Lambda\pi^-$ $[\Xi-]$ and  $\Lambda\to p\pi^-$ $[\Lambda p]$, $A_{\CP}^{[\Xi-]}+A_{\CP}^{[\Lambda p]}$,  was studied. 
Here, the SM prediction amounts to $-0.5\times10^{-4}\leq A_{\CP}^{[\Xi-]}+A_{\CP}^{[\Lambda p]}\leq 0.5\times10^{-4}$~\cite{Tandean:2002vy}.
The published result 
 $A_{\CP}^{[\Xi-]}+A_{\CP}^{[\Lambda p]}=0(7)\times10^{-4}$~\cite{HyperCP:2004zvh}
 is currently considered to be the most precise test of CP symmetry in the hyperon sector.

The prospect of significantly improving the CPV tests in hyperons is due to a novel method where hyperon--antihyperon pairs are produced in  electron--positron collisions at the center-of-mass (c.m.)\ energy corresponding to the $J/\psi$ resonance. The
$J/\psi$ decays into a hyperon--antihyperon pair have relatively large  branching fractions of ${\cal O}(10^{-3})$~\cite{ParticleDataGroup:2020ssz}. The  produced hyperon--antihyperon pair has a well-defined spin-entangled state based on the two possible partial waves (parity symmetry in this strong decay allows for an \textit{s}- and a \textit{d}-wave)~\cite{Cabibbo:1961sz,Dubnickova:1992ii}. The charge-conjugated decay modes of the hyperon and antihyperon can be measured simultaneously, and their properties compared directly.
The uncertainties obtained in the proof-of-concept experiment~\cite{Ablikim:2018zay,Ablikim:2021qkn} based on $1.3\times10^{9}$ $J/\psi$ for the $A_{\CP}^{[\Lambda p]}$, $A_{\CP}^{[\Xi-]}$, and $B_{\CP}^{[\Xi-]}$ observables
are given in the first row of Table~\ref{tab:ALCP}. With the already available data set of $10^{10}$ $J/\psi$ collected at BESIII~\cite{Yuan:2019zfo}, a significantly improved statistical precision is expected, as shown in the second row of the table. However, the uncertainty is still predicted to be two orders of magnitude larger compared to the SM CPV signal.

\begin{table}
  \caption{Illustration of the expected statistical uncertainty for the CPV observables $A_{\CP}^{[\Lambda p]}$, $A_{\CP}^{[\Xi-]}$ and $B_{\CP}^{[\Xi-]}$ at BESIII and the proposed SCTF electron--positron collider. The results of the published BESIII measurements are given in the first row~\cite{Ablikim:2018zay,Ablikim:2021qkn}. The uncertainties given in the two remaining rows are straightforward re-scaling based on the expected number of events.
    The SM prediction for $A_{\CP}^{[\Lambda p]}$ is $\sim(1-5)\times10^{-5}$ while for $B_{\CP}^{[\Xi-]}$ it amounts to ${\cal O}(10^{-4})$~\cite{Tandean:2002vy}.  \label{tab:ALCP}
     }
\begin{ruledtabular}
\begin{tabular}{l|llll|l}
    &$\sigma(A_{\CP}^{[\Lambda p]})$&$\sigma(A_{\CP}^{[\Xi-]})$&$\sigma(B_{\CP}^{[\Xi-]})$&&Comment\\
    \hline
    BESIII\vphantom{\large$^2$}&$1.0\times10^{-2}$~\footnotemark[1]&$1.3\times10^{-2}$&$3.5\times10^{-2}$&&$1.3\times10^{9}$ $J/\psi$ \cite{Ablikim:2018zay,Ablikim:2021qkn}\\
    BESIII&$3.6\times10^{-3}$&$4.8\times10^{-3}$&$1.3\times10^{-2}$&&$1.0\times10^{10}$ $J/\psi$ (projection)\\
    SCTF&$2.0\times10^{-4}$&$2.6\times10^{-4}$&$6.8\times10^{-4}$&&$3.4\times10^{12}$ $J/\psi$ (projection)\\ 
  \end{tabular}\\
    \footnotetext[1]{This result is a combination of the two BESIII measurements.}
\end{ruledtabular}
\end{table}
Crucial improvements are expected at the next-generation electron--positron colliders, the Super Charm-Tau Factories (SCTF) being under consideration~\cite{Levichev:2018cvd,Luo:2018njj}. Their design luminosity is two orders of magnitude larger than the BEPCII collider~\cite{Ye:1987nh,BESIII:2009fln} allowing for data samples of more than $10^{12}$ $J/\psi$ events. The projections for the improved statistical uncertainties of the CPV tests, due to the increased data samples, are shown in Table~\ref{tab:ALCP}.
This will still not be sufficient to observe an effect if it has a magnitude consistent with the SM predictions.  Therefore, besides the increased luminosity, two additional improvements are being discussed to further increase the precision: 1) 
a c.m.\ energy spread $\Delta E$ compensation and 2) an electron beam
polarization. For the first option, a collision scheme is proposed where electrons (positrons) with higher momenta are matched with positrons (electrons) with lower momenta.  This promises a $\Delta E$ reduction to better match the natural width of $J/\psi$ meson of $\Gamma=0.09$~MeV, thus up to an order of magnitude increase of the number of $J/\psi$ events for a given integrated luminosity~\cite{Renieri:1975wt,Avdienko:1988sq,Telnov:2020rxp}.  For the second option, an electron
beam polarization of 80--90\% at $J/\psi$ energies can be obtained with the same beam current~\cite{Koop:2019sct}.

Since the benefits of the first improvement are obvious, we focus on the impact of the use of a polarized electron beam and show that the precision of the CP tests in $\eeLL$ and  $e^+e^-\to J/\psi\to\Xi\overline\Xi$ can be significantly improved. The initial findings for $\eeLL$ have already been reported at the SCTF workshop~\cite{Kupsc:2019sct} and independently in Ref.~\cite{Bondar:2019zgm}. Here we give a detailed explanation of this result and extend it to sequential hyperon weak decays.
In Sec.~\ref{sec:General} we review the phenomenology and the current experimental status of CP tests in two-body weak decays of hyperons. In Sec.~\ref{sec:Form}
we use the formalism based on Jacob--Wick's \cite{Jacob:1959at}
helicity amplitudes~\cite{Perotti:2018wxm} to derive the hyperon--antihyperon production spin-correlation matrix for 
electron--positron collisions with longitudinal polarization of the electron beam. The asymptotic maximum log-likelihood method from Ref.~\cite{Adlarson:2019jtw} used for the analysis of uncertainties for the CPV observables is introduced in Sec.~\ref{sec:AMLL}. The single-step decays are discussed in Sec.~\ref{sec:SSdecay} and the two-step decays in Sec.~\ref{sec:DSdecay}. Further experimental considerations are presented in Sec.~\ref{sec:Exp} and Sec.~\ref{sec:Conclusion} contains an outlook.

\section{CP tests in hyperon decays}
\label{sec:General}
\subsection{General considerations}

There are three independent observables that provide a complete description of a weak decay $D(B\to b+\pi)$ with the amplitude given in Eq.~\eqref{eq:Ampli}.
The first is the partial decay width given by
\begin{equation}
    \Gamma=\frac{|{\bf q}|}{4\pi M_B}(E_{b}+M_{b})|{\cal A}|^2\ ,
    \label{eq:Gamma}
\end{equation}
where $|{\cal A}|^2=|S|^2+|P|^2$ and  $E_{b}=\sqrt{|{\bf q}|^2+M_{b}^2}$. The $M_{B}$ and $M_{b}$ are the masses of the mother and daughter baryon, respectively. The first of the two parameters defined in Eq.~\eqref{eq:decPar}, $-1<\alpha_D<1$,  can be determined from the angular distribution of the daughter baryon when the mother baryon is polarized. For example, the proton angular distribution from the $\Lambda(\Lambda\to p\pi^-)$ decay  in the $\Lambda$ rest frame is given as 
\begin{equation}
    \frac{1}{\Gamma}\frac{\dd\Gamma}{\dd\Omega}=\frac{1}{4\pi}\left(1+\alpha_\Lambda {\bf P}_\Lambda\cdot{{\bf\hat{n}}}\right)\ ,
\end{equation}
where ${\bf P}_\Lambda$ is the $\Lambda$ polarization vector. The 
second independent decay parameter can be chosen as the angle $\phi_D$, $-\pi<\phi_D<\pi$, which gives the rotation of the spin vector between the $B$ and $b$ baryons. To measure $\phi_D$, the polarization of both mother and daughter baryons must be determined. For the decay $\Xi(\Xi^-\!\to\Lambda\pi^-)$, where the cascade is polarized, the $\phi_D$ parameter can be determined from the subsequent $\Lambda\to p\pi^-$ decay, which acts as a polarimeter.
The relation between the initial $\Xi^-$ polarization ${\bf P}_\Xi$ and the daughter $\Lambda$
polarization ${\bf P}_\Lambda$ is given by the Lee--Yang formula \cite{Lee:1957qs}:
\begin{equation} 
    {\bf P}_\Lambda = \frac{(\alpha_\Xi + {\bf P}_\Xi\cdot{\bf\hat n}) {\bf\hat n} + \beta_\Xi {\bf P}_\Xi \times {\bf\hat n} + \gamma_\Xi {\bf\hat n} \times ({\bf P}_\Xi \times {\bf\hat n})}{1+\alpha_\Xi {\bf P}_\Xi \cdot {\bf \hat n}} \ ,  \label{eq:LeePol}
\end{equation}
where 
the $\beta$- and $\gamma$-type decay parameters are expressed as
\begin{align}
    \beta_D & = \sqrt{1-\alpha_D^2}\sin\phi_D \ , ~~~~~
    \gamma_D :=\frac{|S|^2-|P|^2}{|S|^2+|P|^2}= \sqrt{1-\alpha_D^2}\cos\phi_D \ , ~~~ \label{eq:PDGdef}
\end{align}
implying that $\alpha_D^2+\beta_D^2+\gamma_D^2=1$. In Table~\ref{tab:decayproperties} the branching fractions (${\cal B}$) and the values of the $\alpha_D$ and $\phi_D$ parameters for decays of the ground-state octet baryons are listed. When available we report the hyperon--antihyperon average values, defined as
\begin{equation}
\braket{\alpha_D} = \frac{\alpha_D - \overline{\alpha}_D}{2}\ ,  ~~~ \braket{\phi_D} = \frac{\phi_D-\overline\phi_D}{2}\ .
\end{equation}
In most cases, the parameters of the antihyperon decays have not been determined yet.
The $\alpha_D$ parameter is much easier to measure than $\phi_D$, since only the polarization of the initial or final 
baryon has to be determined. Before 2018 the consensus was that the $\alpha_D$ parameters were known accurately. The BESIII measurement~\cite{Ablikim:2018zay,Ablikim:2021qkn} has shown that values for $\Lambda\to p\pi^-$ and $\Xi^-\to \Lambda\pi^-$ were wrong by 17\%.

The use of $\alpha_D$ and $\beta_D$ parameters provides a symmetric description of the real and imaginary parts of the $S$ and $P$ amplitudes. {On the other hand, the preferred choice of the $\alpha_D$ and $\phi_D$ parameters by the Particle Data Group (PDG) is motivated experimentally, as the $\phi_D$ and $\alpha_D$ uncertainties are approximately uncorrelated. However, the $\phi_D$ parameter is not directly related to the relative phase between the $S$ and $P$ amplitudes, since it can be written as
\begin{align}
     \phi_D
     &={\arg}\!\left\{(S+P)(S^* - P^*)\right\} \ .
\end{align}
}
\begin{table}
\begin{center}
  \caption{Properties of two-body hadronic decays of the ground-state octet hyperons.  Branching fractions ${\cal B}$ are rounded to $\pm0.5\%$ accuracy. In bold are the values assumed in this report. The motivation for the selection of the specific values is given in the main text. }
    \label{tab:decayproperties}
\begin{ruledtabular}
\begin{tabular}{lccccl}
    \hfill$D$&${\cal B}$&\multicolumn{1}{c}{$\braket{\alpha_D}$}&$\braket{\phi_D}$ [rad]&$A_{\CP}$&Comment\\
    \hline
    $\Lambda\to p\pi^-$\hfill $[\Lambda p]$&\phantom{0}$64\%$&$\phantom{-}\bm{0.755(03)}$\footnotemark[1]&\boldmath$-0.113(61)$\footnotemark[2]&\boldmath$-0.005(10)$\footnotemark[1]&\\
    &&{$\phantom{-}0.754(3)(2)$}&--&{$-0.006(12)(7)$}&BESIII \cite{Ablikim:2018zay}\\
    &&$\phantom{-}0.721(6)(5)^*$&--&--&CLAS \cite{Ireland:2019uja}\\
    &&{$\phantom{-}0.760(6)(3)$}&--&{$-0.004(12)(9)$}&BESIII \cite{Ablikim:2021qkn}\\ 
    $\Lambda\to n\pi^0$\hfill $[\Lambda n]$&\phantom{0}$36\%$&$\phantom{-}\bm{0.692(17)}$\footnotemark[3]&--&--&BESIII \cite{Ablikim:2018zay}\\
    $\Sigma^+\to p\pi^0$\hfill $[\Sigma p]$&\phantom{0}$52\%$&$\bm{-0.994(04)}$\footnotemark[4]&\phantom{-}\boldmath$0.63(59)$\footnotemark[7]&\boldmath $-0.004(37)$\footnotemark[4]&\\
    $\Sigma^+\to n\pi^+$~\hfill $[\Sigma n]$&\phantom{0}$48\%$&\boldmath$\phantom{-}0.068(13)^*$&\phantom{-}\boldmath$2.91(35)^*$&--&PDG \cite{ParticleDataGroup:2020ssz}\\
    $\Sigma^-\to n\pi^-$\hfill$[\Sigma-]$&$100\%$&\boldmath$-0.068(08)^*$&\phantom{-}\boldmath$0.17(26)^*$&--&PDG \cite{ParticleDataGroup:2020ssz}\\
    $\Xi^0\to \Lambda\pi^0$\hfill$[\Xi0]$&$100\%$&$\bm{-0.345(08)}$\footnotemark[5]&\phantom{-}\boldmath$0.36(21)^*$&--&AVG \cite{Handler:1982gd,Batley:2010bp}\\
    $\Xi^-\to \Lambda\pi^-$\hfill$[\Xi-]$&$100\%$&$\bm{-0.379(04)}$\footnotemark[6]&$-0.042(16)^*$&--&AVG \cite{ParticleDataGroup:2020ssz,Huang:2004jp} \\
    &&{$-0.373(5)(2)$}&{$\phantom{-}0.016(14)(7)$} &{\boldmath$\phantom{-}0.006(13)(6)$}&BESIII \cite{Ablikim:2021qkn}\\ 
\end{tabular}
\end{ruledtabular}
\end{center}
\vspace{-0.5cm}
\noindent\begin{minipage}[t]{\textwidth}
 \footnotetext[0]{\hspace{-3mm}${}^{*}$ Solely based on the result for hyperons (not antihyperons)}
\footnotetext[1]{Weighted average of $\braket{\alpha_{[\Lambda p]}}$ from \cite{Ablikim:2018zay,Ablikim:2021qkn}}
\footnotetext[2]{Weighted average of $\phi_{[\Lambda p]}$ from \cite{Cleland:1972fa,Overseth:1967zz,Cronin:1963zb} the same as in PDG~\cite{ParticleDataGroup:2020ssz}}
\footnotetext[3]{ The $-\overline{\alpha}_{[\Lambda p]}$ value from \cite{Ablikim:2018zay}}
\footnotetext[4]{Value from \cite{Ablikim:2020lpe}}
\footnotetext[5]{From $\alpha_{[\Xi0]}\alpha_{[\Lambda p]}=-0.261(6)$ \cite{ParticleDataGroup:2020ssz} divided by $\alpha_ {[\Lambda p]}$\footnotemark[1]}
\footnotetext[6]{Combination of $\braket{\alpha_{[\Xi-]}}$ \cite{Ablikim:2021qkn} and $\alpha_{[\Xi-]}\alpha_{[\Lambda p]}=-0.294(5)$ \cite{ParticleDataGroup:2020ssz} divided by $\alpha_{[\Lambda p]}$\footnotemark[1]}
\footnotetext[7]{Weighted average of $\phi_{[\Sigma p]}$ from \cite{Harris:1970kq,Lipman:1973mz}} 
\end{minipage}
\end{table}

In general, we do not need to know the exact values of the decay parameters to predict the uncertainties of the CPV observables given in Eq.~\eqref{eq:CPV}. Many of our results can be described using approximate analytic formulas where the dependence on parameters is given explicitly. Furthermore, in the proposed measurements the values of the decay parameters are determined directly together with the CPV observables, and are uncorrelated with each other.
For specific purposes, such as the estimate of the size of the decay amplitudes in Appendix~\ref{sec:Isospin}, we need the most precise values of the decay parameters and branching ratios or life times. We have made a critical evaluation of the available data, and the preferred values which we have selected are given in bold in Table~\ref{tab:decayproperties}. Here, we provide a detailed explanation how some values were determined:
\begin{itemize}
    \item The $\braket{\alpha_{[\Lambda p]}}$ value is the average of the two BESIII measurements \cite{Ablikim:2018zay} and  \cite{Ablikim:2021qkn}. We do not include the result from CLAS experiment~\cite{Ireland:2019uja} since it does not report the measurement of $\braket{\alpha_{[\Lambda p]}}$ and would indicate significant violation of the CP symmetry due to the statistically  inconsistent value with the BESIII measurement of the antihyperon $\overline{\alpha}_{[\Lambda p]}$. The BESIII results for $\alpha_D$ and $\overline\alpha_D$ are correlated and have large uncertainty separately. 
    \item Since the  $\braket{\phi_{[\Xi-]}}$ measured at BESIII \cite{Ablikim:2021qkn} differs by 2.6 standard deviations from $\phi_{[\Xi-]}$ measured by HyperCP \cite{Huang:2004jp}, we do not provide the average value for $\braket{\phi_{[\Xi-]}}$.
\end{itemize}
Finally, we use other results which do not fit to the format of the table, such  as $B_{\CP}^{[\Xi-]}$,   $A_{\CP}^{[\Xi-]}+A_{\CP}^{[\Lambda p]} $ or life times of the cascades. They are introduced and referred to when we need to use them.
For example, for the determination of the contribution of the $\Delta I =3/2$ amplitudes we use more precise values of the branching fractions from Ref.~\cite{ParticleDataGroup:2020ssz}: ${\cal B}(\Lambda\to p\pi^-)=0.639(5)$
    and  ${\cal B}(\Lambda\to n\pi^0)=0.358(5)$.

\subsection{CP violation phenomenology}
\label{ssec:CPpheno}

Isospin is not conserved in weak transitions, meaning that both the isospin vector length and the third component $I_3$ change in the decay process. In our hyperon decays of interest, there is effectively a transition from a strange to a down quark: thus, $I_3$ changes by $-1/2$. For the total isospin, the situation is more involved. It is convenient to classify the weak transition by the isospin $\Delta I$ of the transition operator. Starting with the initial isospin $I_{\rm ini}$ of the decaying hyperon, the isospin $I$ of the final state can take values between $\vert I_{\rm ini} - \Delta I \vert$ and $I_{\rm ini} + \Delta I$. As a result of these considerations, it is practical to characterise the weak process by the isospin of the final state $I$ and by the change of isospin $\Delta I$. To explain this distinction, let us consider the process $\Xi^-\to\Lambda\pi^-$ where the initial and final isospins are $1/2$ and $1$, respectively. This final state can be reached by a transition with $\Delta I=1/2$, where the isospins are aligned, and a transition with $\Delta I=3/2$, where the isospins are anti-aligned. 
Therefore, the transition amplitudes of the decomposition should be labelled by both $I$ and $\Delta I$, and we adopt the notation $S_{2\Delta I,2I}$ and $P_{2\Delta I,2I}$. 

The transition amplitudes $L=S,P$ can be decomposed as~\cite{Donoghue:1986hh}:
\begin{align}
    L&=\sum_j L_j\exp\left\{{i(\xi_{j}^L+\delta_{j}^L})\right\} \ , \label{eq:TL} 
\end{align}
where $j$ represents a possible $\{2\Delta I,2I\}$ combination, while  $\xi_{j}^L$ and $\delta_{j}^L$ denote the weak CP-odd phase and the phase of the combined strong and electromagnetic (e.m.) final state interaction, respectively. Appendices \ref{sec:AppEFT} and \ref{sec:FSI-app} provide a  justification for the decomposition \eqref{eq:TL} where 
the $S_j$ and $P_j$ amplitudes are real numbers. 
The final-state interaction phase is dominated by the phase shifts of the strong elastic rescattering. The isospin breaking effects in the rescattering due to hadron mass differences for different charge states are a few percent. Further contributions can be due to $m_d-m_u$ terms in the amplitudes and e.m. interactions of the hadrons, such as radiative corrections or Coulomb interactions. 
The $\delta_{j}^L$ phase can be written as $\delta_{j}^L=\delta_{2I}^L+\Delta\delta_{j}^L$, where the correction term
$\Delta\delta_{j}^L$ includes the isospin breaking effects due to e.m. interactions in the final state. Here, we will neglect this term, but for future precision studies it should be considered similar to how it was for the kaon to two-pion decays~\cite{Cirigliano:2003gt}. 

For the $N$--$\pi$ final states, the phases-shifts  $\delta_{2I}^L$ are well known. We summarise in Table~\ref{tab:Sphases} the values from Ref.~\cite{Hoferichter:2015hva}  which are relevant for the $\Lambda$ and $\Sigma$ decays. 
The $\Lambda$-$\pi$ scattering phase-shifts, on the other hand, are less precisely determined from experiment. 
In particular, for $\Xi\to\Lambda\pi$ they can be found via the relation
\,$\tan\big(\delta_2^P-\delta_2^S\big)=\sin\phi_\Xi^{}\sqrt{1-\alpha_\Xi^2}/\alpha_\Xi^{}$,\, 
neglecting the weak-phase difference, where $\alpha_\Xi$ and $\phi_\Xi$ are obtainable directly from the sequential decays. In doing so, we note that the current $\phi_\Xi$ data are not all consistent with each other yet, as pointed out in the preceding subsection.  
On the theoretical side, various analyses have produced different results~\cite{Nath1965,Lu:1994ex,Kamal:1998se,Datta:1998pv,Tandean:2000dx,Meissner:2000re,Huang:2017bmx}, the latest one being 
$\delta_2^P-\delta_2^S=8.8(2)^\circ$\,\,~\cite{Huang:2017bmx}, which is compatible with one of the earlier predictions~\cite{Tandean:2000dx} and will be used in updating the $A_{\rm CP}^{[\Xi-]}$ prediction. 

\begin{table}[t]
    \caption{Values of the $N$--$\pi$ scattering phase shifts $\delta_{2I}^L$  relevant
    for $\Lambda$ and $\Sigma$ decays from~\cite{Hoferichter:2015hva}. 
    \label{tab:Sphases}}
    \centering
\begin{ruledtabular}
\begin{tabular}{c|cllll}
    & $|{\bf q}|$&$\delta_{1}^S$&$\delta_{3}^S$&$\delta_{1}^P$&$\delta_{3}^P$\\
    & [MeV/c] & $[{}^\circ]$ & $[{}^\circ]$& $[{}^\circ]$& $[{}^\circ]$\\ \hline
   $\Lambda\to N\pi$ &103  &$6.52(9)$ &$-4.60(7)$ & $-0.79(8)$ & $-0.75(4)$\\
 $\Sigma\to N\pi$  &190 &$9.98(23)$& $-10.70(13)$& $-0.04(33)$&$-3.27(15)$
    \end{tabular}
\end{ruledtabular}
\end{table}

Now we will discuss signatures of CP violations in the hyperon decays. They are based on the comparison of the hyperon decay amplitudes, Eq.~\eqref{eq:TL}, with the ones corresponding to the antihyperon c.c. decay,
\begin{align}
    \overline{S}&=-\sum_j S_j\exp\left\{{i(-\xi_{j}^S+\delta_{2I}^S})\right\} \ ~~{\rm and} \ ~~
    \overline{P}=\sum_j P_j\exp\left\{{i(-\xi_{j}^P+\delta_{2I}^P})\right\}\ , \label{eq:aTL}
\end{align}
where  the real-number parameters ${L}_j$, $\xi_{j}^L$ and $\delta_{2I}^L$, ($L=S,P$), have the same values for the hyperon and antihyperon decays. The isospin-decomposition relations obtained in Appendix~\ref{sec:Isospin} can be applied for the c.c.\ decays of antihyperons. 
{\it A priori}, up to three independent observables can be used to compare properties of a decay to the c.c. one. The first observable is the difference between the partial decay widths 
\begin{equation}
    \Delta_{\CP}:=\frac{\Gamma-{\overline\Gamma}}{\Gamma+{\overline\Gamma}} \ .
\end{equation}
In the $\Delta I =1/2$ limit the $\Delta_{\CP}$ observable is exactly zero and cannot be used to test CP symmetry.
In addition, for $\Xi\to\Lambda\pi$ the isospin of the final $\Lambda$--$\pi$ state is $I=1$ and there is only one strong phase for each of the $S$ and $P$ amplitudes. This implies that the corresponding $\Delta_{\CP}$ is zero even if the weak transition includes $|\Delta I|=3/2$ operators. 
However, the $\Delta_{\CP}$ test is possible for $\Lambda\to N\pi$, as the final state can have $I=1/2$ or 3/2. 
For the two $\Lambda$-decay modes, to lowest order in the $\Delta I=3/2$ amplitudes, we have the relation $2\Delta_{\CP}^{[\Lambda p]}=-\Delta_{\CP}^{[\Lambda n]}=2\sqrt{2}\Delta_{\CP}$ with
\begin{align}
\Delta_{\CP} & =  \frac{P_{1,1}P_{3,3}\sin(\xi_{1,1}^P-\xi_{3,3}^P)\sin(\delta_{1}^P-\delta_{3}^P)
+
S_{1,1}S_{3,3}\sin(\xi_{1,1}^S-\xi_{3,3}^S)\sin(\delta_{1}^S-\delta_{3}^S)}{P_{1,1}^2+S_{1,1}^2}\ .
\end{align}
This requires two weak and two strong phases either in the $S$ amplitude, as in the kaon decays, or in the $P$ amplitude. The precision of the test is suppressed by the small  $|\Delta I|=3/2$ amplitudes and by the term containing sinus of the small strong phases. Therefore, such a test is not competitive and we will not discuss it further.

The remaining two CP tests are based on  the $A_{\CP}^D$ and $B_{\CP}^D$ observables
defined in Eq.~\eqref{eq:CPV}. The  $B_{\CP}^D$ observable can also be expressed as 
\begin{align}
    B_{\CP}^D&=\Phi_{\CP}^D\frac{\sqrt{1-\alpha_D^2}}{\alpha_D}\cos\phi_D-
    A_{\CP}^D\frac{\alpha_D}{\sqrt{1-\alpha_D^2}}\sin\phi_D\ ,
\end{align}
where
\begin{equation}
\Phi_{\CP}^D:=\frac{\phi_D+\overline\phi_D}{2} 
\end{equation}
is based on the spin-rotation decay parameter $\phi_D$.  In a large acceptance experiment, the decay parameters $\alpha$ and $\phi$ are uncorrelated, as well as the CPV tests based on the $A_{\CP}^D$ and $\Phi_{\CP}^D$ variables.

Contrary to the CP violation in $K_{L,S}\to\pi\pi$, where $\Delta I =1/2$ and $\Delta I =3/2$ amplitudes are both consequential, the dominant effect in hyperons can be studied using only the $\Delta I =1/2$ amplitudes. The corrections to the CPV effect studied in this approximation will be a few percent, as given by the size of the $P_3$ and $S_3$ amplitudes. This is sufficient for the precision expected at SCTF. If a better precision is required, one can construct isospin averages of the observables from different isospin modes to recover the results in the $\Delta I =1/2$ limit. {Such averages are constructed from the isospin decomposition of a given decay process (channel) -- for more details, we refer to Appendix~\ref{sec:Isospin}.
For $\Xi$, up-to the linear terms in the $\Delta I =3/2$ amplitudes, they amount to}
\begin{align}
B_{\CP}^{\Xi}:=\frac{2{B_{\CP}^{[\Xi-]}}+ {B_{\CP}^{[\Xi0]}}}{3}&=\phantom{-}\tan(\xi_{1,2}^P-\xi_{1,2}^S)\\
A_{\CP}^{\Xi}:=\frac{2{A_{\CP}^{[\Xi-]}}+ {A_{\CP}^{[\Xi0]}}}{3}&=-\tan(\xi_{1,2}^P-\xi_{1,2}^S)
\tan(\delta_{2}^P-\delta_{2}^S)\label{eq:DXi2}\ ,
\end{align}
and for $\Lambda$
\begin{align}
B_{\CP}^{\Lambda}:=\frac{2{B_{\CP}^{[\Lambda p]}}+ {B_{\CP}^{[\Lambda n]}}}{3}&=\phantom{-}
\tan(\xi_{1,1}^P-\xi_{1,1}^S)\\
A_{\CP}^{\Lambda}:=\frac{2{A_{\CP}^{[\Lambda p]}}+ {A_{\CP}^{[\Lambda n]}}}{3}&=
-\tan(\xi_{1,1}^P-\xi_{1,1}^S)
\tan(\delta_{1}^P-\delta_{1}^S)\label{eq:DLa2}\ .
\end{align}
The leading-order correction for the two isospin states of the cascades is:
\begin{align}
  {B_{\CP}^{[\Xi-]}}- {B_{\CP}^{[\Xi0]}}&=-\frac{3}{2} \left[\frac{P_{3,2}}{P_{1,2}} \sin(\xi_{1,2}^P-\xi_{3,2}^P) - \frac{S_{3,2}}{S_{1,2}} \sin(\xi_{1,2}^S-\xi_{3,2}^S)\right]\label{eq:DXi4}\\
  {A_{\CP}^{[\Xi-]}}- {A_{\CP}^{[\Xi0]}}&=-\left({B_{\CP}^{[\Xi-]}}- {B_{\CP}^{[\Xi0]}}\right)\tan(\delta_{2}^P-\delta_{2}^S)\nonumber
\ ,
\end{align}
which implies that even if the LO $\Delta I=3/2$ corrections are included, the $A$ and $B$ tests are still connected --- giving the same combination of the weak phases. For the $\Lambda$ decays such a relation is not valid and the  
$A$- and $B$-type variables provide independent information on the 
weak-phase combinations. We will not discuss this case, since the $B$-type observables cannot be measured with the standard techniques available at the  electron--positron-collider experiments. 
A combination of the CP tests for the isospin related channels allows for an increased statistical significance of the tests. Such an approach is feasible at SCTF for the $\Xi$ and $\Lambda$ decays, since all the decay parameters for (anti)cascade and the $\alpha$ parameters for $\Lambda$ can be measured. 

A simpler approach is to treat each decay mode separately when  comparing decay parameters for the hyperon and, from the c.c. decay, for the antihyperon. 
In  the $\Delta I=1/2$ approximation we can write
\begin{alignat}{2}
    S&=\sin\!\zeta\exp\!{(i\xi_{S}+i\delta_{S})} \ , ~~~\  \ &&{\overline S} =-\sin\!\zeta\exp({-i\xi_{S}+i\delta_{S}})\label{eq:SP}\ , \\ 
    P&=\cos\!\zeta\exp\!{(i\xi_{P}+i\delta_{P})} \ , ~~\ \ &&{\overline P}=\phantom{-}\cos\!\zeta\exp({-i\xi_{P}+i\delta_{P}}) \nonumber\ , 
\end{alignat}
where $0\le\zeta\le\pi/2$, $\xi_S(\xi_P)$ is the weak CP-odd phase for the  $\Delta I =1/2$ transition and $\delta_S(\delta_P)$ is the strong \textit{s}(\textit{p})-wave baryon--pion  phase-shift at the c.m. energy corresponding to the hyperon mass. 
The structure of Eq.~\eqref{eq:SP} can be justified, if one assumes that the complete decay process can be split up into the decay itself where one does not resolve the intrinsic structure and a final-state interaction that conserves P and C separately. If one does not resolve the space-time structure of the initial decay, then one can use an effective hermitian Lagrangian to describe the decay and one just reads off the relations $\overline S_{\rm ini} = -S_{\rm ini}^*$ and $\overline P_{\rm ini} = P_{\rm ini}^*$. More details are given in Appendix~\ref{sec:AppEFT}. The final-state interaction can be described by a $4\times 4$ Omn\`es-function matrix that is applied to the four initial amplitudes; see also Appendix~\ref{sec:FSI-app}. If P (and baryon number) is conserved, then this matrix is diagonal. If C is conserved, then the entries are pairwise the same for particle and antiparticle. Without inelasticities, Watson's theorem~\cite{Watson:1954uc} identifies the phases with the scattering phase shifts. The decay parameters $\left(\alpha,\beta,\gamma\right)$ and $\left(\overline{\alpha},\overline{\beta},\overline{\gamma}\right)$\footnote{In the remaining part of this section we simplify the notation by omitting subscript $D$ for the decay parameters.} are then given as
\begin{alignat}{4}
    \alpha&=\sin(2\zeta)\cos(\xi_P-\xi_S+\delta_P-\delta_S) \ , \ ~ ~~~&&\overline{\alpha}&&=-&&\sin(2\zeta)\cos(-\xi_S+\xi_P+\delta_S-\delta_P) \ , \label{eq:SPone}\\
     \beta&=\sin(2\zeta)\sin(\xi_P-\xi_S+\delta_P-\delta_S) \ , ~~~&&\overline{\beta}&&=-&&\sin(2\zeta)\sin(-\xi_P+\xi_S+\delta_P-\delta_S) \ , \\
    \gamma&=-\cos(2\zeta) \ ,  &&\overline{\gamma}&&=-&&\cos(2\zeta)
    \ . 
\end{alignat}
Without final-state interactions, $\alpha + \overline\alpha$ is always zero and $A_{\CP}$ does not constitute an observable
that can indicate CP violation, while  $B_{\CP}=\tan(\xi_P-\xi_S)$ does. One needs CP violation and final-state interactions to make $A_{\CP}$ different 
from zero. In the presence of final-state interactions, $\beta \neq 0$ does not necessarily indicate CP violation, but   $B_{\CP}$ still does.
The CPV tests based on the $A_{\CP}$, $B_{\CP}$ (and $\Phi_{\CP}$) observables can be expressed using Eq.~\eqref{eq:SP} as
\begin{align}
    A_{\CP}&=-\frac{\sqrt{1-\alpha^2}}{\alpha}\sin\phi\tan(\xi_P - \xi_S)\label{eq:ACP1} \\
    &=-\tan(\delta_P-\delta_S)\tan(\xi_P - \xi_S)\label{eq:ACP2} \ ,\\ 
    B_{\CP}&=\phantom{-}\tan(\xi_P - \xi_S)\label{eq:BCP1} \ ,\\ 
    \Phi_{\CP}&=\phantom{-}\frac{\alpha}{\sqrt{1-\alpha^2}}{\cos\phi}\tan(\xi_P - \xi_S) \ .
\end{align}
Therefore the tests are not independent as they are related to the same  $\xi_P - \xi_S$ combination of the CP-odd weak phases. For single-step decays of the singly-strange baryons, measurement of the $B_{\CP}(\Phi_{\CP})$ would require a dedicated detector to determine the daughter-nucleon polarization. Therefore,  for the $\Lambda$ and $\Sigma$ hyperon decays, we consider only the $A_{\CP}$ observable measurements. In this case, the weak phases are determined by Eq.~\eqref{eq:ACP2} using the well known values of the strong $N$--$\pi$ phases.
Since the strong phases $\delta_P$ and $\delta_S$, representing the final state interaction between the baryon and pion, are small, the
$B_{\CP}$ observable  provides much better determination of the weak-phase difference than $A_{\CP}$. This statement assumes that the uncertainties of the $A_{\CP}$ and $B_{\CP}$ (or $\Phi_{\CP}$) measurements are comparable. In Sec.~\ref{sec:Exp} we will discuss strategies for the simultaneous measurement of the two observables in the cascade decays.

\subsection{Status of the CPV predictions}

In this subsection, we review the estimates of CPV signals for the decay channels $\Lambda\to p\pi^-$ and $\Xi^-\!\to\Lambda\pi^-$, commonly considered to be the most sensitive modes. In the experimental study of the latter, the former is used as the subsequent process. 
The SM contributions to $\xi_P-\xi_S$ for the two decay modes are shown in the third column of Table~\ref{tab:weak}. 
These predictions are both $\mathcal{O}(10^{-4})$, taking into account the substantial uncertainties which are related to our present lack of ability to explain simultaneously the s- and p-waves of hyperon nonleptonic decays~\cite{Tandean:2002vy}.
The second column of this table contains $\xi_P-\xi_S$ divided by $\eta\lambda^5A^2$, which is a product of the Wolfenstein parameters for the Cabibbo--Kobayashi--Maskawa matrix and has a value of $1.36(7)\times10^{-4}$ according to the most recent PDG report~\cite{ParticleDataGroup:2020ssz}. 
The SM entries in this table are updates of the corresponding numbers found in Ref.~\cite{Tandean:2002vy} and are somewhat modified with respect to the latter, mainly because of our use of the (boldfaced) new $\alpha$ results for $\Lambda\to p\pi^-$ and $\Xi^-\!\to\Lambda\pi^-$ quoted in Table~\ref{tab:decayproperties}. 

\begin{table}[t]
    \caption{Weak-phase differences in hyperon decays. (left) Standard-model predictions and  (right) parameters $C_B$ and $C'_B$ used in Eq.~\eqref{eq:BSM} to relate the weak-phase differences in hyperon decays to the beyond SM (BSM) constraints from kaon CPV observables. The SM and BSM entries are updates of the corresponding numbers obtained in Refs.~\cite{Tandean:2002vy} and~\cite{Tandean:2003fr}, respectively, as explained in the main text.\label{tab:weak}}
    \centering
\begin{ruledtabular}
\begin{tabular}{cccc|cc}
    &\multicolumn{3}{c|}{$\xi_P-\xi_S$}&$C_B$&$C'_B$\\
    &$(\eta\lambda^5A^2)$&[$10^{-4}$ rad]& &&\\
 &\multicolumn{2}{c}{SM}&&\multicolumn{2}{c}{BSM}\\
     \hline
        $\Lambda\to p\pi^-$ & $-0.1\pm1.5$&$-0.2\pm2.2$&&
        $\phantom{-}0.9\pm1.8$ & $0.4\pm0.9$  \\
        $\Xi^-\to\Lambda\pi^-$ &$-1.5\pm1.2$  &$-2.1\pm1.7$&&
        $-0.5\pm1.0$ & $0.4\pm0.7$\\ 
    \end{tabular}
\end{ruledtabular}
\end{table}

To compare the theoretical $A_{\CP}$ with its most precise measurements to date given in Table~\ref{tab:decayproperties} requires multiplication of the calculated $\xi_P-\xi_S$ by the strong-interaction parameters, as indicated in Eqs.~\eqref{eq:ACP1}-\eqref{eq:ACP2}, an extra step which increases the experimental uncertainty and/or decreases the precision of the predictions. 
Nevertheless, from Eq.~\eqref{eq:BCP1}, we expect that future measurements of $B_{\CP}$ can directly determine $\xi_P-\xi_S$ with good precision. 
For $\Lambda\to p\pi^-$ the strong phases pertaining to Eq.~\eqref{eq:ACP2} are $\delta_1^S=0.11(2)$~rad and  $\delta_1^P=-0.014(1)$~rad from Table~\ref{tab:Sphases}.  
For $\Xi^-\to \Lambda\pi^-$ the strong-phase difference can be extracted experimentally using the methods discussed in this report.
However, since $\beta_{[\Xi-]}$ is not yet well measured, the $\alpha_{[\Xi-]}$ data cannot be used to obtain $\delta_2^P-\delta_2^S$ with good precision via Eq.~\eqref{eq:Xi4}.  
To update the prediction for $A_{\rm CP}^{[\Xi-]}$, we adopt instead $\delta_2^P-\delta_2^S=8.8(2)$\,\,deg computed in Ref.~\cite{Huang:2017bmx}.
Putting together the weak and strong phases, we arrive at the SM ranges $-3\times10^{-5} \le A_{\rm CP}^{[\Lambda p]} \le 3\times 10^{-5}$ and $0.5\times10^{-5} \le A_{\rm CP}^{[\Xi-]} \le 6\times10^{-5}$, which are below their respective experimental bounds inferred from Table\,\,\ref{tab:decayproperties} by more than two orders of magnitude.

Measurements on hyperon CPV and its kaon counterpart are complementary to each other because they do not probe the underlying physics in the same way.   
As mentioned above, in the context of the SM, the direct-CPV parameter $\epsilon'$ in the kaon decay $K\to\pi\pi$ arises from both $|\Delta I|=1/2$ and $|\Delta I|=3/2$ transitions, where the CP-odd phases come from the QCD, Fig.~\ref{fig:diag}(a), and electroweak, Fig.~\ref{fig:diag}(b), penguin contributions, respectively, all of which are induced by effective four-quark operators.  There is a delicate balance and cancellation between the two contributions. In the hyperon case, the CPV signal of interest here, such as measured by $A_{\rm CP}$ or $B_{\rm CP}$, mainly comes from $|\Delta I|=1/2$ transitions and is dominated by the QCD penguins.  

In the presence of physics beyond the SM (BSM), there might be new ingredients causing other types of quark operators to generate effects that are enhanced relative to the SM contributions. 
This possibility can be realised, for instance, by the so-called chromomagnetic-penguin operators, which contain a $ds$ quark bilinear coupled to gluon fields and could be influenced by sizeable new physics in various models~\cite{Chang:1994wk,He:1995na,Buras:1999da,He:1999bv,Chen:2001cv,Tandean:2003fr}.
The parity-odd and parity-even portions of the operators contribute to $\epsilon'$ and the CPV parameter $\epsilon$ in neutral-kaon mixing, respectively, and both parts simultaneously affect $\xi_P-\xi_S$. Model independently, one can derive a general relation between the contributions of these operators to the hyperon weak-phase difference and kaon observables~\cite{Tandean:2003fr}: 
\begin{equation}
    (\xi_P-\xi_S)_{\rm BSM} = \frac{C_B'}{B_G}\left(\frac{\epsilon'}{\epsilon}\right)_{\rm BSM}
    + \frac{C_B}{\kappa}\, {\epsilon}_{\rm BSM} \ , \label{eq:BSM}
\end{equation}
which further illustrates the complementarity of hyperon and kaon decays.
The values of $C_B$ and $C_B'$, updated from their counterparts evaluated in Ref.~\cite{Tandean:2003fr}, are given in Table~\ref{tab:weak}, $B_G$ parameterizes the hadronic uncertainty, and $\kappa$ quantifies the contribution of meson poles. The allowed ranges of $\left({\epsilon'}/{\epsilon}\right)_{\rm BSM}$ and ${\epsilon}_{\rm BSM}$ can be estimated by comparing the experimental values of ${\rm Re}({\epsilon'}/{\epsilon})$ and $|\epsilon|$ with the recent SM predictions~\cite{Cirigliano:2019cpi,Brod:2019rzc,Aebischer:2020mkv}. 
Following Ref.~\cite{Aebischer:2020mkv} we impose
\begin{align}
\left|\frac{\epsilon'}{\epsilon}\right|_{\rm BSM} & \leq 1\times10^{-3} \ , ~~~~~ \ |\epsilon|_{\rm BSM} \leq 2\times10^{-4} \ . ~~~~~
\end{align}
Accordingly, using  $0.5 < B_G < 2$ and $0.2 < |\kappa| < 1$~\cite{He:1999bv}, we find that the kaon data imply the limits $|\xi_P-\xi_S|_{\rm BSM}^{[\Lambda p]}\leq5.3\times10^{-3}$ and $|\xi_P-\xi_S|_{\rm BSM}^{[\Xi-]}\leq3.7\times10^{-3}$. 
Additionally, we arrive at $|A_{\rm CP}^{[\Lambda p]}+A_{\rm CP}^{[\Xi-]}|_{\rm BSM} \le 11\times10^{-4}$, and therefore the upper end of this range is already in tension with the aforementioned HyperCP limit~\cite{HyperCP:2004zvh}. 
Clearly, hyperon CPV measurements with much improved precision will provide an independent constraint on the BSM contributions in the strange quark sector. 
However, a lot also remains to be done on the theory side, as the predictions presently suffer from considerable uncertainties.
It is hoped that lattice QCD analyses~\cite{Beane:2003yx} in the future could help solve this problem. 

\subsection{Experimental status of CPV tests}

The dedicated CPV experiment HyperCP (E871) at Fermilab~\cite{White:1999dv}, operating between 1996 and 1999, has set the world’s best upper limits on hyperon CP violation using the $\Xi^-\to\Lambda\pi\to p\pi^-\pi^-$ decay sequence. A secondary cascade beam was produced by having 800 GeV/c primary protons interacting with a copper target. The sum of the asymmetries $A_{\CP}^{[\Xi-]}+A_{\CP}^{[\Lambda p]}=0(5)(4)\times10^{-4}$~\cite{HyperCP:2004zvh} was determined with a data sample of $117\times10^6$ $\Xi^-$ and $41\times10^6$ $\overline\Xi^+$ using unpolarized cascades. A preliminary result $A_{\CP}^{[\Xi-]}+A_{\CP}^{[\Lambda p]}=-6(2)(2)\times10^{-4}$ based on the full data sample of $862\times10^6 \ \Xi$ and  $230\times 10^6 \ \overline\Xi$ was presented at the BEACH2008 conference~\cite{Materniak:2009zz}. Since the final result was never published, one can suspect that an inherent problem 
to understand the systematic effects at the level of $4\times10^{-4}$ was found. The HyperCP has also measured the most precise value of $\phi_{[\Xi-]}$, see Table~\ref{tab:decayproperties}, using $144\times10^6$ $\Xi^-$ events with average polarization of $\sim5\%$~\cite{Huang:2004jp}. The drawback of the HyperCP experimental method is the  charge-conjugation-asymmetric production mechanism and the need to use separate runs with different settings for the baryon and antibaryon measurements. Furthermore, the accuracy of the $\phi_{[\Xi-]}$ parameter determination was limited by the low value of the $\Xi^-$-beam polarization.

The most recent results, marked by bold fonts in Table~\ref{tab:decayproperties}, come from the proof-of-concept measurements~\cite{Ablikim:2018zay,Ablikim:2020lpe,Ablikim:2021qkn} at BESIII using a novel method~\cite{Faldt:2017kgy,Perotti:2018wxm,Adlarson:2019jtw}. These results have been obtained using collisions of unpolarized electron and positron beams at the c.m. energy corresponding to the $J/\psi$ resonance. The relevant properties of the $J/\psi\to B\overline{B}$ processes are given in Table~\ref{tab:Prod}. Given the relatively large branching fractions and low hadronic background, these $e^+e^-$ experiments are well suited for CPV tests. Two different analysis methods can be used: exclusive measurement (double tag, {DT}) where the decay chains of the baryon and antibaryon are fully reconstructed; inclusive measurement (single tag, {ST}) where only the decay chain of the baryon or antibaryon is reconstructed. For the ST analysis, the two-body production process is uniquely identifiable, and its kinematics fully determined using missing energy/mass technique. Of importance for all single-step weak decays, \textit{e.g.} $\Lambda\to p\pi^-$, is that the $\Lambda$ and $\overline{\Lambda}$ are produced with a transverse polarization. The polarization and the spin correlations allow for a simultaneous determination of $\alpha$ and $\overline{\alpha}$, with the method proposed in Ref.~\cite{Faldt:2017kgy}. 
The currently available results for $J/\psi\to \Lambda\overline{\Lambda}$~\cite{Ablikim:2018zay}, $J/\psi\to \Sigma^+\overline{\Sigma}\vphantom{X}^-$~\cite{Ablikim:2020lpe}  and  $J/\psi\to \Xi^-\overline{\Xi}\vphantom{X}^+$~\cite{Ablikim:2021qkn} use $1.3\times10^9$ $J/\psi$ data with $4.2\times10^5$ (background 400 events), $8.8\times10^4$ (background $4.4\times10^3$ events) and $7.3\times10^4$ (background 200 events) selected DT candidates, respectively. The final state charged particles are measured in the main drift chamber (and the calorimeter for the photons from the $\Sigma^+\to p\pi^0(\to\gamma\gamma)$ decay), where a superconducting solenoid provides the magnetic field for momentum determination of the pions and (anti)protons with an accuracy of 0.5\% at 1.0 GeV/$c$~\cite{BESIII:2009fln}. The pions and protons have distinctly different momentum ranges, making particle identification straightforward in the DT-type measurements.
The analyses of the already collected  $10^{10}$ $J/\psi$ data by BESIII, have not been finished yet, but one can expect a threefold reduction of the statistical uncertainties as shown in Table~\ref{tab:ALCP}.

\begin{table}
  \caption{Properties of the $e^+e^-\to J/\psi\to B\overline B$ decays to the pairs of ground-state octet hyperons. \label{tab:Prod}}
\begin{ruledtabular}
\begin{tabular}{lllll}
  Final state&${\cal B}(\times 10^{-4})$& $\alpha_\psi$&$\Delta\Phi (\text{rad})$&Comment\\ \hline
$\Lambda\overline{\Lambda}$&$19.43(3)$&$\phantom{-}0.461(9)$&$\phantom{-}0.740(13)$&\cite{Ablikim:2018zay,Ablikim:2017tys}\\
$\Sigma^+\overline\Sigma\vphantom{X}^-$&$15.0(24)$&$-0.508(7)$&$-0.270(15)$ &\cite{BES:2008hwe, Ablikim:2020lpe}\\
$\Sigma^-\overline\Sigma\vphantom{X}^+$&\multicolumn{4}{c}{--- no data ---}\\
$\Sigma^0\overline\Sigma^0$&$11.64(4)$&$-0.449(20)$&--&\cite{Ablikim:2017tys}\\
$\Xi^0\overline\Xi\vphantom{X}^0$&$11.65(43)$&$\phantom{-}0.66(6)$&--&\cite{Ablikim:2016sjb}\\
$\Xi^-\overline\Xi\vphantom{X}^+$&$\phantom{0}9.7(8)$&$\phantom{-}0.586(16)$&$\phantom{-}1.213(48)$&\cite{ParticleDataGroup:2020ssz,Ablikim:2021qkn}\\
  \end{tabular}
\end{ruledtabular}
\end{table}

\section{Formalism}
\label{sec:Form}
\subsection{Production process}
\begin{figure}
\centering
\includegraphics[width=1.\columnwidth]{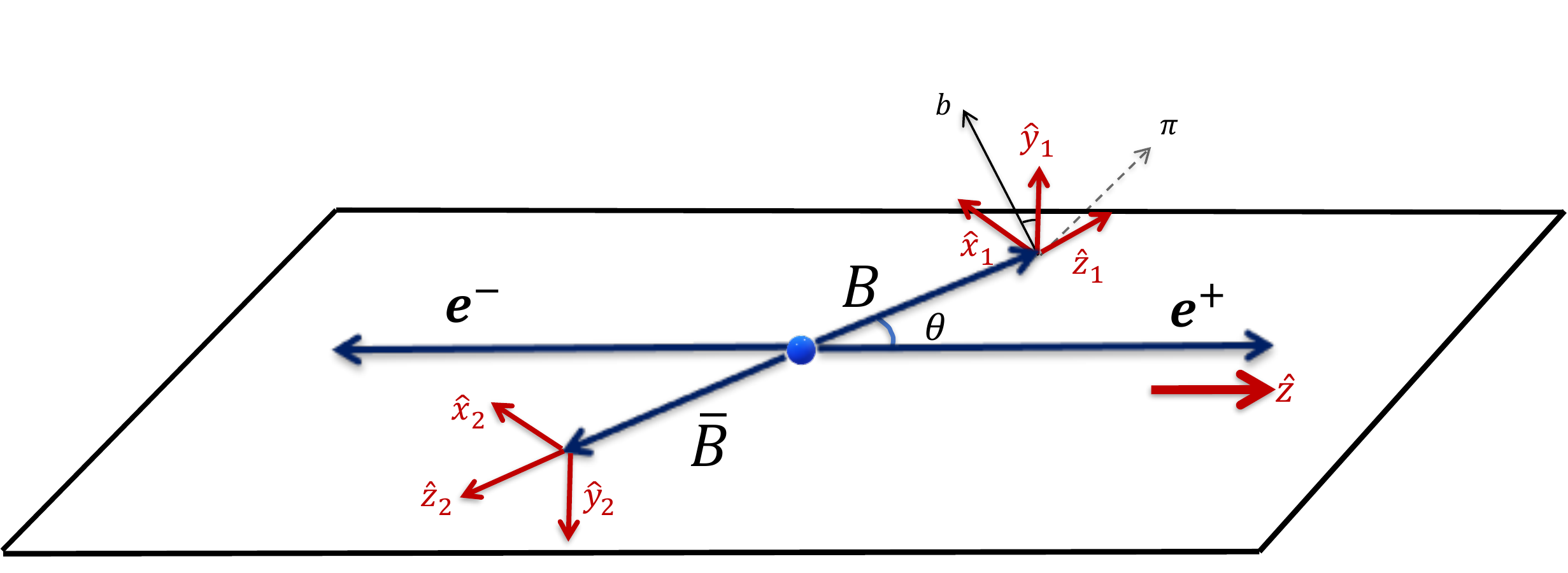}
\caption[]{Orientation of the three coordinate systems used in the
  analysis. The axes in the baryon $B$ and antibaryon $\overline B$ rest (helicity)
  frames are $({\bf\hat x}_1,{\bf\hat y}_1,{\bf \hat z}_1)$ and
  $({\bf\hat x}_2,{\bf\hat y}_2,{\bf\hat z}_2)$, respectively. They are related as 
  $({\bf\hat x}_2,{\bf\hat y}_2,{\bf\hat z}_2)=({\bf\hat x}_1,-{\bf\hat y}_1,-{\bf \hat z}_1)$. In the
  overall c.m. frame, the ${\bf \hat z}$ axis is along the positron
  momentum.}
  \label{fig:axes}
\end{figure}

We start from a description of baryon--antibaryon production in electron--positron annihilations with a polarized electron beam.
The production process $e^+e^-\to B\overline B$, viewed in the c.m.\ frame, defines the $z$ axis which is chosen along the positron momentum shown in Fig.~\ref{fig:axes}. We consider production of  spin-1/2 baryon--antibaryon pair in electron--positron annihilation with longitudinally polarized electron beam. 
Neglecting the electron mass and assuming the one-photon approximation, the helicity of the electron ($\lambda$) and positron ($\overline\lambda$) has to be opposite since the photon only couples right-handed particles to left-handed antiparticles and vice versa. 
The number of right-handed ($n_R$) and left-handed ($n_L$) electrons in the beam with longitudinal polarization $P_e$ is:
\begin{equation}
  n_R=n_-\cdot\frac{1+P_e}{2}\ {\rm and}\   n_L=n_-\cdot\frac{1-P_e}{2}\ ,
\end{equation}
where $n_-=n_R+n_L$ is the total number of electrons. 
The two helicity configurations where the annihilation is possible are $\lambda=+1/2,\overline\lambda=-1/2$ ($\lambda_z=-1$)
and $\lambda=-1/2,\overline\lambda=+1/2$ ($\lambda_z=1$). For the collisions with unpolarized positrons, the relative weights  of the two configurations are $(1+P_e)/2$
and  $(1-P_e)/2$, respectively.
Therefore, the spin density of the initial electron--positron system can be written as:
\begin{eqnarray}
  \rho_1^{i,j}(\theta) := 
  \frac{1+P_e}{2}{d}_{-1,i}^{1*}(\theta) \, {d}^1_{-1,j}(\theta)
+ \frac{1-P_e}{2}{d}_{1,i}^{1*}(\theta) \, {d}^1_{1,j}(\theta)
  \, 
  \label{eq:defrho0init}  
\end{eqnarray}
where the quantization axis along the $B$ momentum. 
 The density matrix for the production process is the sum of the contributions from the two helicities,
see Eq.~(14) in Ref.~\cite{Perotti:2018wxm}: 
\begin{eqnarray}
  \rho^{\lambda_1,\lambda_2;\lambda_1',\lambda_2'}_{B\overline B} \propto A_{\lambda_1,\lambda_2} \, A^*_{\lambda'_1,\lambda'_2} \, \rho_1^{\lambda_1-\lambda_2,\lambda'_1-\lambda'_2}(\theta) 
  \label{eqn:amp}  
\end{eqnarray}
with the reduced density matrix $\rho_1$ given by
\begin{equation}
  \frac{1}{2}\left(
\begin{array}{ccc}
 \frac{1+\cos^2\!\theta}{2}\!-\!P_e\cos\!\theta & 
\frac{(P_e-\cos\theta)\sin\theta}{\sqrt{2}} & \frac{\sin ^2\!\theta}{2} \\[0.5em]
\frac{(P_e-\cos\theta)\sin\theta}{\sqrt{2}} & 
\sin ^2\!\theta & \frac{(P_e+\cos\theta)\sin\theta}{\sqrt{2}} \\[0.5em]
\frac{\sin ^2\!\theta}{2} & \frac{(P_e+\cos\theta)\sin \theta}{\sqrt{2}}  &
\frac{1+\cos^2\!\theta}{2}\!+\!P_e\cos\!\theta  \\
\end{array}
\right)  \,.
\label{eq:explicit-init}
\end{equation}
The four \textit{a priori} possible helicity amplitudes reduce to only two,  $\F_1:=A_{-1/2,-1/2}=A_{1/2,1/2}$ and  $\F_2:=A_{1/2,-1/2}=A_{-1/2,1/2}$. 
Disregarding the overall normalisation the magnitude of the two form factors can be represented as  $|\F_1|=\cos\!\chi$ and $|\F_2|=\sqrt{2}\sin\!\chi$, where $0\le\chi\le\pi/2$. In addition,
the relative phase between the form
factors is $\Delta\Phi:=\arg(\F_1/\F_2)$. 
The general expression for the joint density matrix of the $B\overline B$ pair is:
\begin{equation}
\rho_{B\overline B}=\sum_{\mu,\nu=0}^{3}C_{\mu\nu}\, \sigma_\mu^{B}\otimes{\sigma}_{\nu}^{\overline B}\ ,
\label{eqn:sig12}
\end{equation}
where a set of four Pauli matrices $\sigma_\mu^{B}(\sigma_\nu^{\overline B})$ in the  $B({\overline B})$ rest frame is used and $C_{\mu\nu}$ is a $4\times 4$ real matrix representing polarizations and spin correlations of the baryons. The orientation of the coordinate systems in the baryon rest frames is defined in Fig.~\ref{fig:axes}.
The axes are denoted
${\bf\hat x}_1,{\bf\hat y}_1,{\bf \hat z}_1$ and ${\bf\hat x}_2,{\bf\hat y}_2,{\bf\hat z}_2$.
The elements of  the $C_{\mu\nu}$ matrix
  are functions of the production angle $\theta$ of the $B$ baryon: 
\begin{widetext}
\begin{equation}
\frac{3}{3+\alpha_\psi}
\cdot\left(
\begin{array}{cccc}
 1\!+\!\alpha_\psi  \cos ^2\!\theta & \gamma_\psi P_e\sin\theta & {\beta_\psi  {\sin\theta\cos\theta}} & (1+\alpha_\psi) P_e\cos\theta\\
 \gamma_\psi P_e\sin\theta& \sin ^2\!\theta & 0 & {\gamma_\psi  {\sin\theta\cos\theta}}  \\
 -{\beta_\psi  {\sin\theta\cos\theta}}  & 0 & \alpha_\psi  \sin ^2\!\theta  & -{\beta_\psi  P_e\sin\theta} \\
 -(1+\alpha_\psi) P_e\cos\theta & -{\gamma_\psi  {\sin\theta\cos\theta}}  &  -{\beta_\psi  P_e\sin\theta}& -\alpha_\psi\!-\!\cos ^2\!\theta  \\
\end{array}
\right),\label{eqn:c1212p}
\end{equation}
where the real parameters $\alpha_\psi$, $\beta_\psi$ and $\gamma_\psi$ are defined in terms of the parameters $\chi$ and $\Delta\Phi$ as:
\begin{equation}
\alpha_\psi:= -\cos(2\chi) 
 \ , \ \beta_\psi: = \sin(2\chi) \sin(\Delta\Phi) \ , \ \gamma_\psi: = \sin(2\chi) \cos(\Delta\Phi)
 \end{equation}
and $\alpha_\psi^2+\beta_\psi^2+\gamma_\psi^2=1$. The $B$-baryon  angular distribution is
\begin{equation}
    \frac{1}{\sigma}\frac{\dd\sigma}{\dd\Omega_B}=
    \frac{3}{4\pi}\frac{ 1+\alpha_\psi\cos^2\!\theta}{3+\alpha_\psi}\ . \label{eq:dSigdOm}
\end{equation}
This relation determines the normalisation factor in Eq.~\eqref{eqn:c1212p}.
The $B$-baryon polarization vector ${\bf P}_B$ defined in the rest frame of baryon $B$, coordinates $({\bf\hat x}_1,{\bf\hat y}_1,{\bf \hat z}_1)$, is:
\begin{equation}
  {\bf P}_B:=\frac{C_{10}{\bf\hat x}_1+C_{20}{\bf\hat y}_1+C_{30}{\bf\hat z}_1}{C_{00}}=\frac{\gamma_\psi P_e\sin\theta{\bf\hat x}_1 - {\beta_\psi  {\sin\theta\cos\theta}}{\bf\hat y}_1-(1+\alpha_\psi) P_e\cos\theta{\bf\hat z}_1}{1+\alpha_\psi\cos^2\theta} \ . \label{eq:PolB}
\end{equation}
In the chosen helicity frames one has $C_{01}=C_{10}$, $C_{02}=-C_{20}$, $C_{03}=-C_{30}$ and  ${\bf P}_{\overline{B}}=({C_{01}{\bf\hat x}_2+C_{02}{\bf\hat y}_2+C_{03}{\bf\hat z}_2})/C_{00}$. Therefore, the polarization vectors of the baryon and the antibaryon are equal and have the same direction, ${\bf P}_{\overline{B}}={\bf P}_B$.
In the limit of large c.m.\ energies (HE), where $\alpha_\psi= 1$ and $\beta_\psi=\gamma_\psi= 0$ \cite{Brodsky:1981kj}, the baryon can only have the longitudinal polarization component ${\bf P}_B{\bf\hat z}_1={2P_e\cos\theta}/{(1+\cos^2\theta)}$. In the low energy (LE) limit (close
to threshold) $\alpha_\psi= 0$ and $\Delta\Phi=0$, implying $\beta_\psi= 0$, $\gamma_\psi=1$ and ${\bf P}_B=P_e({\sin\theta{\bf\hat x}_1+\cos\theta{\bf\hat z}_1})$. Therefore, the value of the baryon polarization is equal to the initial electron beam polarization in this case. 
Fig.~\ref{fig:polar} shows the production-angle dependence of the baryon-polarization magnitude  in the $\eeLL$,  $e^+e^-\to J/\psi\to\Xi^-\overline\Xi\vphantom{X}^+$ and $e^+e^-\to J/\psi\to\Sigma^+\overline\Sigma\vphantom{X}^-$ processes for three different values of the electron-beam polarization. 
The values of the $\alpha_\psi$ and $\Delta\Phi$ parameters from Table~\ref{tab:Prod} are used.

For the determination of the uncertainties of the CPV tests, the following tensor $\braket{ C^2}_{\mu\nu}$ representing properties of the production process will be needed:
\begin{align}
     \braket{ C^2}_{\mu\nu}&:=\frac{1}{4\pi}\int \frac{C_{\mu\nu}^2}{C_{00}}\dd\Omega_B
     =\frac{1}{2}\int_{-1}^{1}\frac{C_{\mu\nu}^2}{C_{00}}\dd\!\cos\theta\ .\label{eq:ProdTensor}
\end{align}
The production tensor is symmetric and positively defined. In addition $\braket{ C^2}_{00}=1$. For example, it can be used to express the mean-squared polarization $\braket{{\bf P}_B^2}$ of the $B$-baryon defined as:
\begin{equation}
\begin{split}
   \braket{{\bf P}_B^2} =& \int{\bf P}_B^2 \left(\frac{1}{\sigma}\frac{\dd\sigma}{\dd\Omega_B}\right)\dd\Omega_B=\sum_{i=1}^3\braket{ C^2_{i0}} \ . \label{eq:polAVG}
\end{split}
\end{equation}
This integral can be calculated exactly, and the result expressed as a linear function of the electron polarization squared $P_e^2$
\begin{equation}
   \braket{{\bf P}_B^2} =\mathbb{p}_0+\mathbb{p}_2P_e^2\ , \label{eq:PBparam}
\end{equation}
where the expression for coefficients $\mathbb{p}_0$ and $\mathbb{p}_2$ are given in Appendix~\ref{app:avpol}. As we will show later, $\braket{{\bf P}_B^2}$ determines the uncertainty of the $A_{\CP}$ and $\Phi_{\CP}$ measurement. The results for $\sqrt{\braket{{\bf P}_B^2}}$ are shown in Fig.~\ref{fig:apolar}.
\begin{figure}
\centering
\includegraphics[width=\textwidth]{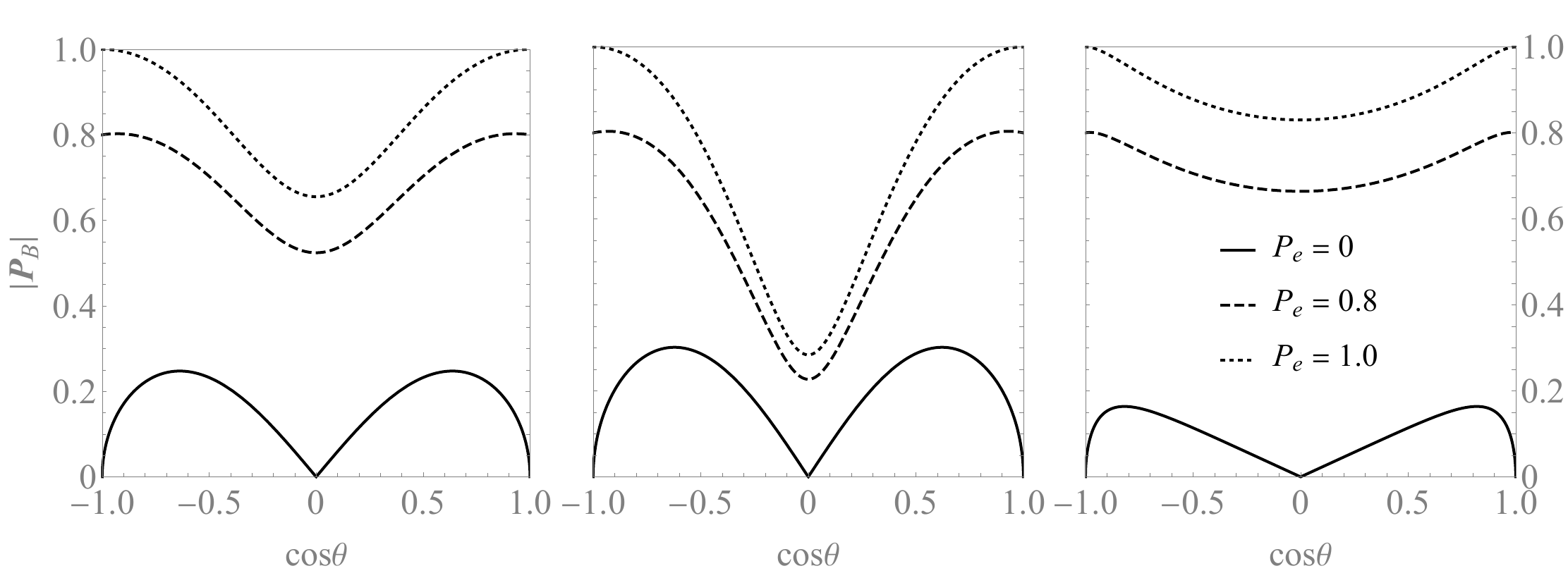}
\caption[]{ Magnitudes of the hyperon polarization as a function of the production angle for:
(a) $\Lambda$, (b) $\Xi^-$ and (c) $\Sigma^+$ for the electron beam polarizations $P_e=0$, $0.8$, $1$ (solid, dashed and dotted lines, respectively).
  The  $\alpha_\psi$ and $\Delta\Phi$ values are taken from Table~\ref{tab:Prod}.
  \label{fig:polar}}
\end{figure}
\begin{figure}
\centering
\includegraphics[width=1.0\textwidth]{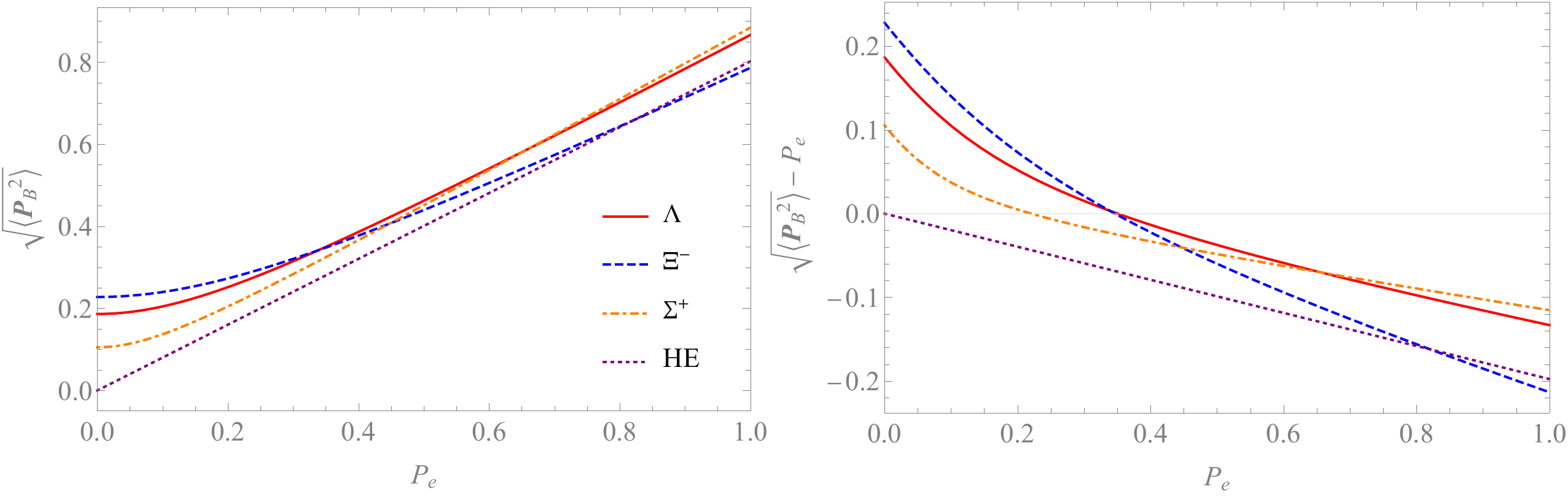}
\put(-400,120){\large(a)}   
\put(-120,120){\large(b)}    
\caption[]{ Average  polarization 
$\sqrt{\braket{{\bf P}_B^2}}$ for: 
 $\Lambda$ (solid line),  $\Xi^-$ (dashed line), $\Sigma^+$ (dot-dashed line) and high-energy limit (dotted line) as a function of electron beam polarization. In panel (b) the quantity $\sqrt{\braket{{\bf P}_B^2}}-P_e$ is plotted to facilitate a more precise comparison. The low-energy limit corresponds to $P_B=P_e$.
}
  \label{fig:apolar}
\end{figure}
We will use the following notation for the  polarization and spin-correlation contributions of the production-process tensor:
\begin{align}
  \braket{\mathbb{P}^2_B}&:=\sum_{i=1}^3\left(\braket{ C^2}_{i0}+\braket{ C^2}_{0i}\right)=2\braket{\mathbf{P}^2_B}\nonumber \\ 
  \braket{\mathbb{S}_{B\overline B} ^2}&=\sum_{i,j=1}^3\braket{ C^2}_{ij}\ .\label{eq:PolandSc}
\end{align}
The values of the $\braket{\mathbb{P}^2_B}$ and
$\braket{\mathbb{S}_{B\overline B} ^2}$ terms as function of $P_e$ are shown in Fig.~\ref{fig:PandSterms} for some processes which are discussed later. The dependence on the $P_e$ is much stronger for the polarization terms  than for the spin-correlation terms. As we will show in Secs.~\ref{sec:SSdecay} and~\ref{sec:DSdecay} the sizes of the contributions determine the precision of the CPV observables.
\begin{figure}
    \centering
\includegraphics[width=0.32\textwidth]{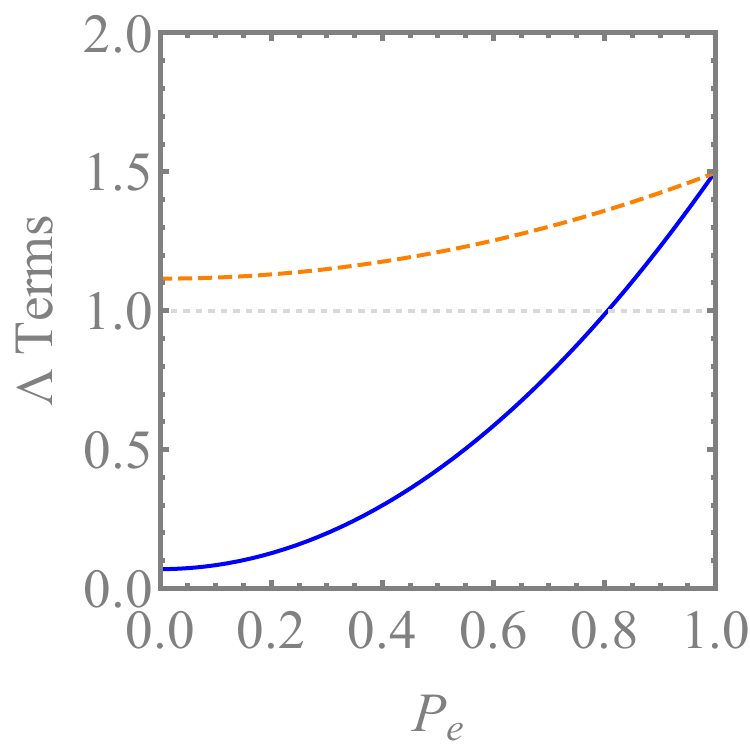}
\put(-115,120){\large(a)}    \includegraphics[width=0.32\textwidth]{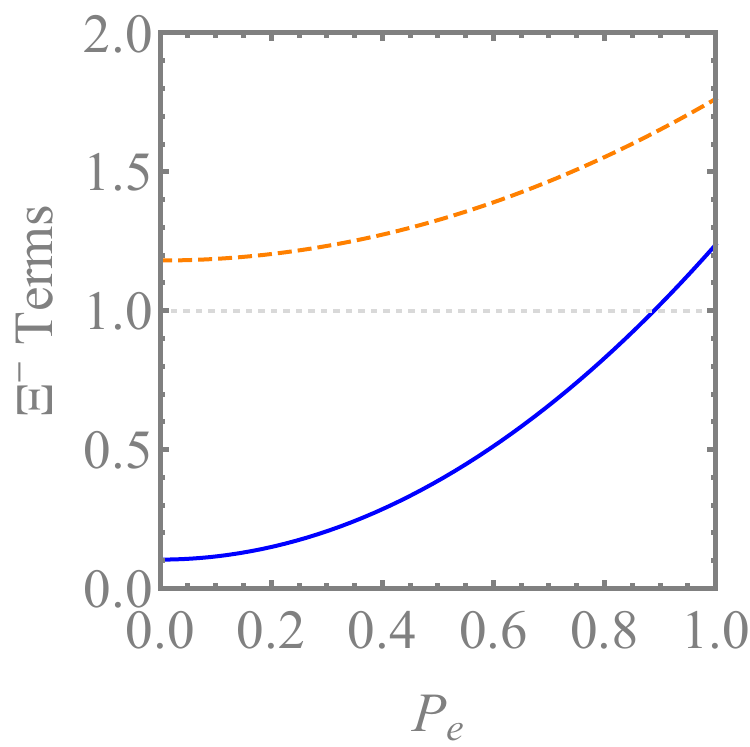}
\put(-115,120){\large(b)}    \includegraphics[width=0.32\textwidth]{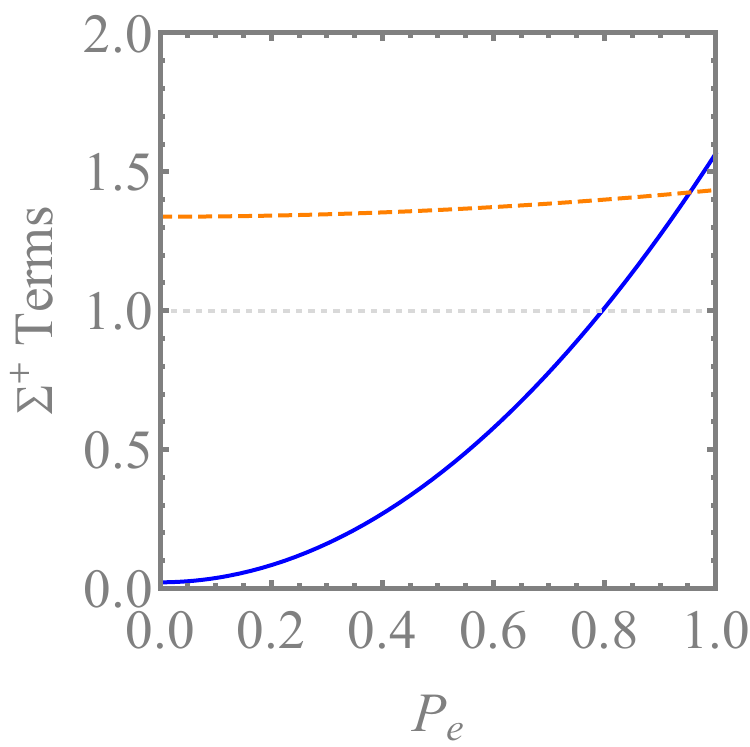}
\put(-115,120){\large(c)}    

\includegraphics[width=0.32\textwidth]{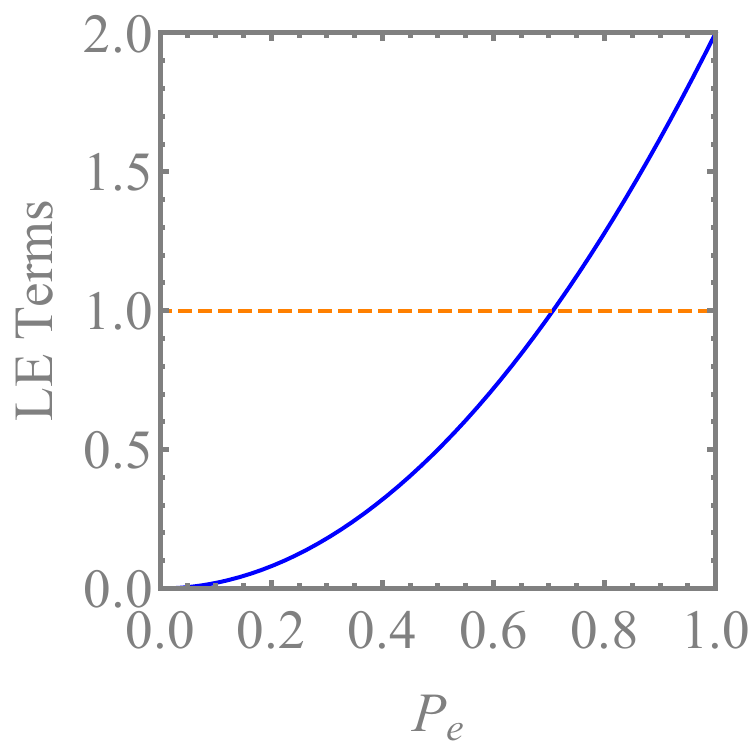}
\put(-110,100){\large(d)}    \includegraphics[width=0.44\textwidth]{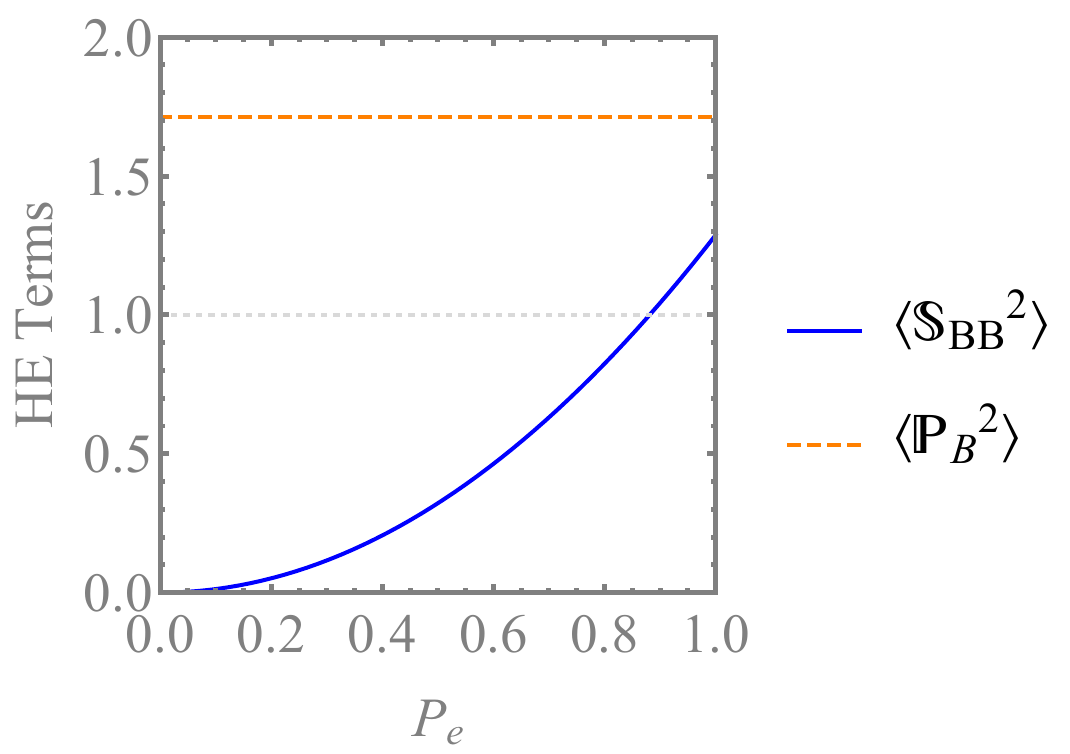}
\put(-110,100){\large(e)}    
\caption{Polarization $\braket{\mathbb{P}_B^2}$ (solid lines) and spin-correlation terms $\braket{\mathbb{S}_{B\overline B}^2}$ (dashed lines) of the $e^+e^-\to B\overline B$ processes: (a) $J/\psi\to\Lambda\overline\Lambda$, (b) $J/\psi\to\Xi\overline\Xi$, (c) $J/\psi\to\Sigma\overline\Sigma$, (d) low-energy limit and (e) high-energy limit.    }
    \label{fig:PandSterms}
\end{figure}

\subsection{Joint angular distributions}

The complete joint angular distributions for a production process $e^+e^-\to B\overline{B}$ followed by weak two-body decays of the  hyperon $B$ and the antihyperon $\overline B$ can be obtained using the modular framework from  Ref.~\cite{Perotti:2018wxm}.
For a single-step decay
$D(B\!\to\! b\pi)$ and the corresponding c.c. decay mode $\overline{D}(\overline B\!\to\! \overline b\overline\pi)$, like $\eeLL$ with $\Lambda \to p \pi^-$ and 
$\overline{\Lambda} \to \overline{p} \pi^+$, the joint angular distribution,
\begin{equation}
    {\cal{P}}^{D\overline D}(\boldsymbol{\xi};\boldsymbol{\omega}):=\frac{1}{\Gamma}\frac{\dd\Gamma}{\dd\boldsymbol{\xi}}\ ,
\end{equation}
is 
\begin{equation}
  {\cal{P}}^{D\overline D}(\boldsymbol{\xi};\boldsymbol{\omega})=
  \frac{1}{(4\pi)^3}\sum_{\mu,\nu=0}^{3}C_{\mu\nu}(\Omega_B;\alpha_\psi,\Delta\Phi,P_e) a_{\mu0}^{D}(\Omega_{b}; \alpha_D)
    a_{\nu0}^{\overline D}(\Omega_{\overline{b}}; \overline\alpha_{D})\ .\label{eq:LaLa}
\end{equation}
The production is described by the
spin-correlation matrix $C_{\mu\nu}(\Omega_B;\alpha_\psi,\Delta\Phi,P_e)$ in  Eq.~\eqref{eqn:c1212p} and the $4\!\times\!4$ decay matrices $a_{\mu0}^{D}:=
a_{\mu0}^D(\Omega_{b}; \alpha_D)$ and
$a_{\nu0}^{\overline{D}}:= a_{\nu0}^{\overline{D}}(\Omega_{\overline{b}}; \overline\alpha_D)$. 
The decay matrices $a^D_{\mu\nu}$ represent the transformations of the spin operators (Pauli matrices) $\sigma^{B}_\mu$ and  $\sigma_\nu^{b}$ defined in the $B$
and $b$ baryon helicity frames, respectively~\cite{Perotti:2018wxm}:
\begin{equation}
\sigma^{B}_\mu\to\sum_{\nu=0}^3a_{\mu\nu}^{D}\sigma_\nu^{b}\ . \label{eq:decay}
\end{equation}
The helicity reference frame for the daughter-baryon $b$ is defined in the following way. In the $B$ rest frame with the $z$ axis defined by the unit vector ${\bf \hat z}_B$, the  direction of the $b$ momentum  is  denoted as ${\bf\hat p}_b$. The $b$-baryon helicity system is the $b$ rest frame
where the orientation of the Cartesian coordinate system is given by the unit vectors:
\begin{align}
   {\bf \hat x}_b&= \left.\frac{{\bf \hat z}_B \times {\bf\hat p}_b}{|{\bf \hat z}_B \times {\bf\hat p}_b|}\right.\times {\bf\hat p}_b,\ \
    {\bf \hat y}_b=  \frac{{\bf \hat z}_B \times {\bf\hat p}_b}{|{\bf \hat z}_B \times {\bf\hat p}_b|} \ \ {\rm and}\ \
   {\bf \hat z}_b=  {\bf \hat p}_b\ . \label{eq:helicity}
\end{align}
The explicit form of the 
$a^D_{\mu\nu}(\Omega;\alpha_{D},\beta_D,\gamma_D)\leftrightarrow a^D_{\mu\nu}(\left\{\theta,\varphi\right\};\alpha_{D},\beta_D,\gamma_D)$ matrix, representing the polarization vector transformation from Eq.~\eqref{eq:LeePol} in our framework, is:
\begin{equation}
\left(
\begin{array}{cccc}
 1 & 0 & 0 & \alpha_D \\
 \alpha_D \sin\theta \cos\varphi & \gamma_D  \cos\theta \cos\varphi-\beta_D \sin\varphi& -\beta_D \cos\theta \cos\varphi-\gamma_D  \sin\varphi & \sin \theta \cos\varphi \\
 \alpha_D \sin\theta \sin\varphi& \beta_D \cos\varphi+\gamma_D  \cos\theta \sin\varphi & \gamma_D  \cos\varphi-\beta_D \cos\theta \sin\varphi & \sin \theta\sin\varphi \\
 \alpha_D \cos\theta & -\gamma_D  \sin\theta & \beta_D \sin\theta & \cos \theta \\
\end{array}
\right)  \ .
\end{equation}
For the single-step processes only the first column $a_{\mu0}(\Omega;\alpha_{D})$ is used and it depends only on the decay parameter $\alpha_D$. The vector $\boldsymbol{\xi}:=(\Omega_B,\Omega_{b},\Omega_{\overline{b}})$
represents a complete set of the kinematic variables describing a single-event configuration in the six-dimensional phase space.
We use {\it helicity angles} to parameterize the multidimensional phase space. These are spherical coordinates defined in the helicity systems in Eq.~\eqref{eq:helicity}. 
There are five global parameters to describe the complete angular distribution, and they are represented by the vector $\boldsymbol{\omega}:=(\alpha_\psi,\Delta\Phi,P_e,\alpha_{D}, \overline{\alpha}_{D})$.

For the processes with two-step decays
like $e^+e^-\to \Xi\overline{\Xi}$ with $\Xi \to\Lambda \pi$,
$\Lambda \to p \pi^-$ $+$ c.c. the joint angular distribution reads:
\begin{equation}
 {\cal{P}}^{\Xi\overline \Xi}(\boldsymbol{\xi}_{\Xi\overline\Xi};\boldsymbol{\omega}_\Xi)=
  \frac{1}{(4\pi)^5}\sum_{\mu,\nu=0}^{3}C_{\mu\nu}\left(\sum_{\mu'=0}^{3} a_{\mu\mu'}^{\Xi} a_{\mu'0}^{\Lambda}\right)\left(
\sum_{\nu'=0}^{3}    a_{\nu\nu'}^{\overline \Xi} 
  a_{\nu'0}^{\overline \Lambda}\right)\ ,\label{eqn:XiXi}
\end{equation}
where 
$\boldsymbol{\xi}_{\Xi\overline\Xi}:=(\Omega_\Xi,\Omega_{\Lambda},\Omega_{\overline{\Lambda}},\Omega_{p},\Omega_{\overline{p}})$ and  $\boldsymbol{\omega}_\Xi:=(\alpha_\psi,\Delta\Phi,P_e,\alpha_{\Xi}, \overline{\alpha}_{\Xi},\phi_{\Xi}, \overline{\phi}_{\Xi},\alpha_{\Lambda}, \overline{\alpha}_{\Lambda})$ --- the phase space has 10 dimensions and there are 9 global parameters.
\end{widetext}

The single tag (ST) distributions are obtained by integrating out the unmeasured variables. For example, the ST angular distribution of the $B$ baryon measurement for single sequence decays Eq.~\eqref{eq:LaLa} is:
\begin{equation}
  \begin{split}
    {\cal{P}}^{D}(\boldsymbol{\xi}_b;\boldsymbol{\omega})=\frac{1}{(4\pi)^2}
    \sum_{\mu=0}^{3}C_{\mu0}\cdot a_{\mu0}^{D}
    =\frac{1}{(4\pi)^2}C_{00}\cdot(1+\alpha_D{\bf {P}}_B\cdot {\bf \hat{p}}_b)\ ,\label{eqn:La}
\end{split}
\end{equation}
where $\boldsymbol{\xi}_B:=(\Omega_B,\Omega_b)$ and ${\bf {P}}_B$ is given by Eq.~\eqref{eq:PolB}. 
{As reference for comparing the ST uncertainties to the DT measurements with $N$ reconstructed events, we will use a set of two independent ST experiments where the baryon and antibaryon decays are analysed with $N$ reconstructed events each.}

%
\subsection{Asymptotic maximum likelihood method}
\label{sec:AMLL}
The importance of the individual parameters $\omega_k$ in the joint angular probability density functions (p.d.f.s) of Eqs.~\eqref{eq:LaLa} and \eqref{eqn:XiXi} and their correlations are studied using an ideal asymptotic maximum likelihood method (MLL), discussed in Ref.~\cite{Adlarson:2019jtw}. The method allows one to reliably estimate the statistical accuracy of the determined global parameters in experiments with large acceptance detectors.

The asymptotic expression of the inverse covariance matrix element $kl$ between parameters $\omega_k$ and $\omega_l$ of the  parameter vector $\boldsymbol{\omega}$  is given by the Fisher information matrix~\cite{Fisher:1922aa}:
\begin{equation}
  {\cal I}({\omega_k,\omega_l}):=N\int \frac{1}{ {\cal P}}\frac{\partial {\cal P}}{\partial \omega_k}\frac{\partial {\cal P}}{\partial \omega_l}\dd\boldsymbol{\xi}\ ,
\label{eq:VarMLL}
\end{equation}
where $N$ is the number of events in the final selection.
The calculated values are used to construct the matrix, which is inverted to obtain the covariance matrix $V={\cal I}^{-1}$ for the parameters. 
Since asymptotically, in the case of negligible background, the statistical uncertainties given by the standard deviations (s.d.), $\sigma(\omega_k)$, are inversely proportional to the square root of the number of the reconstructed signal events $N$ we will use  the product
\begin{eqnarray}
  \sigma_C(\omega_k):=\sigma(\omega_k)\times \sqrt{N}\ ,\label{eq:SDCdef}
\end{eqnarray}
and call it {\it s.d. coefficient} or normalised statistical uncertainty.
It allows for a comparison of the precision of different estimators for a given number of reconstructed events.
In most cases, the integral Eq.~\eqref{eq:VarMLL} has to be calculated numerically. However, in this approach the explicit dependence on the production and decay parameters is hidden, and the calculations have to be repeated for each parameter set. Therefore, we have constructed analytic approximations, which are presented and discussed in the two following sections.

\section{Single-step decays}
\label{sec:SSdecay}
We derive an approximate analytic solution for standard deviation of the $A_{\CP}$ measured in a single-step processes described by the p.d.f. in Eq.~\eqref{eq:LaLa}. The straightforward method is to determine all elements of the $5\times5$ inverse covariance matrix corresponding to the parameter vector $\boldsymbol{\omega}=(\alpha_\psi,\Delta\Phi,P_e,\alpha_{D}, \overline{\alpha}_{D})$, invert the matrix and use error propagation to determine the variance $\text{Var}(A_{\CP})$. 
If the parameter vector can be changed to include
the $A_{\CP}$ observable and to have the remaining parameters uncorrelated,
then the variance $\text{Var}({A_{\CP}})$ will be simply given as the inverse of the corresponding information matrix element
\begin{equation}
  \frac{1}{\text{Var}({A_{\CP}})}={\cal I}(A_{\CP}):=N\int \frac{1}{ {\cal P}^{D\overline D}}\left(\frac{\partial {\cal P}^{D\overline D}}{\partial A_{\CP}}\right)^2\dd\boldsymbol{\xi}\ .
\end{equation}
Such parameterization 
can be constructed using  the $\braket{\alpha_{D}}$ and $A_{\CP}$ parameters and expressing  $\alpha_{D}=\braket{\alpha_D}(1+A_{\CP})$ and $\overline\alpha_{ D}=-\braket{\alpha_D}(1-A_{\CP})$. The new  parameter set leads to the following expression for the partial derivative of ${\cal{P}}^{D\overline D}$ with respect to  $A_{\CP}$ (taken at $A_{\CP}=0$)
\begin{align}
  \frac{\partial {\cal{P}}^{D\overline D}}{\partial A_{\CP}}
  &= \frac{\braket{\alpha_D}}{{\cal V}} \sum_{\mu,\nu=0}^{3}C_{\mu\nu}\left( 
  \frac{\partial a_{\mu0}^{D}}{\partial \alpha_D}    a_{\nu0}^{\overline D}+a_{\mu0}^{D} 
 \frac{\partial a_{\nu0}^{\overline D}}{\partial \overline\alpha_{ D}}
\right)\\
  &    = \frac{{\alpha_D}}   {{\cal V}}C_{00}\left( {\bf {P}}_B\cdot {\bf \hat{p}}_b- {\bf {P}}_{\overline B}\cdot {\bf \hat{p}}_{\overline b}\right)\ , \label{eq:V0b}
\end{align}
where ${\cal V}:=\int \dd\boldsymbol{\xi}=\int\dd\Omega_B\dd\Omega_{b}\dd\Omega_{\overline{b}}=(4\pi)^3 $ and  $\braket{\alpha_D}=\alpha_D$ in the $A_{\CP}=0$ limit.
In order to calculate the information ${\cal I}(A_{\CP})$, we will use the following representation for the ${\cal P}$ p.d.f.
\begin{align}
   {\cal{P}}
   (\boldsymbol{\xi};\boldsymbol{\omega})&:=
   C_{00}\frac{1+{\cal G}(\boldsymbol{\xi};\boldsymbol{\omega})}
   {\cal V}\ ,
\end{align}
where $\int{\cal G}\dd\boldsymbol{\xi}=0$ and ${\cal G}\ge-1$. In addition, all terms 
included in the function ${\cal G}$ are multiplied by $\pm\alpha_D$ and for 
small values of $|\alpha_D|$ are suppressed. Therefore, it is not unreasonable to use the 
expansion of  $1/(1+{\cal G})$ to approximate ${1}/{{\cal{P}}}$: 
\begin{align}
    \frac{1}{{\cal{P}}}=\frac{\cal V}{C_{00}}\frac{1}{1+{\cal G}}=\frac{\cal V}{C_{00}}\sum_{i=0}^\infty(-{\cal G})^i
\end{align}
and 
\begin{align}
{\cal I}(\omega_k,\omega_l):={\cal I}_0(\omega_k,\omega_l)+\sum_{i=1}^\infty(-{1})^i\Delta {\cal I}_i(\omega_k,\omega_l)\ 
\end{align}
with
\begin{align}
{\cal I}_0(\omega_k,\omega_l)&:=N\int \frac{\cal V}{C_{00}}\frac{\partial {\cal P}}{\partial \omega_k}\frac{\partial {\cal P}}{\partial \omega_l}\dd\boldsymbol{\xi},\\
    \Delta {\cal I}_i(\omega_k,\omega_l)&:=N\int \frac{\cal V}{C_{00}}{\cal G}^i\frac{\partial {\cal P}}{\partial \omega_k}\frac{\partial {\cal P}}{\partial \omega_l}\dd\boldsymbol{\xi}.
\end{align}

We can always compare this analytic result using one or more terms of the expansion with the full numerical calculations. The hope is that the analytic approximation reproduces main features of the exact solution. If it does, it will facilitate understanding how the uncertainties depend on the global parameters. We start by considering the 
0-th term of the above expansion, ${\cal V}/{C_{00}}$, that  leads to the following information: 
\begin{align}
    {\cal I}_0({A_{\CP}})&=
 {N}\int\frac{{\cal V}}{C_{00}}\left(\frac{\partial {\cal{P}}^{D\overline D}}{\partial A_{\CP}}\right)^2\dd\Omega_B\dd\Omega_{b}\dd\Omega_{\overline{b}}
    \label{eq:V0a}\\
    &= {N}\frac{\alpha_D^2}{{\cal V}}\int C_{00}\left( {\bf {P}}_B\cdot {\bf \hat{p}}_b- {\bf {P}}_{\overline B}\cdot {\bf \hat{p}}_{\overline b}\right)^2\dd\Omega_B\dd\Omega_{b}\dd\Omega_{\overline{b}}\ .
\end{align}
Integration over $\Omega_{b}$ and $\Omega_{\overline{b}}$ simplifies due to orthonormality 
\begin{align*}
    \int\left( {\bf {P}}_B\cdot {\bf \hat{p}}_b- {\bf {P}}_{\overline B}\cdot {\bf \hat{p}}_{\overline b}\right)^2 \frac{\dd\Omega_{b}}{4\pi}\frac{\dd\Omega_{\overline{b}}}{4\pi}&=
  \int ({\bf {P}}_B\cdot {\bf \hat{p}}_b)^2\frac{\dd\Omega_{b}}{4\pi}
  + \int ({\bf {P}}_{\overline B}\cdot {\bf \hat{p}}_{\overline b})^2
  \frac{\dd\Omega_{\overline{b}}}{4\pi}  \\ 
    &=\frac{{\bf {P}}_{B}^2}{3}+\frac{{\bf {P}}_{\overline B}^2}{3}=\frac{2}{3}{\bf {P}}_{B}^2 \ .
\end{align*}
Inserting the result into Eq.~\eqref{eq:V0b} and Eq.~\eqref{eq:V0a} we have:
\begin{align}
       {\cal I}_0({A_{\CP}})&=
 \frac{N}{{4\pi}}{\alpha_D^2}\frac{2}{3}   
 \int{\bf {P}}_B^2C_{00}\ \dd\Omega_B\nonumber\\
 &=\frac{2N}{3}{\alpha_D^2}
\int{\bf P}_B^2 \left(\frac{1}{\sigma}\frac{\dd\sigma}{\dd\Omega_B}\right) \dd\Omega_B=\frac{2N}{3}{\alpha_D^2}
\braket{{\bf P}_B^2}
 \ .
\end{align}
Therefore in this approximation the information is proportional to the $B$-baryon average squared polarization, as defined in Eq.~\eqref{eq:polAVG}.
Since $A_{\CP}$ is not correlated with other variables, the  0-th approximation for the uncertainty is
\begin{equation}
    \sigma(A_{\CP})\sqrt{N}=\sigma_C(A_{\CP})\approx \sqrt{\frac{3}{2}}\frac{1}{\alpha_D\sqrt{\braket{{\bf P}_B^2}}} \label{eq:VACP01}\ .
\end{equation}
\begin{figure}
    \centering
    \includegraphics[width=1.0\textwidth]{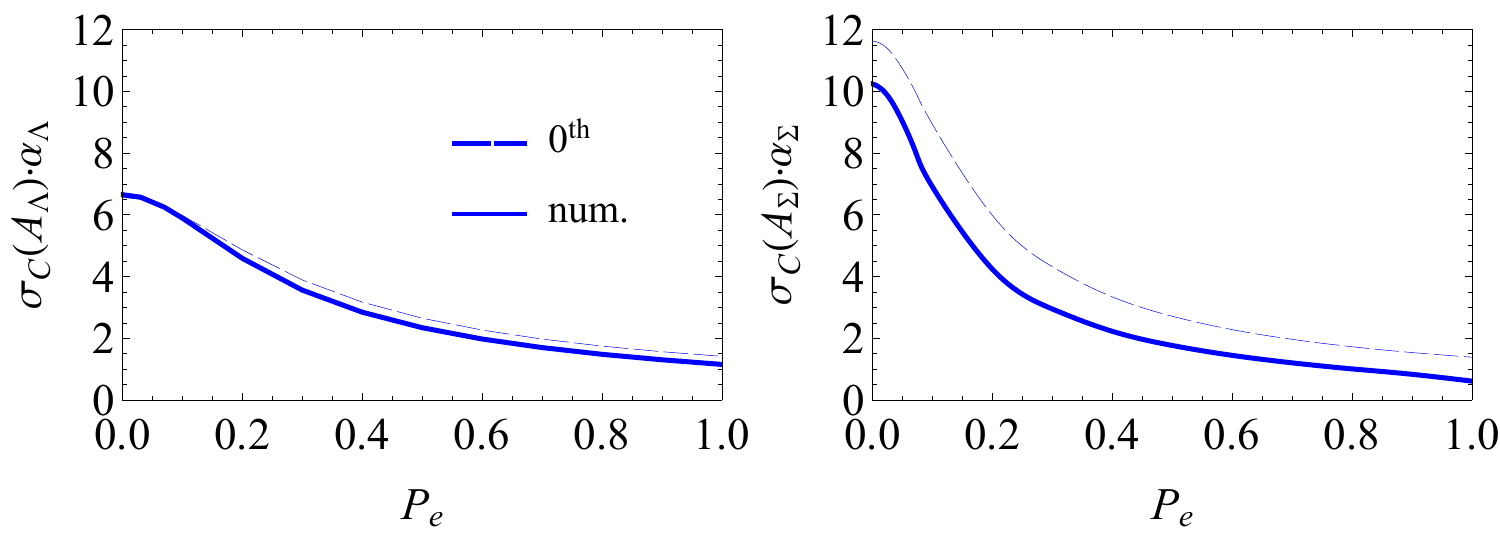}
    \put(-420,120){\large(a)}   
    \put(-140,120){\large(b)}  
    \caption[]{Standard deviation coefficients for $A_{\CP}$, $\sigma_C(A_{\CP})$, multiplied by the  decay parameter value $\alpha_D$ for DT measurements.
    (a) $\eeLL$ with decay $\Lambda(\Lambda\to  p\pi^-)$, (b) $e^+e^-\to J/\psi\to\Sigma^+\overline\Sigma\vphantom{\Sigma}^-$ with decay $\Sigma p(\Sigma^+\to  p\pi^0)$. Dashed lines are the approximations using Eq.~\eqref{eq:VACP01} and solid lines are the exact numerical results. }
    \label{fig:LaSi}
\end{figure}
\noindent Fig.~\ref{fig:LaSi}(a) shows the s.d. coefficients, $\sigma_C(A_{\CP}^{[\Lambda p]})$, multiplied by the $\alpha_\Lambda$ parameter value for the $\eeLL$ processes. The 0-th order result (hereafter we will call it also the analytic approximation) is close to the numerical full result in Eq.~\eqref{eq:VarMLL}, even if $\alpha_\Lambda$ is relatively large ($0.75$). This shows that  the influence of the higher order terms is low for the $A_{\CP}^{[\Lambda p]}$ determination.

We also compare the approximate analytic formula to the full numerical calculations for the $e^+e^-\to J/\psi\to\Sigma^+\overline{\Sigma}\vphantom{\Sigma}^-$  process, where both $\alpha_\psi$ and $\Delta\Phi$ have been measured by BESIII \cite{Ablikim:2020lpe}. The two
$\Sigma^+$ decay modes  $\Sigma n(\Sigma^+\to  n\pi^+)$ and
$\Sigma p(\Sigma^+\to  p\pi^0)$ are interesting as the limiting cases for the expansion since 
$\alpha_{\Sigma n}=0.068\approx0$ and $\alpha_{\Sigma p}=-0.994\approx-1$, respectively. It is worth noting that in the $\Delta I=1/2$ limit $|\alpha_{\Sigma p}|<\cos(\delta^P_1-\delta^S_1)\approx0.980$ (see  Eq.~\eqref{eq:SPone} and the values of the strong phase shifts in Table~\ref{tab:Sphases}). We note that the recent BESIII value  $\braket{\alpha_{\Sigma p}}=-0.994(4)$ (Table~\ref{tab:decayproperties}) violates this bound. 
A proper interpretation of this result requires that all isospin contributions to the $\Sigma^+$ decays are considered, but such discussion  is beyond the scope of this report. The 0-th approximation for $\sigma_C(A_{\CP}^{\Sigma p})\cdot|\alpha_{\Sigma p}|$ is 
given by the dashed line in Fig.~\ref{fig:LaSi}(b). The 
full numerical result (given by the solid line) differs significantly. The difference comes from the spin-correlation contributions, but the analytic approximation is able to describe the overall trend.
From Eq.~\eqref{eq:VACP01} it is clear that the approximation for $\sigma_C(A_{\CP}^{\Sigma n})\alpha_{\Sigma n}$ is 
also given by the same dashed line. As expected, the full numerical result coincides with the 0-th approximation in this case. 
Comparing the trends for $\Lambda$ and $\Sigma^+$, the faster decrease of the uncertainty for $\Sigma^+$ is mainly due to the low value of the $\Delta\Phi$ phase for this reaction. In principle, this would make $\Sigma p$ an attractive decay mode for testing CP symmetry with a polarized electron beam. However, we will not discuss further the $\Sigma$-baryon decays in this report. The reason is that the predicted CPV effects are significantly smaller, $A_{\CP}^{\Sigma p}\cdot\alpha_{\Sigma p}\approx 3.5\times 10^{-6}$  and $A_{\CP}^{\Sigma n}\cdot\alpha_{\Sigma n}\approx 2.7\times 10^{-5}$
\cite{Tandean:2002vy}, and the isospin structure of the amplitudes is more complicated (since also $\Delta I=5/2$ transitions contribute). 

The result for $\sigma_C{(A_{\CP}^D})$ in the DT and ST cases is the same when the ST analysis is done under assumption that the $\braket{\alpha_D}$ value is known and fixed. In a single-step decay, an ST measurement only allows for a determination of the products
$\alpha_D\sqrt{\braket{P^2_B}}$ and $\overline\alpha_D\sqrt{\braket{P^2_{\overline{B}}}}$. Therefore, a $A_{\CP}^D$ determination using a combination of baryon--antibaryon ST measurements requires knowledge of the polarization through some other means or use a production process where ${\braket{P^2_B}}={\braket{P^2_{\overline B}}}$ is assured. For an $e^+e^-\to B\overline{B}$ experiment with an electron beam polarization $P_e$ where the ST data are collected simultaneously and with c.c. symmetric detector acceptance, this condition is fulfilled automatically.

Related to this discussion is a proposal given in Ref.~\cite{Bigi:2017eni} where it is suggested that one could use a triple vector product to determine $A_{\CP}$ even if $\Delta\Phi=0$ and $P_e=0$, \textit{i.e.} the baryons are  unpolarized. For a general baryon--antibaryon state with polarization terms set to zero, the angular distribution after single-step decays reads:
\begin{equation}
  {\cal P}^{D\overline{D}} 
  \propto
  C_{00}+\alpha_D\overline\alpha_D\sum_{i,j=1}^3
  C_{ij} \left[\frac{a_{i0}^{D\vphantom{\overline{D}}}}{\alpha_D}\right]
    \left[\frac{a_{j0}^{\overline D}}{\overline\alpha_D}\right]=:
  C_{00}+\alpha_D\overline\alpha_D {\cal F}(\Omega_B,\Omega_{b},\Omega_{\overline{b}}) \ ,
\end{equation}
where $ {\cal F}(\ldots)$ is a function of the kinematic variables only.
Therefore, the p.d.f.\ is described by a single global parameter $\alpha_{D\overline{D}}:=\alpha_D\overline\alpha_D=-\braket{\alpha_D}^2(1-A_{\CP}^2)$. The parameter is related to $A_{\CP}$ and can  in principle be used to test CP symmetry, but the method has several drawbacks. The information for $\alpha_{D\overline{D}}$ measurement is
$
{\cal I}_0(\alpha_{D\overline{D}})={N}/{9}\braket{\mathbb{S}^2}
$
and the  uncertainty of $A_{\CP}$ from the error propagation  is:
\begin{equation*}
\sigma(A_{\CP})=\frac{1}
{A_{\CP}}\sqrt{\sigma^2(\braket{\alpha_D})+\frac{\sigma^2(\alpha_{D\overline{D}})}{4\braket{\alpha_D}^2}}\ ,
\end{equation*}
which requires an independent determination
of $\braket{\alpha_D}$. 
A meaningful CP test is possible only if $\sigma(A_{\CP})<1$. This requires that the  $\sigma(\braket{\alpha_D})$ precision is better than ${\cal O}(10^{-5})$, since $A_{\CP}\sim{\cal O}(10^{-5})$ in the SM.  If $\sigma(\braket{\alpha_D})$ is not small enough, 
the $A_{\CP}\ne0$ value can be interpreted as a $A_{\CP}$ null result but with 
the decay parameters $\alpha_D$ and $\overline{\alpha}_{D}$ reduced by the  factor $\sqrt{1-A_{\CP}^2}$. 

\section{Two-step decays}
\label{sec:DSdecay}

In order to study uncertainties of the CP asymmetries in $e^+e^-\to\Xi^-\overline{\Xi}^+$, we 
rewrite Eq.~\eqref{eqn:XiXi} as
\begin{equation}
 {\cal{P}}^{\Xi\overline \Xi}(\boldsymbol{\xi}_{\Xi\overline\Xi};\boldsymbol{\omega}_\Xi)=
  \frac{1}{\cal V}\sum_{\mu,\nu=0}^{3}C_{\mu\nu} {\cal D}^\mu_\Xi \overline{\cal D}^\nu_\Xi
\end{equation}
using the following  notation:
\begin{align*}
    {\cal D}^\mu_\Xi&:={\cal D}^\mu(\Omega_{\Lambda},\Omega_{p};\alpha_{\Xi},\phi_{\Xi}, \alpha_{\Lambda}):=\sum_{\mu'=0}^{3} a_{\mu\mu'}^{\Xi} a_{\mu'0}^{\Lambda},\\
    \overline{\cal D}^\mu_\Xi&:={\cal D}^\mu(\Omega_{\overline\Lambda},\Omega_{\overline p};\overline{\alpha}_{\Xi},\overline{\phi}_{\Xi}, \overline{\alpha}_{\Lambda}):=\sum_{\mu'=0}^{3}a_{\mu\mu'}^{\overline \Xi} 
  a_{\mu'0}^{\overline \Lambda},\\
  {\cal V}&:=\int\dd\boldsymbol{\xi}_{\Xi\overline\Xi}={(4\pi)^5}\ .
\end{align*}
We use a modified parameter set where $\alpha_D$ and $\overline{\alpha}_D$ are expressed by $A_{\CP}^D$ and $\braket{\alpha_D}$. For $A_{\CP}^{[\Xi-]}$ and $A_{\CP}^{[\Lambda p]}$, we use a simplified notation  $A_\Xi$ and $A_\Lambda$, respectively. Similarly, we use $\Phi_{\CP}^{[\Xi-]}$ (denoted as $\Phi_{\CP}$) to represent $\phi_\Xi=\Phi_{\CP}+\braket{\phi_\Xi}$ and  $\overline\phi_\Xi=\Phi_{\CP}-\braket{\phi_\Xi}$.
The vector of the parameters related to the $\Xi$ and $\Lambda$ decays is $\boldsymbol{\omega}:=(\braket{\alpha_{\Xi}},
\braket{\phi_\Xi},\braket{\alpha_\Lambda},A_{\Xi},\Phi_{\CP},A_{\Lambda})$.
Therefore, the partial derivative \textit{e.g.} with respect to $\Phi_{\CP}$ is 
\begin{align}
   \frac{\partial {\cal{P}}^{\Xi\overline \Xi}}{\partial \Phi_{\CP}}=   \frac{1}{\cal V}\sum_{\mu,\nu=0}^{3}C_{\mu\nu}
   \left( \frac{\partial{\cal D}^\mu_\Xi}{\partial \phi_{\Xi}}
   \overline{\cal D}^\nu_\Xi+{\cal D}^\mu_\Xi\frac{\partial\overline{{\cal D}}^\nu_\Xi}{\partial \overline{\phi}_{\Xi}} \right)\nonumber\ .
\end{align}
Due to the orthonormality of the decay and production functions the information matrix elements related to the decay parameters 
$\omega_i$ and $\omega_j$ can be written as
\begin{align}
{\cal I}_0(\omega_i,\omega_j)&=N\sum_{\mu,\nu=0}^{3} \braket{ C^2}_{\mu\nu} \braket{\Delta_{\omega_i}\Delta_{\omega_j}}^{\mu\nu}\ .
\end{align}
We have checked these orthonormality relations in the explicit  calculations. The production tensor is defined in Eq.~\eqref{eq:ProdTensor}.
The decay tensor is 
\begin{align}
   \braket{\Delta_{\omega_i}\Delta_{\omega_j}}^{\mu\nu}&:=\frac{1}{(4\pi)^4}\int \frac{\partial({\cal D}^\mu_\Xi \overline{\cal D}^\nu_\Xi)}{\partial \omega_i}
\frac{\partial({\cal D}^\mu_\Xi \overline{\cal D}^\nu_\Xi)}{\partial \omega_j}
 \dd\Omega_\Lambda\dd\Omega_p \dd\Omega_{\overline\Lambda}\dd\Omega_{\overline p}\ .
\end{align}
For example ${\cal I}_0(\Phi_{\CP})$ can be expressed as
\begin{align}
     {\cal I}_0(\Phi_{\CP})&=N\int\frac{{\cal V}}{C_{00}}\left(\frac{\partial {\cal{P}}^{\Xi\overline \Xi}}{\partial \Phi_{\CP}}\right)^2\dd\boldsymbol{\xi}\nonumber\\
   &=N\sum_{\mu,\nu=0}^{3} \left[\frac{1}{4\pi}\int \frac{C_{\mu\nu}^2}{C_{00}}\dd\Omega_\Xi\right]\left[\frac{1}{(4\pi)^4}\int \left(\frac{\partial({\cal D}^\mu_\Xi \overline{\cal D}^\nu_\Xi)}{\partial \Phi_{\CP}}\right)^2
 \dd\Omega_\Lambda\dd\Omega_p \dd\Omega_{\overline\Lambda}\dd\Omega_{\overline p}\right]\nonumber\\
   &=:N\sum_{\mu,\nu=0}^{3} \braket{ C^2}_{\mu\nu} \braket{\Delta^2_{\Phi_{\CP}}}^{\mu\nu}\nonumber\ .
\end{align}
The information matrix elements for the decay parameters 
can be obtained as
\begin{align}
{\cal I}_{0}(\omega_i,\omega_j)&=N\left[\mathbb{a}_{ij}+
    \mathbb{b}_{ij}\braket{\mathbb{P}^2_\Xi}+
    \mathbb{c}_{ij}\braket{\mathbb{S}_{\Xi\overline\Xi} ^2}
    \right]\ ,\label{eq:FIMdeco}
\end{align}
where $\braket{\mathbb{P}^2_\Xi}(=2\braket{\mathbf{P}^2_\Xi})$ and $\braket{\mathbb{S}_{\Xi\overline\Xi} ^2}$ are the sums of the $\braket{ C^2}_{\mu\nu}$-matrix polarization and spin-correlation elements, respectively, defined in Eq.~\eqref{eq:PolandSc} (and shown for few production processes in Fig.~\ref{fig:PandSterms}(b) as the function of electron-beam polarization). Such representation is possible since the decay tensor elements have only three different values $\mathbb{a}_{ij}$, $\mathbb{b}_{ij}$ and $\mathbb{c}_{ij}$. It turns out that the only nonzero elements of the information matrix involving the CP-odd variables for the two-step process  are
\begin{align}
   {\cal I}_0(\Phi_{\CP})&=\frac{2N}{27} \left(1-\alpha_\Xi^2\right) \alpha_\Lambda^2\left[
    \left(3 + {\alpha_\Xi^2 \alpha_\Lambda^2}\right)\braket{\mathbb{P}^2_\Xi}+\frac{2}{3} \left(\alpha_\Xi^2 \left(3-2 \alpha_\Lambda^2\right)+3 \alpha_\Lambda^2\right)\braket{\mathbb{S}^2_{\Xi\overline{\Xi}}}
\right]\label{eq:FIM1},\\
{\cal I}_0({A_\Xi})&=\frac{2N}{3}\alpha_\Lambda^2\alpha_\Xi^2\left[ 1+
   \frac{3 \left(\alpha_\Lambda^4+3\right)-\alpha_\Xi^2 \left(3-\alpha_\Lambda^2\right)^2}{18 \left(1-\alpha_\Xi^2\right)\alpha_\Lambda^2} \braket{\mathbb{P}^2_\Xi}+\frac{\alpha_\Xi^2 \left(2 \alpha_\Lambda^2-3\right)+9}{27 \left(1-\alpha_\Xi^2\right)}\braket{\mathbb{S}^2_{\Xi\overline{\Xi}}}
\right]\label{eq:FIM2},\\
{\cal I}_0({A_\Lambda})&=\frac{2N}{3}\alpha_\Lambda^2 \alpha_\Xi^2\left[1+\frac{\alpha_\Xi^4-2 \alpha_\Xi^2+3}{6\alpha_\Xi^2}
    \braket{\mathbb{P}^2_\Xi}+\frac{1}{9} (3 - 2 \alpha_\Xi^2)\braket{\mathbb{S}^2_{\Xi\overline{\Xi}}}
\right]\label{eq:FIM3},\\
{\cal I}_0({A_\Lambda},{A_\Xi})&=\frac{2N}{3}\alpha_\Lambda^2\alpha_\Xi^2\left[1-\frac{1}{3}
    \left(\braket{\mathbb{P}^2_\Xi}+\braket{\mathbb{S}^2_{\Xi\overline{\Xi}}}\right)
\right]\label{eq:FIM4}\ .
\end{align}
These information matrix elements allows one to determine s.d. and correlations between the CPV observables. The uncertainty for 
$\Phi_{\CP}$ is
$\sigma(\Phi_{\CP})=1/\sqrt{{\cal I}(\Phi_{\CP})}$, since the variable is uncorrelated with any other variable.  The  $A_\Xi$ and $A_\Lambda$ variables are only correlated with each other and the covariance matrix is obtained by inverting two-dimensional information matrix
\begin{equation}
\left(\begin{array}{cc}
        \sigma^2({A_\Xi}) & \text{Cov}({A_\Lambda,A_\Xi}) \\
     \text{Cov}({A_\Lambda,A_\Xi})    & \sigma^2({A_\Lambda})
    \end{array}\right)    =\left(\begin{array}{cc}
        {\cal I}({A_\Xi}) & {\cal I}({A_\Lambda,A_\Xi}) \\
     {\cal I}({A_\Lambda,A_\Xi})    & {\cal I}({A_\Lambda})
    \end{array}\right)^{-1}\ .
\end{equation}

The expressions in Eqs.~\eqref{eq:FIM1}--\eqref{eq:FIM4} have some interesting properties which are valid for any two-step process that can be studied {by allowing the $\alpha_\Lambda$ and $\alpha_\Xi$ parameters to vary}. We discuss these properties using a generic notation, where the first decay process is $B\to b\pi$ and the baryon $b$ decays in the sequential weak two-body non-leptonic process.
\begin{itemize}
\item The $\Phi_{\CP}$ uncertainty is not correlated with any other variable,
and none of the information matrix elements depend on the $\braket{\phi_B}$ value. This because $\phi_B$ represents the shift in the $\varphi_b$ azimuthal angle of the $b$-baryon, which is integrated out. A dependence on $\braket{\phi_B}$ might appear in experiments where 
the acceptance in the $\varphi_b$ variable is limited.
\item For $\alpha_b=0$ only   ${\cal I}_0({A_B})=\frac{1}{3}\alpha_B^2 \braket{\mathbb{P}^2_B}$ is nonzero and the CPV test is the same as in a  single-step decay.
\item For $\alpha_B\to0$ two terms are nonzero
 ${\cal I}_0({\Phi_{\CP}})=\frac{2}{27} \alpha_b^2\left[
    3\braket{\mathbb{P}^2_B}+{2} \alpha_b^2\braket{\mathbb{S}^2_{B\overline{B}}}\right]$ and
${\cal I}_0({A_b})=\frac{1}{3}\alpha_b^2 \braket{\mathbb{P}^2_B}$. Therefore, both $\Phi_{\CP}$ and $A_b$ can be measured. In particular, due to the nonzero $\mathbb{c}$-type term for ${\cal I}_0({\Phi_{\CP}})$ the polarization of the $B$ baryon is not needed. {This is an attractive scenario for CPV tests for any baryon decaying into $\Lambda$.}
\item The term ${\cal I}_0({A_B})$ is divergent for $|\alpha_B|\to 1$ indicating that $\sigma({A_B})$ vanish in this limit. This is a consequence of the $\sqrt{1-\alpha_B^2}$ terms in the angular distribution. The validity of such expressions requires that the boundary $|\alpha_B|\le 1$ must be strictly fulfilled and in the $|\alpha_B|\to 1$ limit there is no linear term in the expansion of the $\alpha_B$ parameter (\textit{i.e.} the linear error is 0). To get a meaningful result, one should use a 
parameterization which respects this boundary, such as Eq.~\eqref{eq:SPone} from Sec.~\ref{ssec:CPpheno}.  In principle, one can directly investigate the uncertainty of the weak phase difference  $\xi_P-\xi_S$.
However, as seen from Eq.~\eqref{eq:ACP1} this will introduce correlation 
with the $\Phi_{\CP}$ observable (due to the term $\sin\phi$). Instead, 
one can present results for $\Delta\zeta_B:={\alpha_B} /\sqrt{1-\alpha_B^2}A_{B}$, which do not introduce such correlation.
The information matrix elements are modified due to the Jacobian of the variable transformation to
\begin{align*}
    {\cal I}_0({\Delta\zeta_B})&=\frac{ (1-\alpha_B^2)}{\alpha_B^2}{\cal I}_0({A_B}),\\
    {\cal I}_0({A_b},{\Delta\zeta_B})&=\frac{ \sqrt{1-\alpha_B^2}}{\alpha_B}{\cal I}_0({A_b},{A_B})\ .
\end{align*}
\end{itemize}

We first discuss uncertainties for the production tensors corresponding to the simplest cases. 
Unpolarized and uncorrelated sources of $B$ and $\overline{B}$ correspond to $\braket{ C^2}_{\mu\nu}= {\rm diag}(1,0,0,0)$. The only nonzero elements of the information matrix are
$${\cal I}_0({A_B})={\cal I}_0({A_b})={\cal I}_0({A_b,A_B})=N\frac{2}{3}\alpha_b^2\alpha_B^2\ ,
$$
where we have assumed samples of $N$ events each for the cascade and anticascade decays. Since the information matrix corresponding to the $A_\Xi$ and $A_\Lambda$ is singular, the asymmetries are fully correlated and cannot be determined separately, but the sum $A_\Xi+A_\Lambda$ can and the uncertainty is $\sigma_C(A_\Xi+A_\Lambda)=\sqrt{3/2}/(\alpha_\Lambda\alpha_\Xi)$.

\begin{figure}
    \centering
    \includegraphics[width=0.49\textwidth]{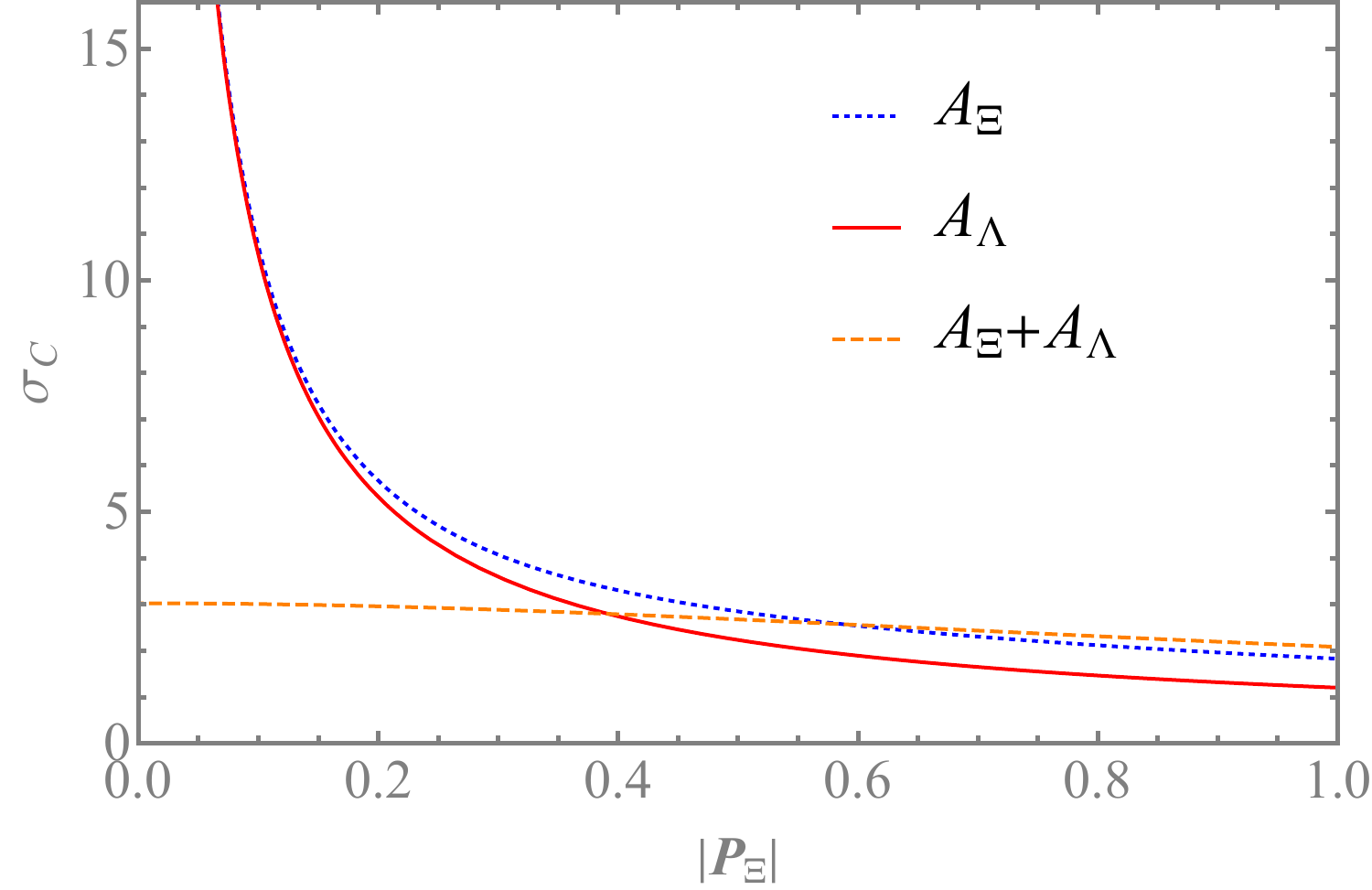}
\put(-155,120){\large(a)}       
\includegraphics[width=0.49\textwidth]{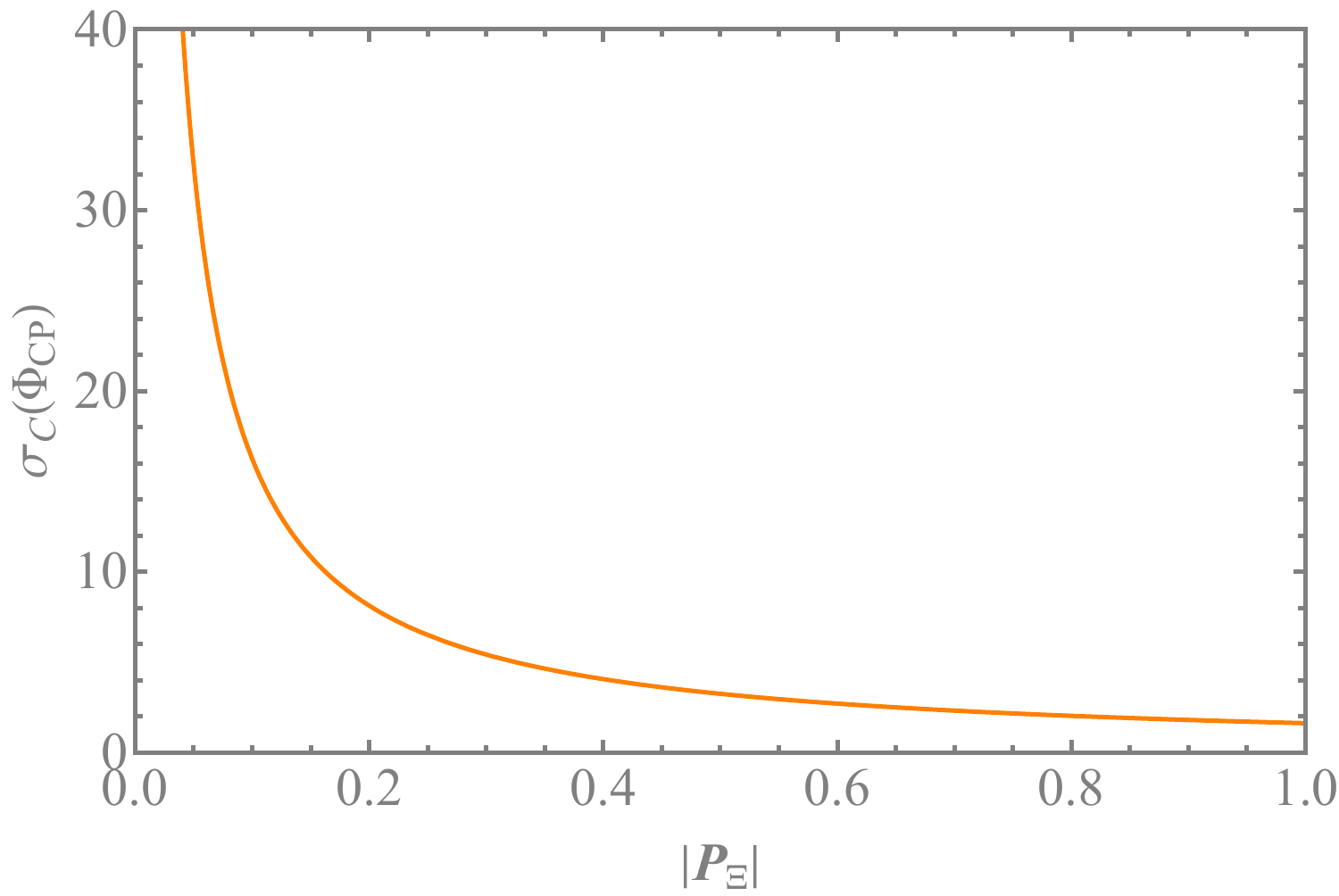}
\put(-165,120){\large(b)}   
    \caption{Uncertainties,  $\sigma_C$, for CP tests in  HyperCP-type experiment using analytic approximation: (a) $A_\Lambda$ (solid line), $A_\Xi$ (dotted line) and $A_\Lambda+A_\Xi$  (dashed line), (b) $\Phi_{\Xi}$.}
    \label{fig:HyperCPA}
\end{figure}

As the next example, we consider two independent ST experiments with $N$ events using polarized cascades and anticascades having the same average  polarization $\braket{\mathbf{P}_B^2}$.   {In the 0-th approximation, the expressions for the uncertainties depend on the production mechanism only via the average $\braket{\mathbf{P}_B^2}$. For example, in the HyperCP-type experiments, where the initial hyperon polarization is considered to be a fixed vector that does not depend on the kinematic variables of the production process, the average reduces to the square of the vector $\sqrt{\braket{\mathbf{P}_B^2}}\to|\mathbf{P}_B|$.} 
The Fisher information matrix is the sum of the matrices for the two ST experiments 
\begin{align}
 {\cal I}_0({\omega_i,\omega_j})&= 2{N}\left[\mathbb{a}_{ij}+
    \mathbb{b}_{ij}\braket{\mathbf{P}^2_B}
    \right]\label{eq:STinfo}
\end{align}
and the elements of the information matrix for the CP-test observables read
\begin{align*}
    {\cal I}_0({\Phi_{\CP}})&=N\frac{{4}}{27} \left(1-\alpha_B^2\right) \alpha_b^2
    \left(3 + {\alpha_B^2 \alpha_b^2}\right)\braket{\mathbf{P}_{B}^2},
\\
 {\cal I}_0({A_B})&=N\frac{4}{3}\alpha_b^2\alpha_B^2\left[ 1+
   \frac{3 \left(\alpha_b^4+3\right)-\alpha_B^2 \left(3-\alpha_b^2\right)^2}{{18} \left(1-\alpha_B^2\right)\alpha_b^2} \braket{\mathbf{P}_{B}^2}
\right],\\
 {\cal I}_0({A_b})&=N\frac{4}{3}\alpha_b^2\alpha_B^2\left[1 +\frac{\alpha_B^4-2 \alpha_B^2+3}{6\alpha_B^2}
    \braket{\mathbf{P}_{B}^2}
\right],\\
{\cal I}_0({A_b,A_B})&=N\frac{4}{3}\alpha_b^2\alpha_B^2\left[1-{\frac{1}{3}}
    \braket{\mathbf{P}_{B}^2}
\right]\ .
\end{align*}
For two HyperCP experiments with $|\mathbf{P}_{B}|=|\mathbf{P}_{\overline B}|$ and $N$ events each the formulas are the same. 
The resulting uncertainties $\sigma_C=\sigma\sqrt{N}$ for the $A_\Lambda$, $A_\Xi$ and  $A_\Lambda+A_\Xi$ observables measured using the $\Xi^-/\overline{\Xi}\vphantom{X}^+$ decay chains are shown in Fig.~\ref{fig:HyperCPA}(a), while for $\Phi_{\CP}$ in Fig.~\ref{fig:HyperCPA}(b). The uncertainty of the sum $A_\Lambda+A_\Xi$ is nearly independent of the average polarization. 

\begin{figure}
    \centering
    \includegraphics[height=0.26\textheight]{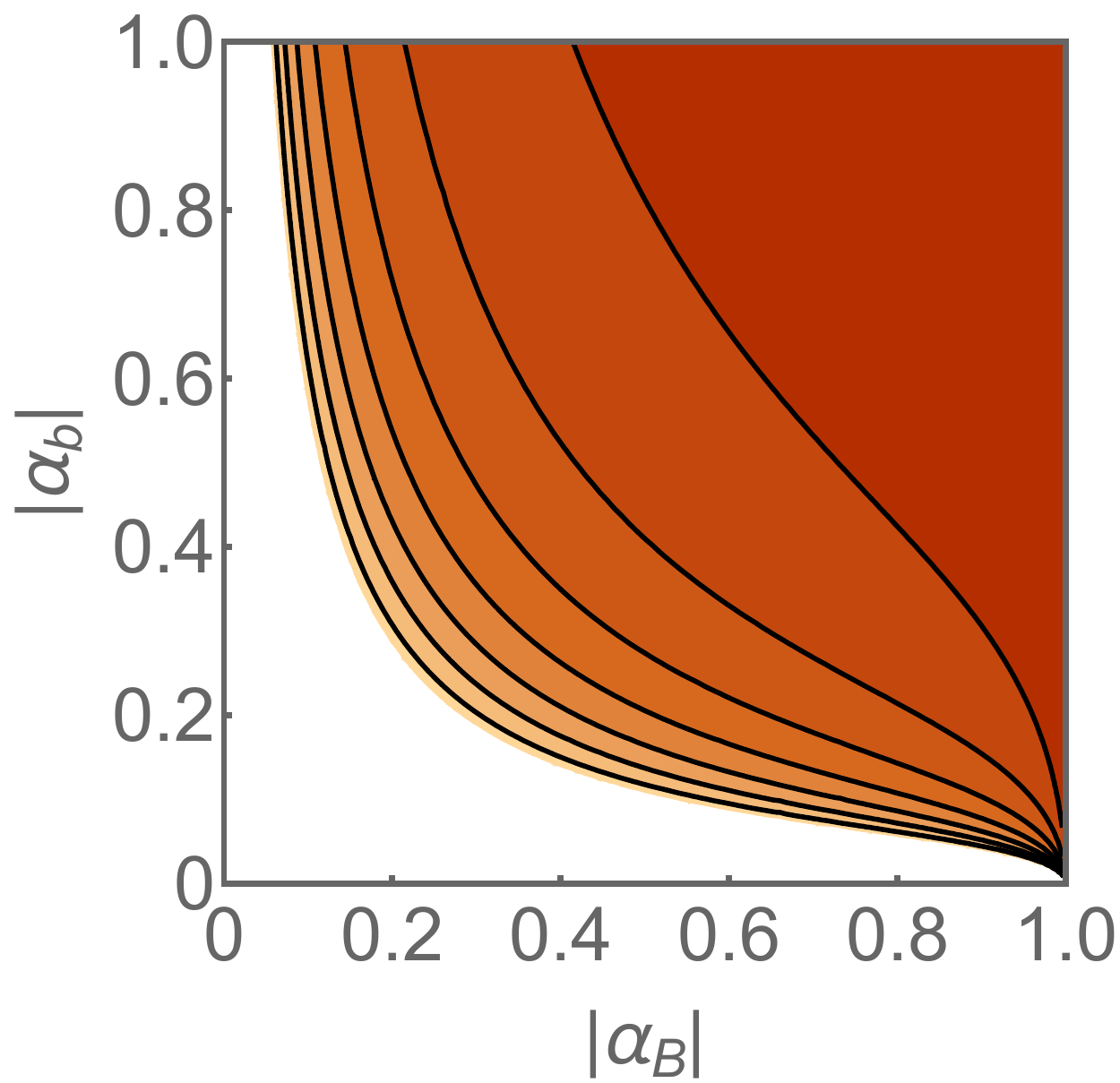}
\put(-105,150){\large(a)}    \includegraphics[height=0.26\textheight]{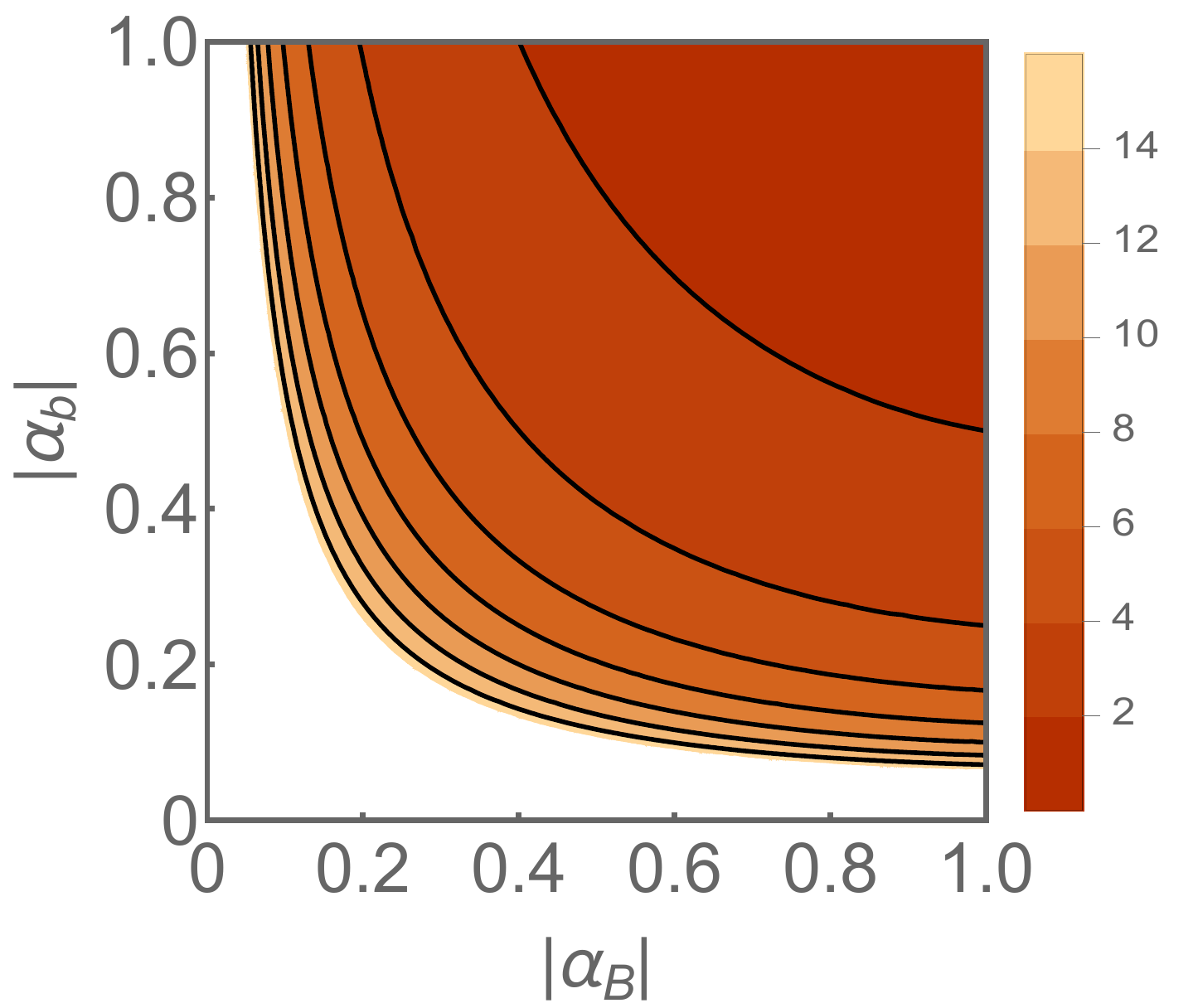}
\put(-135,150){\large(b)}        

\includegraphics[height=0.26\textheight]{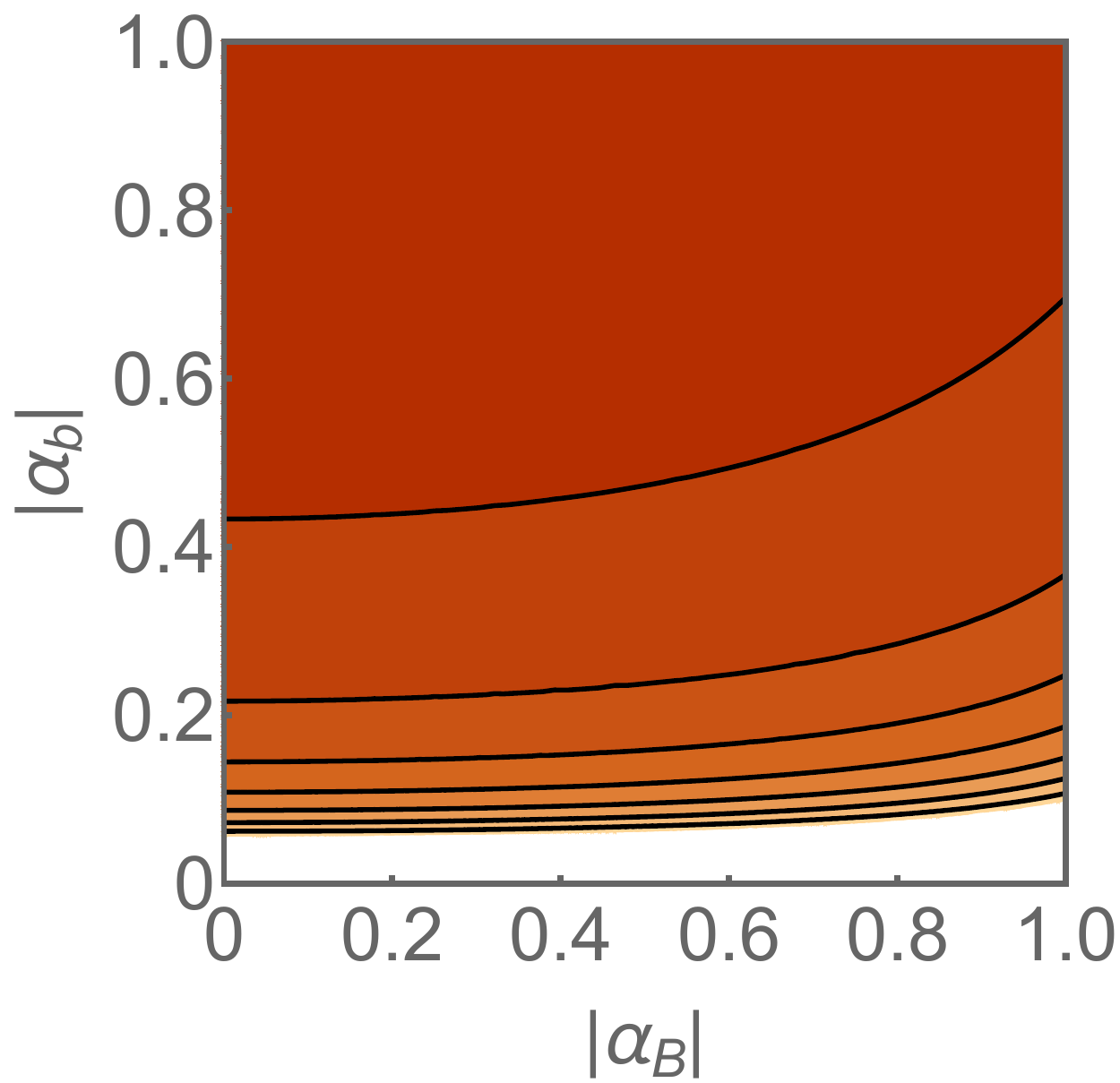}
\put(-105,135){\large(c)}        \includegraphics[height=0.26\textheight]{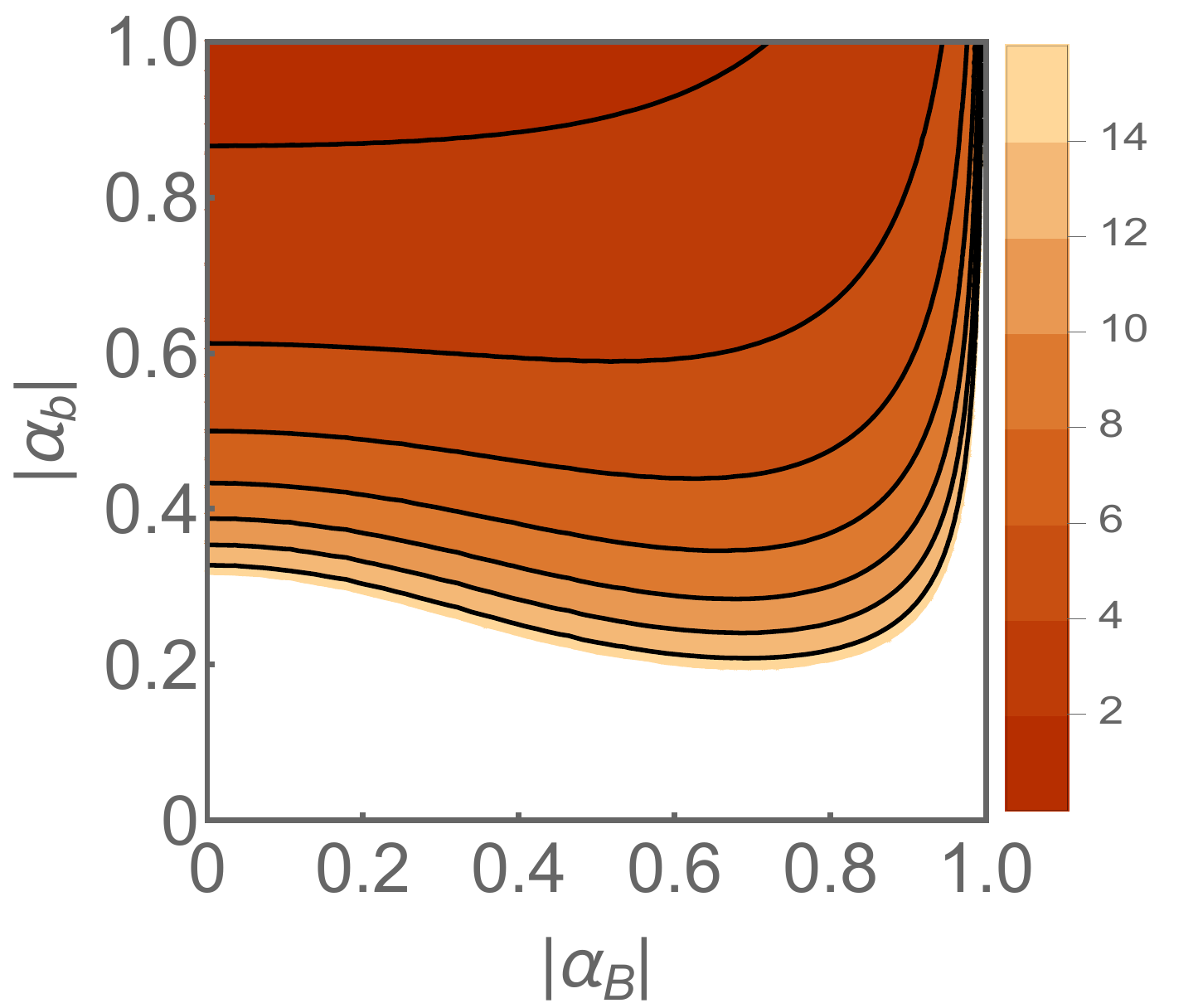}
\put(-135,135){\large(d)}    
\caption{(colour online) Statistical uncertainties $\sigma_C$ of (a) $A_B$, (b) $A_b$, (c) $\Delta\zeta_B$ and (d) $\Phi_{\CP}$ measurement in a (pseudo)scalar meson decay to $B\overline B$ as a function of $\alpha_B$ and $\alpha_b$ treated as free parameters. The white regions in the bottom of the plots correspond to the uncertainties $\sigma_C(\ldots)>15$.}
    \label{fig:etacBBPhi}
\end{figure}

Our next case is the decay of a (pseudo)scalar meson like $\eta_c$ or $\chi_{c0}$ into a
$B\overline{B}$ pair with the production tensor $\braket{ C^2}_{\mu\nu}= {\rm diag}(1,1,1,1)$. There are no polarization terms, $\braket{\mathbb{P}^2_{B}}=0$, and  $\braket{\mathbb{S}^2_{B\overline{B}}}=3$. The information matrix element ${\cal I}_0({A_b,A_B})$ is zero,  which means that all three CPV observables are uncorrelated. The diagonal terms of the information matrix are functions of $\alpha_b$ and $\alpha_B$ only: 
\begin{align*}
   {\cal I}_0(\Phi_{\CP})&=N\frac{4}{27} \left(1-\alpha_B^2\right) \alpha_b^2\left[
    \alpha_B^2 \left(3-2 \alpha_b^2\right)+3 \alpha_b^2
\right],\\
{\cal I}_0({A_B})&=N\frac{2}{3}\alpha_b^2\alpha_B^2\left[ 1+\frac{\alpha_B^2 \left(2 \alpha_b^2-3\right)+9}{9 \left(1-\alpha_B^2\right)}
\right],\\
{\cal I}_0({A_b})&=N\frac{2}{3}\alpha_b^2 \alpha_B^2\left[1+\frac{1}{3} (3 - 2 \alpha_B^2)
\right]\ .
\end{align*}
Fig.~\ref{fig:etacBBPhi} shows the uncertainties $\sigma_C$ for this case as a function of $\alpha_B$ and $\alpha_b$ decay parameters.
This case is interesting since all production parameters are 
fixed and CP-test uncertainties depend only on $\alpha_b$ and $\alpha_B$. 

To understand relative importance of the polarization and spin-correlation terms for the CP tests, one compare the two above extreme cases. For example, the polarization in two ST experiments with $N$ events that would lead to the same uncertainty of the $\Phi_{\CP}$ measurement as in the DT approach with $N$ events is:
$$
   |\mathbf{P}_{B}|^2=\frac{
    \alpha_B^2 \left(3-2 \alpha_b^2\right)+3 {\alpha_b^2}}{3 + {\alpha_B^2 \alpha_b^2}}\ . 
$$
For $\Xi\to\Lambda\pi$ this gives  $|\mathbf{P}_{B}|=0.80$.

\begin{figure}
    \centering
    \includegraphics[width=0.5\textwidth]{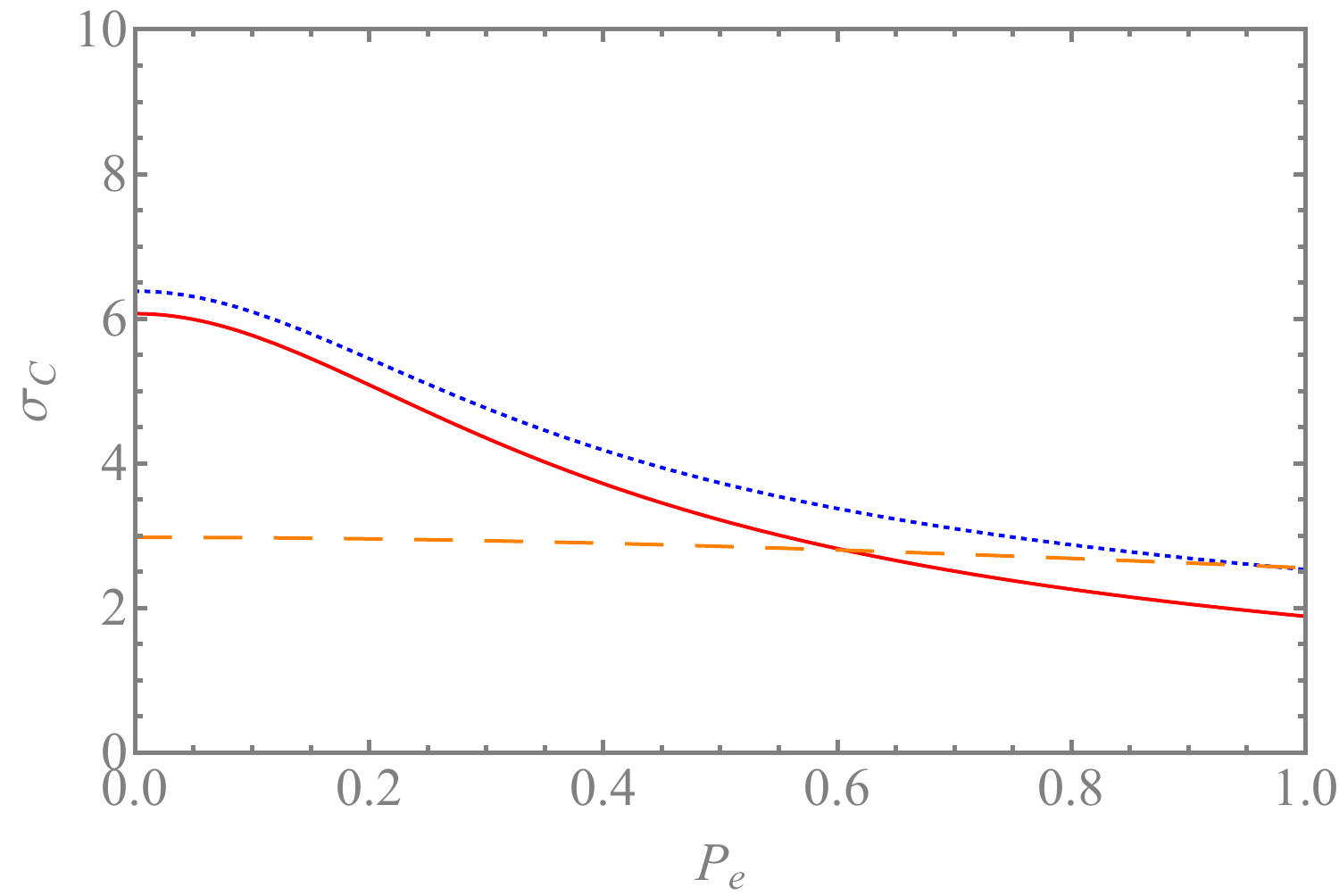}
    \put(-44,120){\large(a)}
    \includegraphics[width=0.49\textwidth]{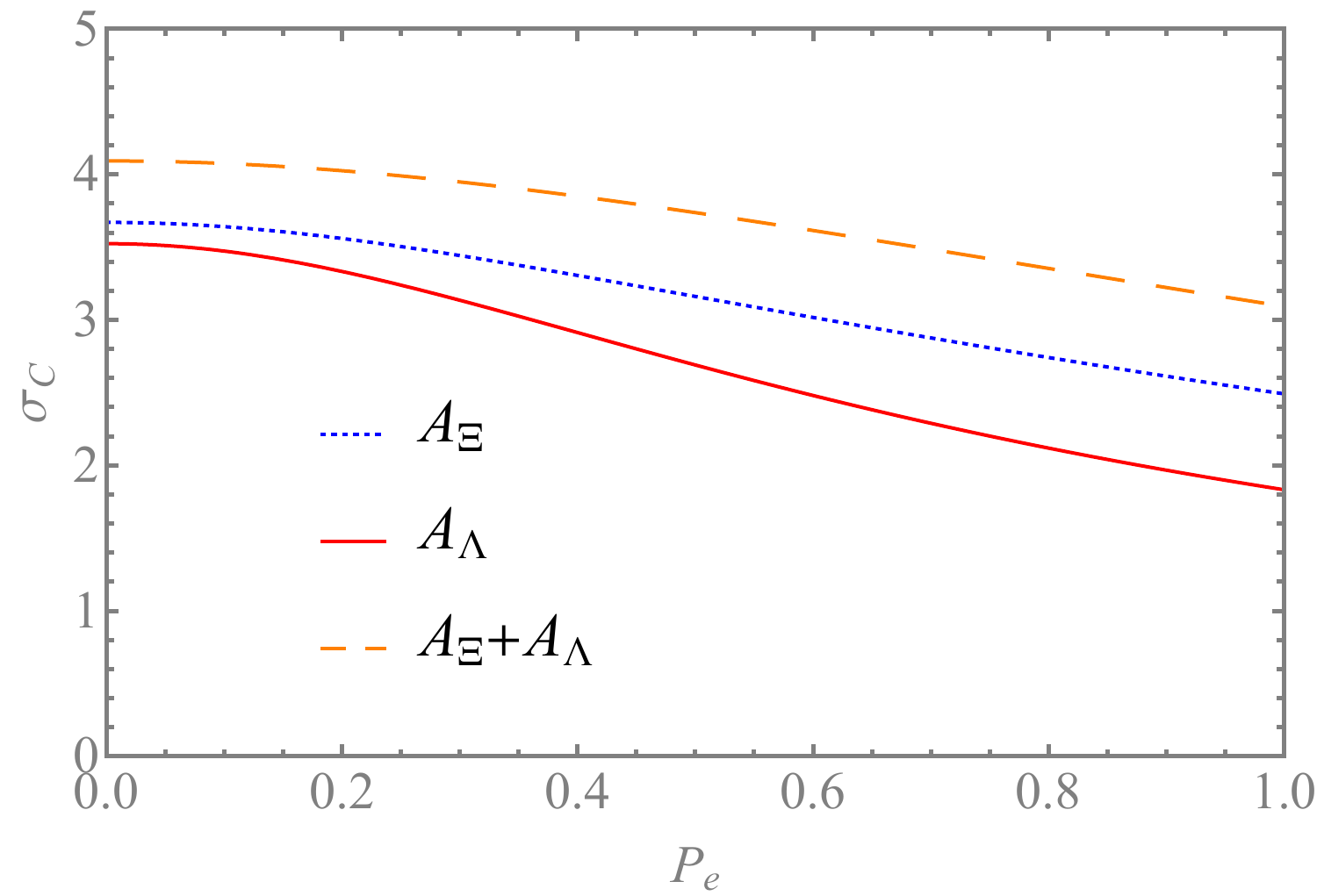}
    \put(-90,130){\large(b)}
    \caption{Uncertainties, $\sigma_C$, for the $e^+e^-\to J/\psi\to\Xi\overline\Xi$ (a) two ST and (b) DT experiments with $N$ events each: $A_\Lambda$ (solid line), $A_\Xi$ (dotted line) and $A_\Lambda+A_\Xi$ (dashed line).  \label{fig:ST-DT-ACP}}
\end{figure}

\begin{figure}
    \centering
    \includegraphics[width=0.95\textwidth]{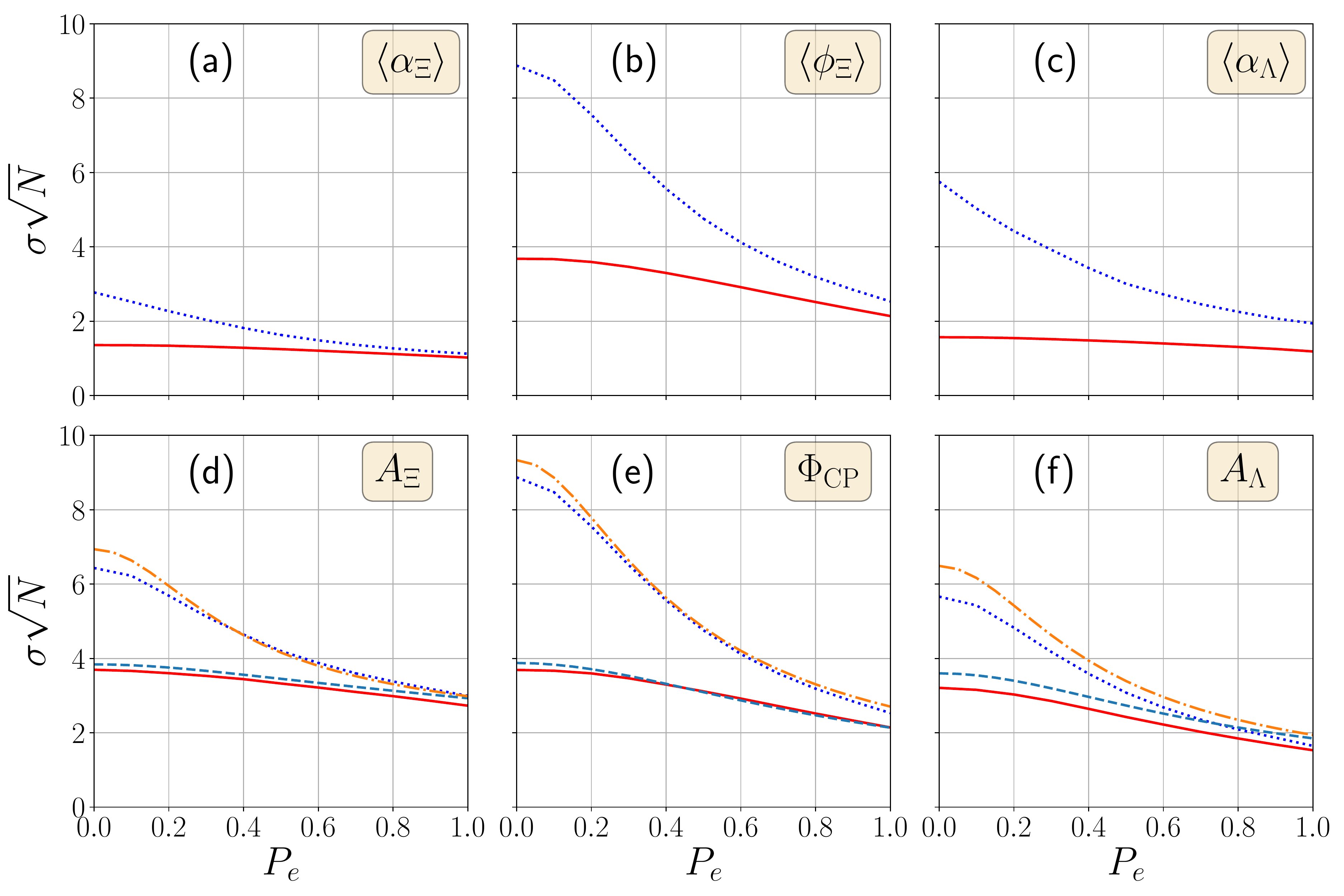}
     \caption[]{
    Numerical estimate of the uncertainty $\sigma\sqrt{N}$ of (a)--(c) average decay parameters and (d)--(f) CPV observables in 
$e^+e^-\to J/\psi\to\Xi^{-}\overline{\Xi}\vphantom{X}^{+}$. 
    The dotted lines and the solid lines are the results for ST and DT experiments, respectively. For the asymmetries $A_\Xi$, $\Phi_{\CP}$ and $A_\Lambda$ also the analytic approximation is given: dashed-dotted lines and dashed are ST and DT results, respectively. 
}
    \label{fig:Xi-results-num}
\end{figure}

Now we will discuss the results specific for the $e^+e^-\to J/\psi\to \Xi\overline \Xi$ process.
The relations for the ST experiment,  realised as two independent measurements with $N$ events each \footnote{Of course this is not the way one does the experiment since both the baryon and antibaryon decays can be measured simultaneously.}, are still valid.  The only difference is that now the results can be represented as a function of the electron-beam polarization $P_e$, and the average cascade polarization $|\mathbf{P}_\Xi|$ is calculated using Eq.~\eqref{eq:PBparam}. The results in the analytic approximation for the $A$-type observables corresponding to the ones in Fig.~\ref{fig:HyperCPA}(a) are shown in Fig.~\ref{fig:ST-DT-ACP}(a). Since even for the $P_e=0$ the average polarization of the cascades is not zero, all the three CP tests are possible. For the average values of the decay parameters we do not provide approximate analytic results since the corresponding information matrix elements are correlated and in general multidimensional matrices have to be inverted to obtain uncertainties. Therefore, likely such analytic solution will not provide better understanding of the interrelations between the parameters. The numeric results for uncertainties of $\braket{\alpha_\Xi}$, $\braket{\phi_\Xi}$ and  $\braket{\alpha_\Lambda}$ are shown in Fig.~\ref{fig:Xi-results-num}(a)--(c) both for ST- and DT-type experiments. For the ST experiments, the uncertainty improves much more than for DT experiments. It is understood by the fact that the spin-correlation terms contribute only to the DT experiments and the dependence on the $P_e$ is weaker.
The numerical results for the $A_\Xi$, $\Phi_{\CP}$ and  $A_\Lambda$ are given in Fig.~\ref{fig:Xi-results-num}(d)--(f) and compared to the analytic approximations, which represent well the results specially for the $\Xi$ decay CPV tests. As a cross-check of the calculations, we provide in Table~\ref{tab:corr2} the full correlation matrix of all parameters using the full numerical calculations for $P_e=0.8$.

\begin{figure}
    \centering
\includegraphics[width=0.99\textwidth]{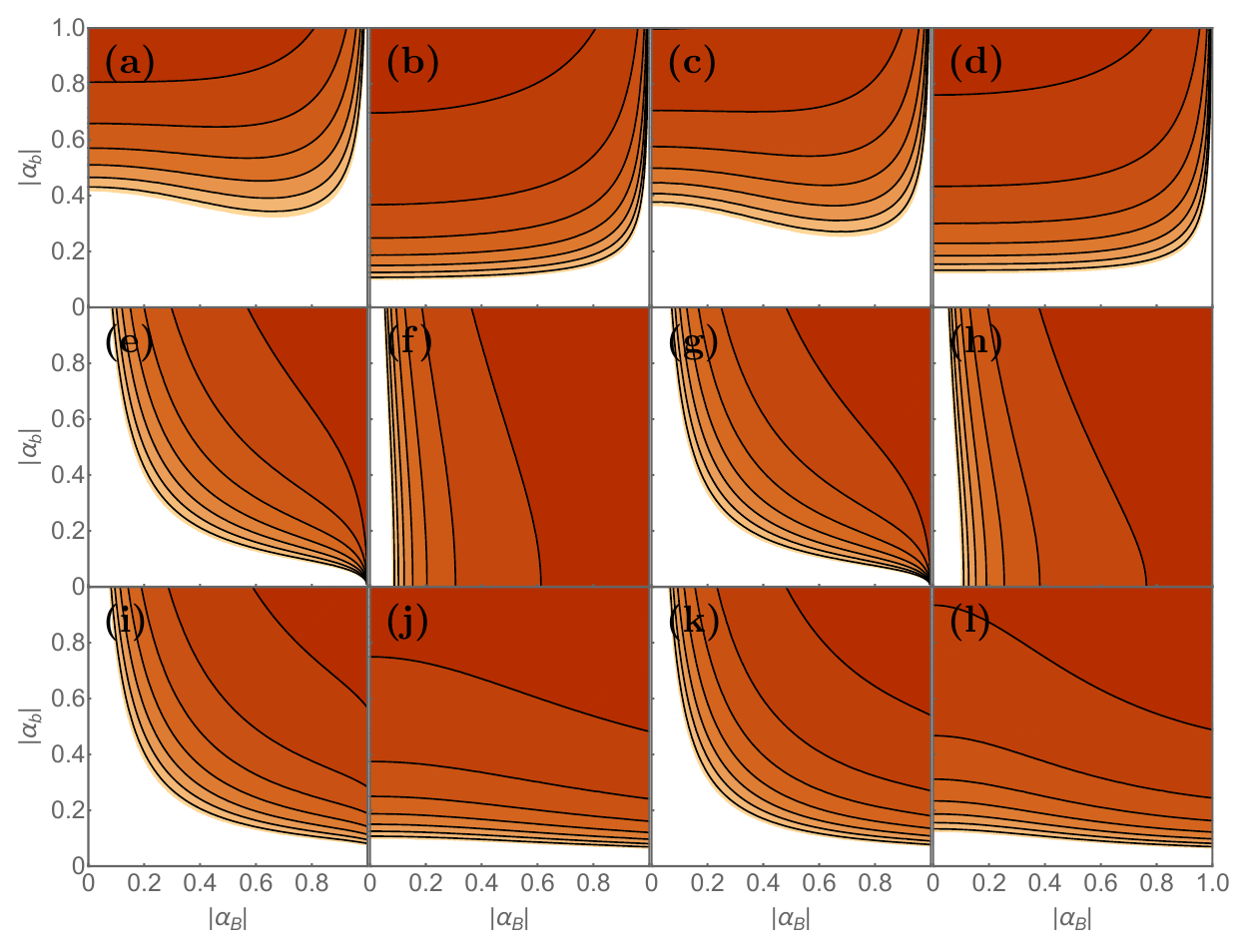}
     \caption{(colour online) Statistical uncertainties $\sigma_C$ of: (a)--(d)  $\Phi_{\CP}$, (e)--(h) $A_B$ and (i)--(l) $A_b$ measurements in the $e^+e^-\to\gamma^*\to B\overline{B}$ process with two-step $B$-baryon decays in the low-energy (LE) limit ($\alpha_\psi=0$ and $\Delta\Phi=0$) and high-energy (HE) limit ($\alpha_\psi=1$) as the function of $\alpha_B$ and $\alpha_b$ treated as free parameters.  The columns from left are: (LE limit, $P_e=0$), (LE limit, $P_e=1$), (HE limit, $P_e=0$) and  (HE limit, $P_e=1$). The same colour scale as in Fig.~\ref{fig:etacBBPhi} is used. The white regions in the plots correspond to uncertainties $\sigma_C(\ldots)>15$. }
    \label{fig:HLABP}
\end{figure}

\begin{table}
  \caption{Correlation matrix for the asymmetries and averages in the $e^+e^-\to J/\psi\to\Xi\overline{\Xi}$ process with $P_e=0.8$ for DT. Input parameters are $\braket{\alpha_{\Xi}}=-0.373$, $\braket{\phi_{\Xi}}=0.016$ and $\braket{\alpha_{\Lambda}}=0.760$. The error is the last significant digit unless specified explicitly, and only the results statistically different from zero are shown. \label{tab:corr2}}
\begin{ruledtabular}
\begin{tabular}{c|ccccccccc}
&&$\Delta\Phi$&$\braket{\alpha_{\Xi}}$&$\left<\phi_{\Xi}\right>$&$\left<\alpha_{\Lambda}\right>$&$A_{\Xi}$&$A_{\Lambda}$&$\Phi_{\CP}$  &$P_e$ \\
    \hline
$\alpha_\psi$ & & ${-0.128}$ & -- & 0.011 & -0.008 & -- & -0.017(2) & -- & -0.031  \\
$\Delta\Phi$ &  & & 0.009 & 0.009 & -0.071(2) & -- & -- & -- & ${0.191(3)}$ \\   
$\left<\alpha_{\Xi}\right>$ &  &  &  & -0.021(4) & 0.078(3) & -- & -- &-- & 0.037 \\    
$\left<\phi_{\Xi}\right>$ &  &  &  &  & -0.032 & -- & -- & -- & $-0.005$ \\
$\left<\alpha_{\Lambda}\right>$ &  &  &  &  &  & -- & -- & --& ${-0.455}$ \\  
  \end{tabular}
\end{ruledtabular}
\end{table}
Finally, it is interesting to consider a general two-step process $B\to b\pi$ in the low- and high-energy limits (LE- and HE limits, respectively, introduced in Sec.~\ref{sec:Form}) for a single photon  $e^+e^-\to B\overline B$ annihilation process.
These cases  might be of interest for close to threshold charm baryon studies or baryon--antibaryon production experiments at high energies.
 In the LE limit  ($\alpha_\psi=0$, $\beta_\psi=0$, $\gamma_\psi=1$) the terms $\braket{\mathbb{P}^2_B}$ and 
$\braket{\mathbb{S}^2_{B\overline B}}$ are ${2}  P_e^2$ and $1$, respectively. In the HE limit ($\alpha_\psi=1$, $\beta_\psi=0$, $\gamma_\psi=0$) they are equal to  ${6} \left(1-{\pi/4 }\right) P_e^2$ and $ 3(\pi/2 -1)$, respectively. In both cases the spin-correlation terms do not depend on the electron polarization and the $\braket{\mathbb{P}^2_B}$ terms are proportional to $P_e^2$. A comparison of the  uncertainties for $P_e=0$ and $P_e=1$  in the DT-experiment setting is presented in Fig.~\ref{fig:HLABP}. 
The conclusion is that the polarization helps to reach better precision in both cases, and the improvement is qualitatively similar. 

\section{Experimental considerations}
\label{sec:Exp}
 The benefits of a large electron-beam polarisation for CP-violation studies should be clear by now. Here we discuss three additional aspects related to the detection technique which should be considered when planning such an experiment,
\begin{itemize}
    \item[(a)] Combination of the ST and DT data sets including detection efficiency and background aspects.
    \item[(b)] Polar angle dependence of uncertainty and the detection efficiency.
    \item[(c)] Implications of the discussed collision scheme with large-crossing angle. 
\end{itemize}

\begin{figure}
\centering
\includegraphics[width=0.33\columnwidth]{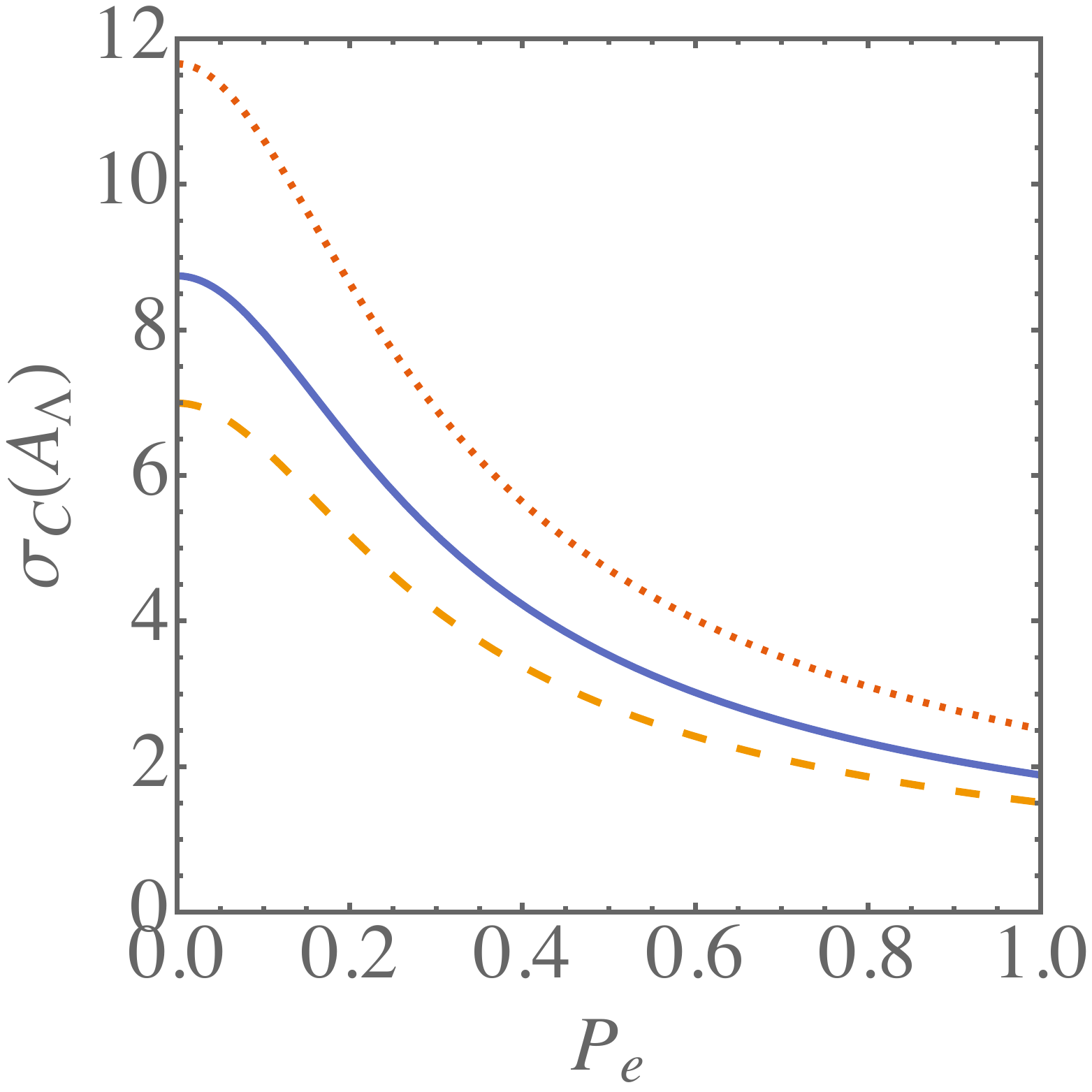}
\put(-40,120){\large (a)}
\includegraphics[width=0.33\columnwidth]{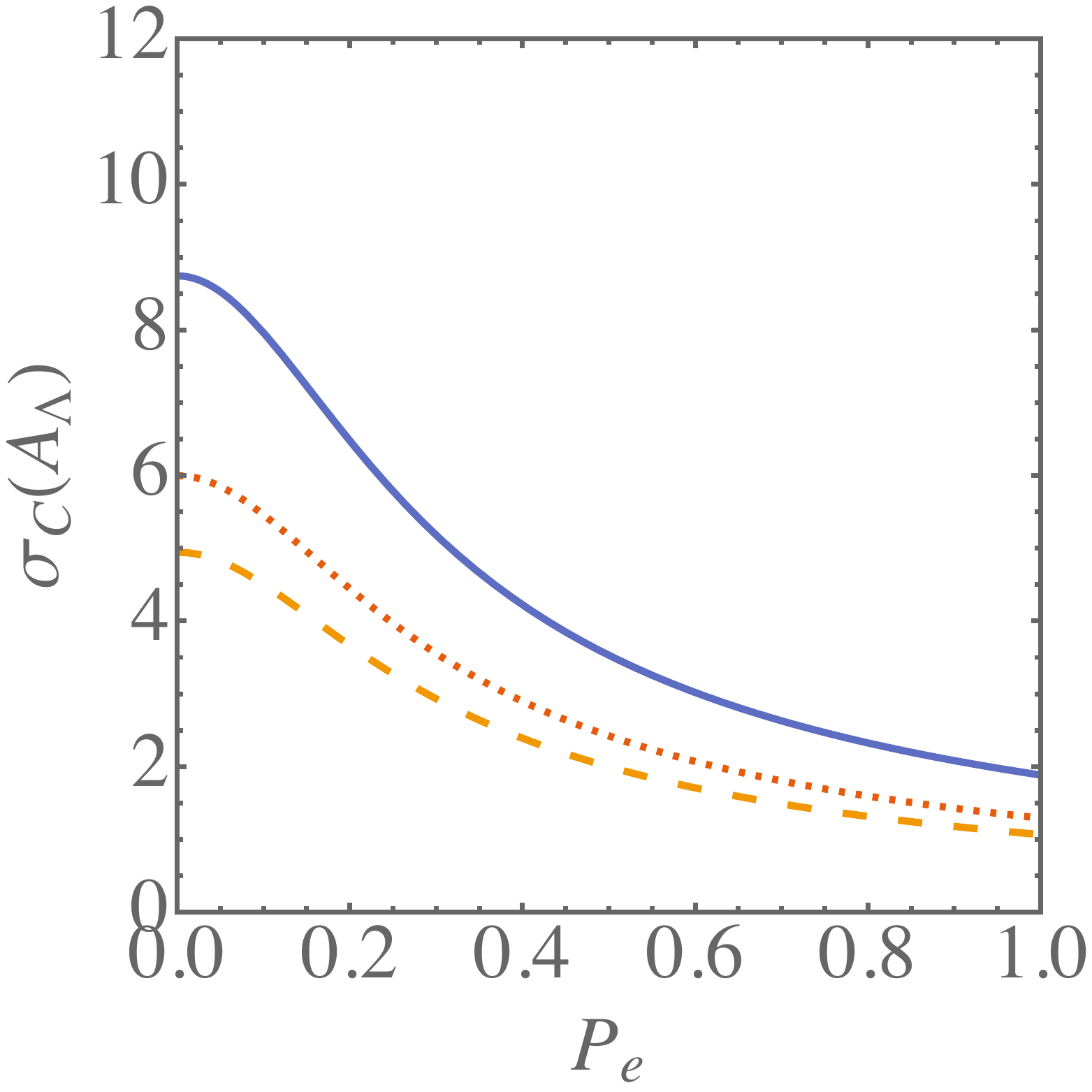}
\put(-40,120){\large (b)}
\caption[]{Statistical uncertainties $\sigma_C(A_{\Lambda})$ for the $e^+e^-\to J/\psi\to\Lambda\overline\Lambda$ process as a function of the electron-beam polarization $P_e$. The solid-blue  lines represent DT measurement.  The dotted-red lines represent contribution from ST events which do not contribute to the DT event class (statistically independent ST events). The orange-dashed lines represent the result from the combination of the two event classes. The decay branching fraction is ${\cal B}=0.64$~\cite{ParticleDataGroup:2020ssz}. The detection efficiency of the $\Lambda$ decay was assumed to be  (a) $\epsilon_\Lambda=\overline{\epsilon}_\Lambda=1$ and (b) $\epsilon_\Lambda=\overline{\epsilon}_\Lambda=0.5$. The results are normalised to the number of the DT events.}
  \label{fig:senALL}
\end{figure}

\paragraph{Combination of ST and DT measurements}
In general the best precision can be achieved by combining three non-overlapping event sets. The first set includes the DT events where both the $B$ and $\overline B$ decay chains are reconstructed. The remaining events can be divided into two ST sets where $B(\overline B)$ decay is fully reconstructed but not the corresponding $\overline B(B)$.
The efficiencies of the $B$, $\overline{B}$ and $B\overline{B}$ sets are denoted as $\epsilon_B$, $\epsilon_{\overline{B}}$ and $\epsilon_{B\overline{B}}$, respectively. The efficiencies 
can depend on the vector $\boldsymbol{\xi}$ of the kinematic variables, but not on the global reaction parameters given by the $\boldsymbol{\omega}$ vector. Since we discuss improvements with respect to the DT-type experiment, $\epsilon_B$ is given by the ratio between the detection efficiencies of the DT and ST cases. We also neglect any efficiency dependence on the kinematic variables.
We recollect that the information in the DT experiment, based on $N$ reconstructed events, is given by Eq.~\eqref{eq:FIMdeco}:
\begin{align*}
{\cal I}_{0}^{\rm DT}(\omega_i,\omega_j)&=N\left[\mathbb{a}_{ij}+
    \mathbb{b}_{ij}\braket{\mathbb{P}^2_B}+
    \mathbb{c}_{ij}\braket{\mathbb{S}_{B\overline B} ^2}
    \right]\ .
\end{align*}
For the two-step process $\mathbb{a}_{ij}$, $\mathbb{b}_{ij}$ and $\mathbb{c}_{ij}$ can be read from Eq.~\eqref{eq:FIM1}--\eqref{eq:FIM4}.
For the single-step process only $A_B$ can be measured
$\mathbb{a}_{A_B}=\mathbb{c}_{A_B}=0$ and $\mathbb{b}_{A_B}=1/3$: 
$$
{\cal I}_0(A_B)=N\frac{1}{3} \braket{\mathbb{P}^2_B}\ .
$$
The information provided by the two additional ST sets is 
\begin{align}
    {\cal I}^{\rm ST}_0({\omega_i,\omega_j})&=N\frac{1-\epsilon_B{\cal B}}{\epsilon_B{\cal B}}\left[2\mathbb{a}_{ij}+
    \mathbb{b}_{ij}\braket{\mathbb{P}^2_B}
    \right]    \ ,
\end{align}
where the branching fraction product of the decay sequence is ${\cal B}$ and equal detection efficiencies   $\epsilon_B=\epsilon_{\overline{B}}$ are assumed. The interpretation of the above equation is that an additional 
$2N/({\epsilon_B{\cal B}})$ events are added from the two ST sets.
Therefore, the information of the combined ST and DT experiment (ST\&DT) is the sum of the two independent measurements 
\begin{align}
    {\cal I}^{\rm ST\&DT}_0({\omega_i,\omega_j})&=N\left[\frac{2-\epsilon_B{\cal B}}{\epsilon_B{\cal B}}\mathbb{a}_{ij}+
    \frac{1}{\epsilon_B{\cal B}}\mathbb{b}_{ij}\braket{\mathbb{P}^2_B}+
    \mathbb{c}_{ij}\braket{\mathbb{S}_{B\overline B} ^2}
    \right]    \ .
\end{align}
In the single-step decays the $\sigma_C(A_{\CP})$ dependence on the electron-beam polarization for both ST and DT experiments is approximately given by Eq.~\eqref{eq:VACP01}. The $A_\Lambda$ uncertainties for ST, DT and the combined $e^+e^-\to J/\psi\to\Lambda\overline\Lambda$ measurement are plotted in Fig.~\ref{fig:senALL} as the function of $P_e$. Two cases of the detection efficiencies $\epsilon_B=1$ and $\epsilon_B=0.5$ are considered and ${\cal B}({\Lambda\to p\pi^-})=0.64$ is used. 
For the case with the reconstruction efficiency of 0.5 a two-times improvement of $\sigma_C$ is achieved for the combination, compared to the DT measurement only.
Of course, a detailed feasibility study which includes the detector response will be needed to determine the efficiency which can be obtained for the combined DT and ST measurement.  

An important background contribution which should be considered for the ST analysis of the $J/\psi\to\Lambda\overline{\Lambda}$ events is $J/\psi\to p K^- \overline{\Lambda}$ + c.c with ${\cal B} = (8.6\pm1.1)\times 10^{-4}$~\cite{ParticleDataGroup:2020ssz} as it will have a similar final state topology as the signal channel. Similar experimental considerations will also hold for the $J/\psi\to\Sigma\overline{\Sigma}$ two-body decay channels.

\begin{figure}
    \centering
\includegraphics[width=0.33\columnwidth]{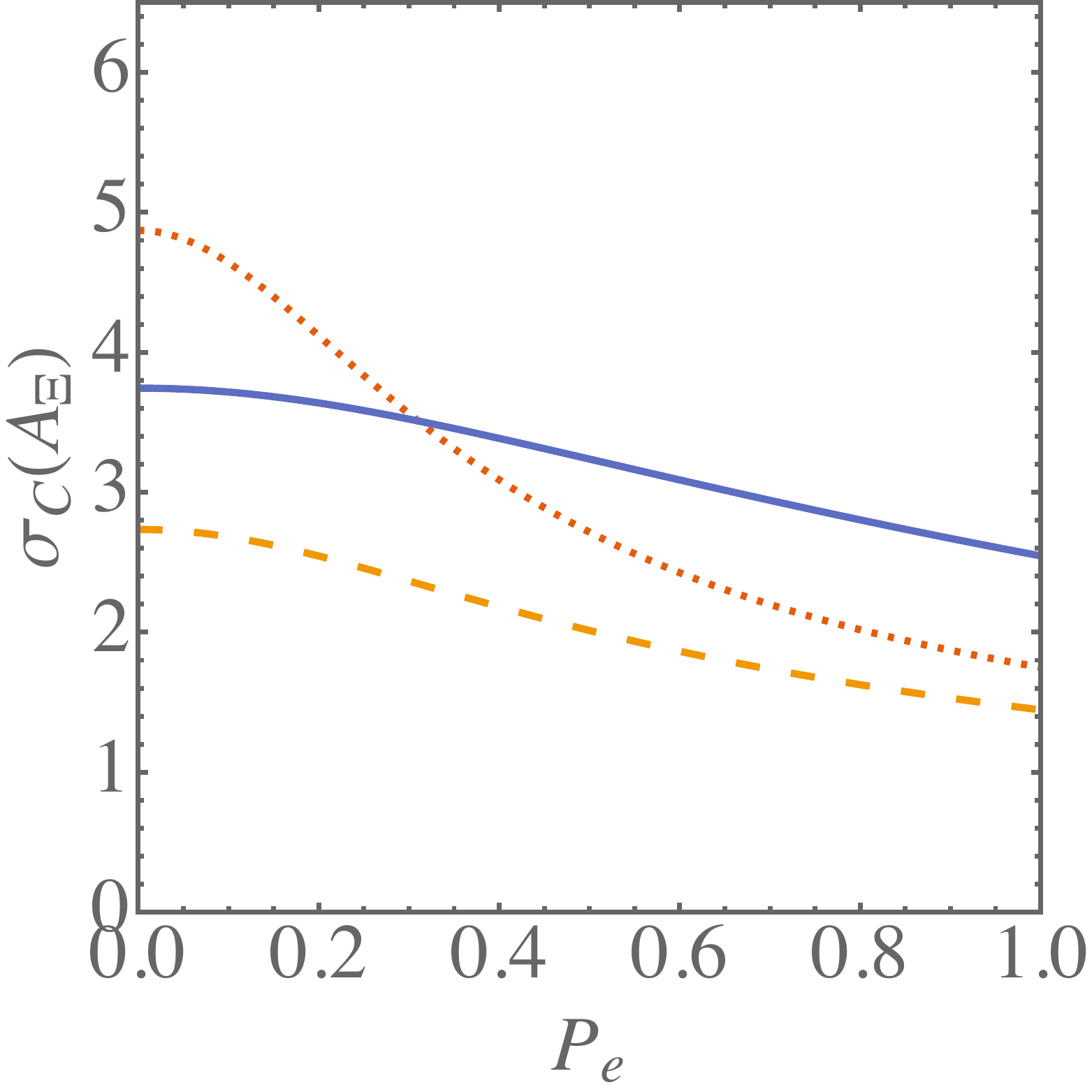}\put(-40,120){\large(a)}\includegraphics[width=0.33\columnwidth]{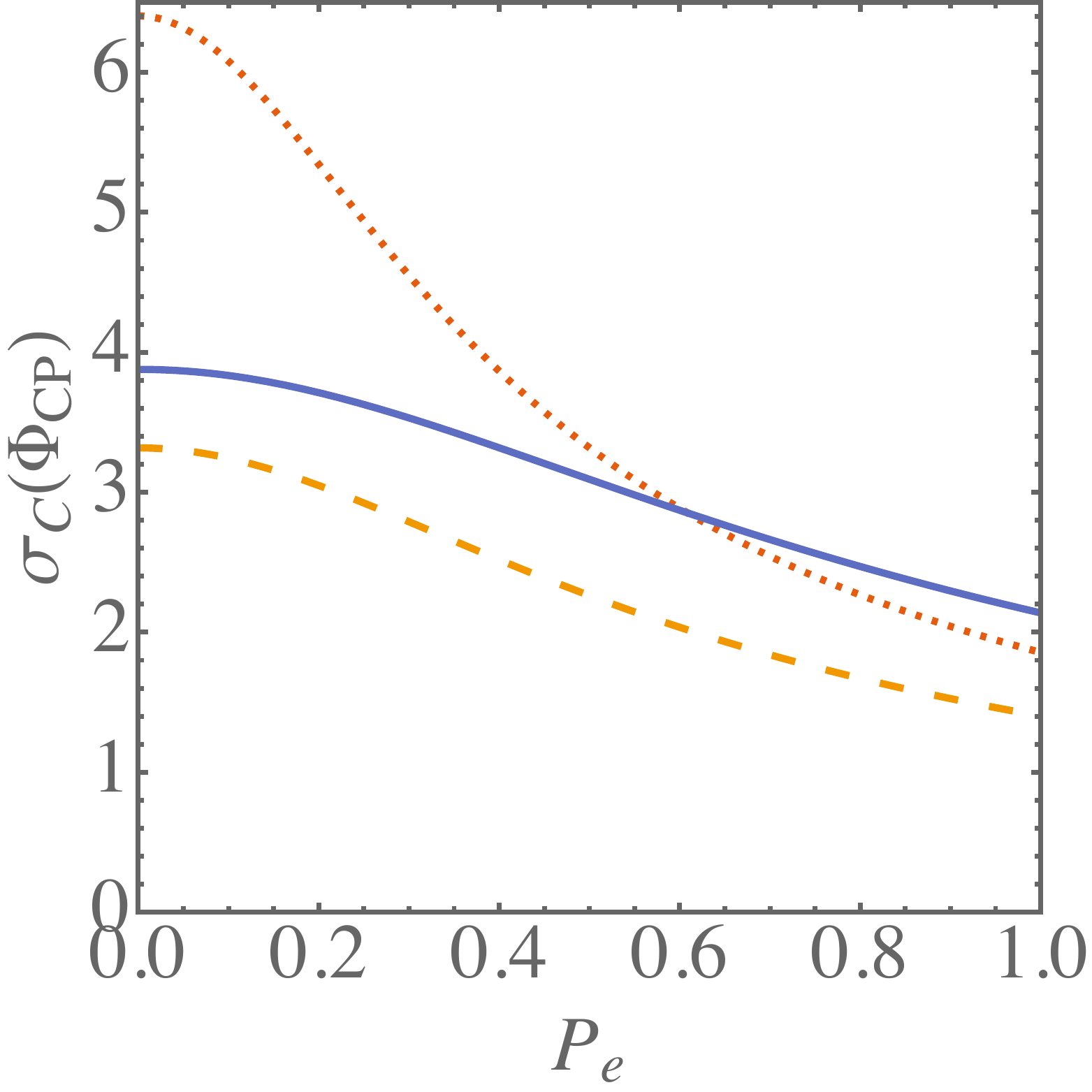}\put(-40,120){\large(b)}\includegraphics[width=0.33\columnwidth]{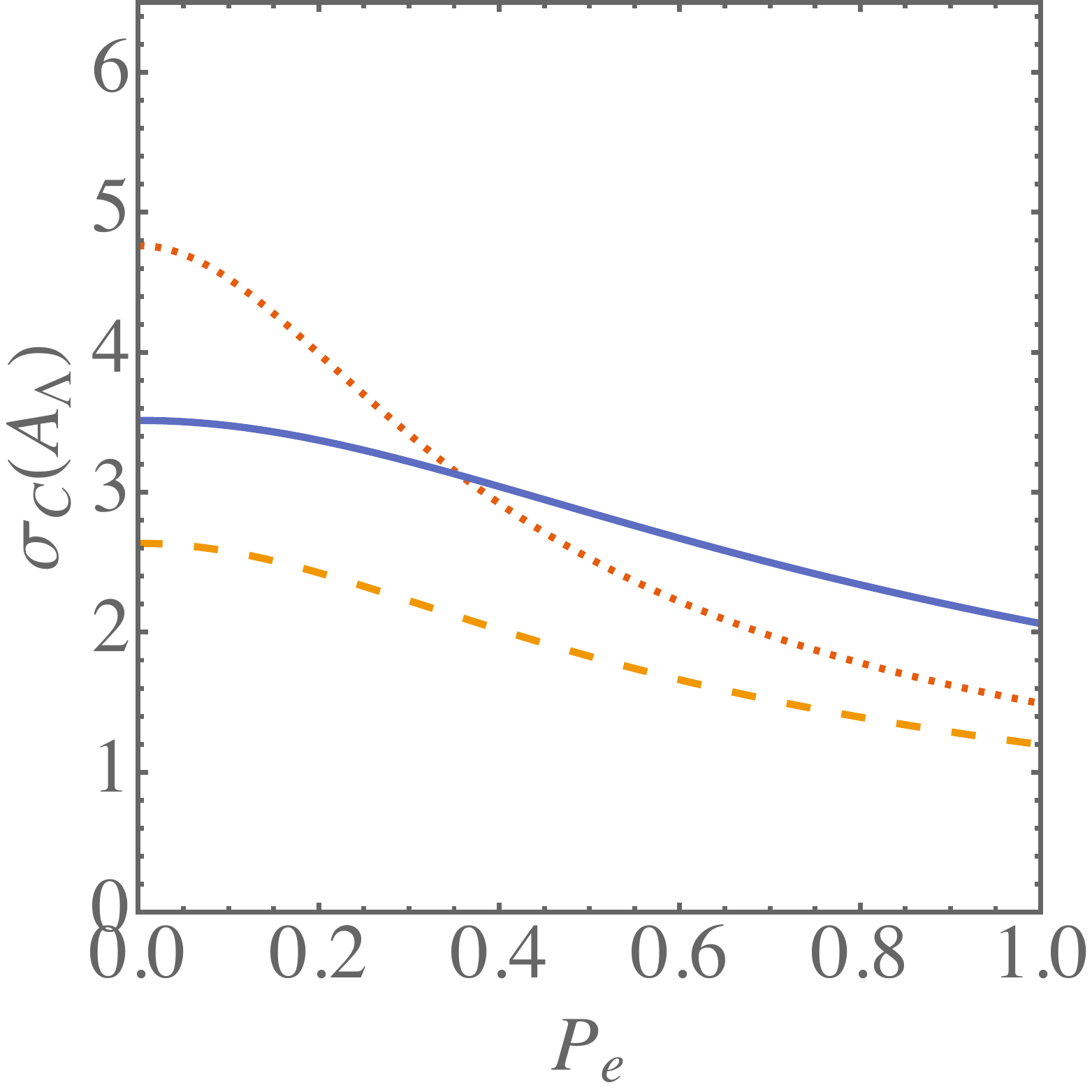}\put(-40,120){\large(c)}
    \caption{Statistical uncertainties, $\sigma_C$, for the CP-violation observables in the $\Xi^-\to \pi^-\Lambda(\to p\pi^-)+\text{c.c.}$ decay sequences from the $e^+e^-\to J/\psi\to\Xi\overline{\Xi}$ process: (a) $\sigma_C(A_\Xi)$, (b) $\sigma_C(\Phi_{\CP})$ 
    and (c) $\sigma_C(A_\Lambda)$ as a function of electron beam polarization $P_e$. The solid-blue  lines represent DT measurement.  The dotted-red lines represent contribution from ST events which do not contribute to the DT event class (statistically independent ST events). The orange-dashed lines represent the result from the combination of the two event classes. The detection efficiency of the $\Xi$-decay sequence was assumed $\epsilon_\Xi=\overline{\epsilon}_\Xi=0.5$ and branching fraction of the complete decay chain ${\cal B}=0.64$. The results are normalised to the number of the DT events.}
    \label{fig:senXi}
\end{figure}

The results for $A_{\Xi}$ and $B_{\Xi}$ in the $e^+e^-\to J/\psi\to\Xi\overline\Xi$ are shown in Fig.~\ref{fig:senXi}.  For the two-step decays the increasing beam polarization improves the ST uncertainties much faster compared to the corresponding DT uncertainties. For the polarization of $P_e=0.8$ the uncertainty of the ST experiment is better if we assume realistic
efficiency of $50\%$ deduced from a comparison of the BESIII ST \cite{Ablikim:2016iym} and DT \cite{Ablikim:2021qkn} analyses. 

Furthermore, the non-reducible background for the ST event samples are also expected to be low. The background channels to be considered are $J/\psi\to \gamma\eta_{c}(\to \Xi^- \overline{\Xi}^+)$, $J/\psi\to \Xi(1530)^{-}\overline{\Xi}^+\to \Xi^{-}\pi^0\overline{\Xi}^+$ and $J/\psi\to \Lambda \pi^- \overline{\Lambda} \pi^+$. While the first two channels can be suppressed using event kinematics variables, the third can be reduced by requiring a non-zero decay length for the $\Xi\to\Lambda \pi$ decay candidates. For the DT method the background contribution is 0.25\% and for ST the background is at the percent level, while roughly three times more ST events can be reconstructed compared to DT.

\paragraph{Polar-angle efficiency dependence}
Detectors at electron-positron colliders experiments have approximate cylindrical symmetry with axis along the beam directions (considerations for large-crossing angle are discussed in a separate paragraph) and uniform detection efficiency in the azimuthal angle.
However, the polar-angle coverage is limited. For example in the BESIII experiment  $|\cos\theta|<0.93$ for tracks of charged particles. 
The hyperons decay some centimetres away from the interaction point and the final state particles with large $|\cos\theta|$ values have low transverse momenta, which are more difficult to reconstruct.  These effects reduce the reconstruction efficiency at large values of $|\cos\theta|$. 

\begin{figure}
    \centering
    \includegraphics[width=0.49\textwidth]{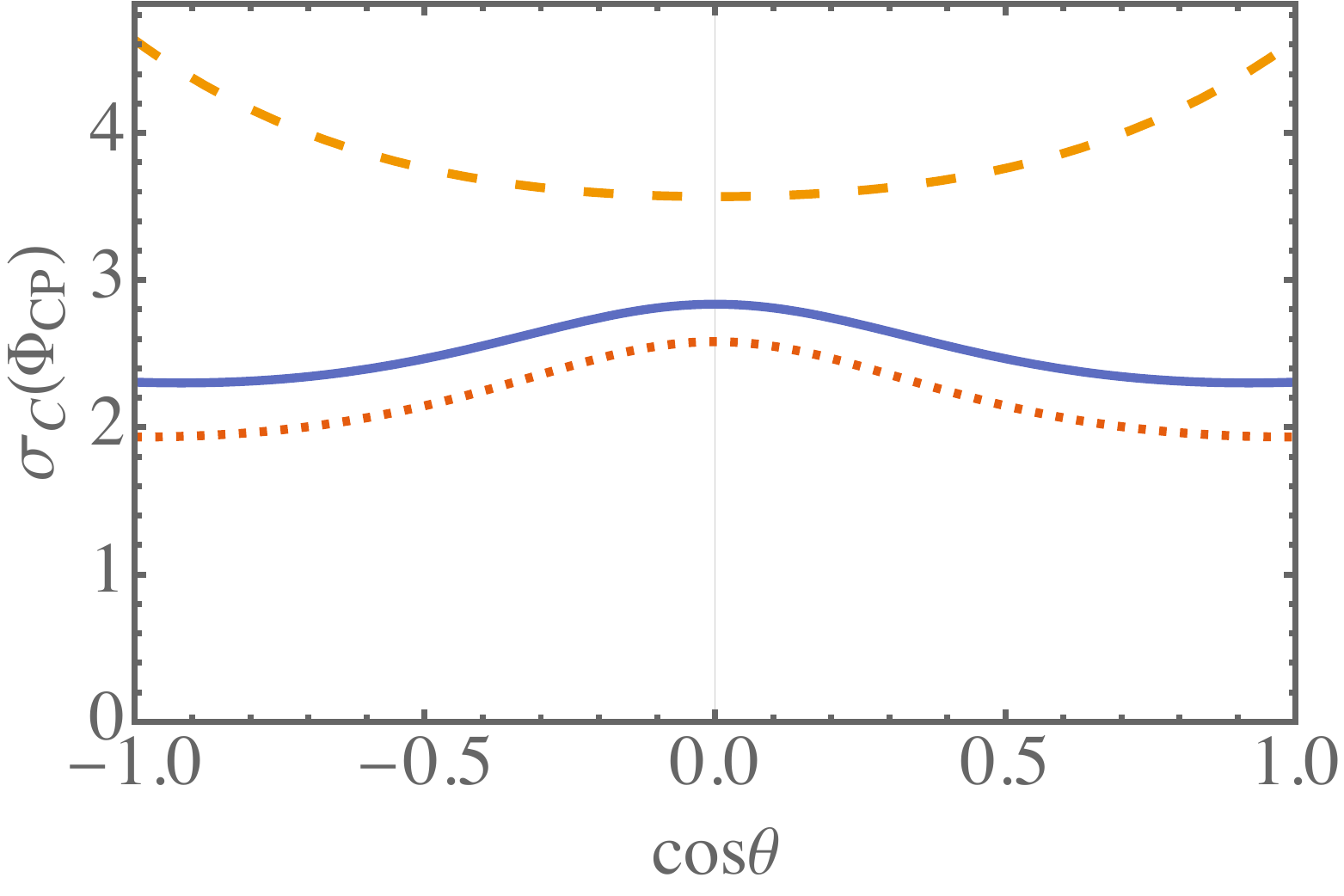}
\put(-180,120){\large(a)}        \includegraphics[width=0.49\textwidth]{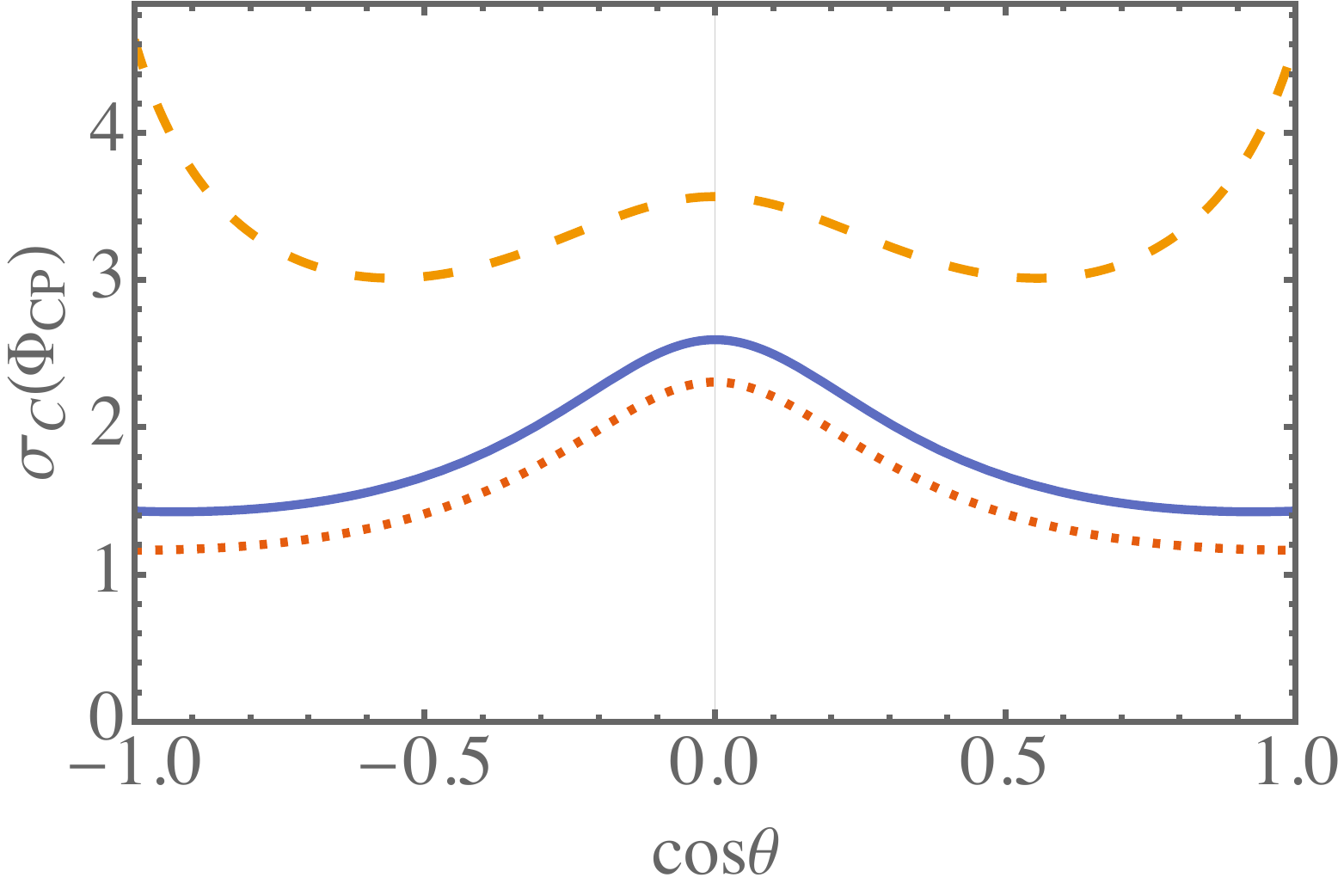}
\put(-180,120){\large(b)}    
\caption{Uncertainties $\sigma_C(\Phi_{\CP})$ as a function of the production angle $\cos\theta$ for (a) DT and (b) DT\&ST with 50\% efficiency experiment where dashed line (orange) shows $P_e=0$, solid line (blue) is for $P_e=0.8$ and dotted line (red) represents $P_e=1$.}
    \label{fig:PhiCPtheta}
\end{figure}

The event yield is a product of the efficiency and 
the differential cross section  of the $e^+e^-\to B\overline B$ process $\dd\Gamma /\dd\Omega \propto (1 + \alpha_{\psi}\cos^{2}\theta)$
as shown in Eq.~\eqref{eq:dSigdOm}. Since both $J/\psi\to\Lambda\overline{\Lambda}$ and  $J/\psi\to\Xi^{-}\overline{\Xi}^{+}$ have $\alpha_{\psi}>0$ (Table~\ref{tab:Prod}) the (anti)hyperons and the decay (anti)nucleons are more likely emitted in the forward and backward directions. 
The uncertainty as a function of the production angle $\cos\theta$ can be obtained by replacing the production tensor $\braket{ C^2}_{\mu\nu}$ by the normalised spin correlation matrix ${C_{\mu\nu}^2}/{C^2_{00}}$. The numerical expressions for the functions $\mathbb{P}^2_B(\cos\theta)$ and $\mathbb{S}^2_{B\overline{B}}(\cos\theta)$ are given in Appendix~\ref{app:avpol}. The results are shown in Fig.~\ref{fig:ALth} for $\sigma_C(A_\Lambda)$ in DT experiments in $e^+e^-\to J/\psi\to\Lambda\overline\Lambda$ and $e^+e^-\to J/\psi\to\Xi\overline\Xi$ for different values of the electron beam polarization. Corresponding plots for $\sigma_C(\Phi_{\CP})$ in the $e^+e^-\to J/\psi\to\Xi\overline\Xi$ DT and combined DT\&ST measurements are shown in Fig.~\ref{fig:PhiCPtheta}.

\begin{figure}
    \centering
    \includegraphics[width=0.49\textwidth]{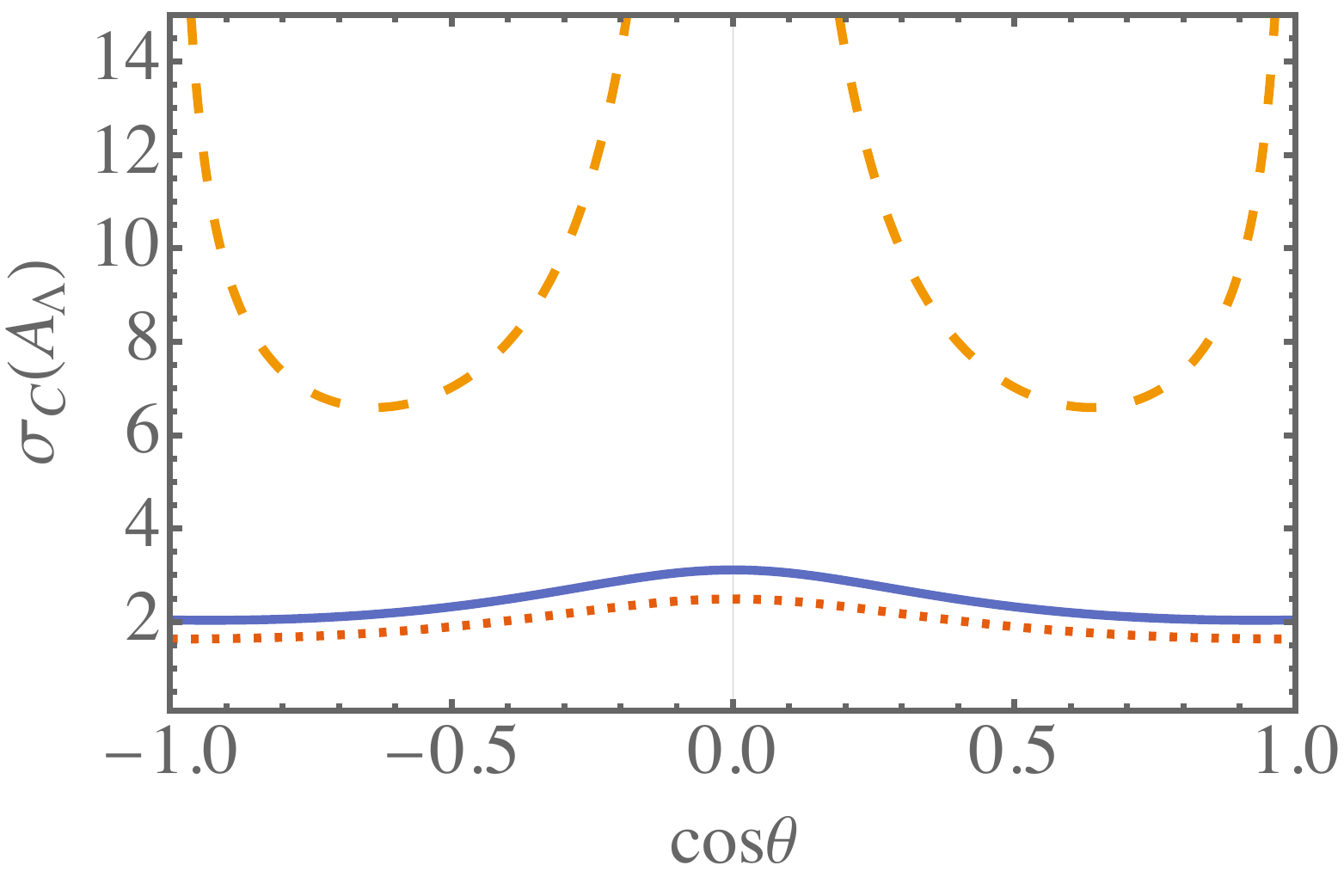}
\put(-180,120){\large(a)}    \includegraphics[width=0.49\textwidth]{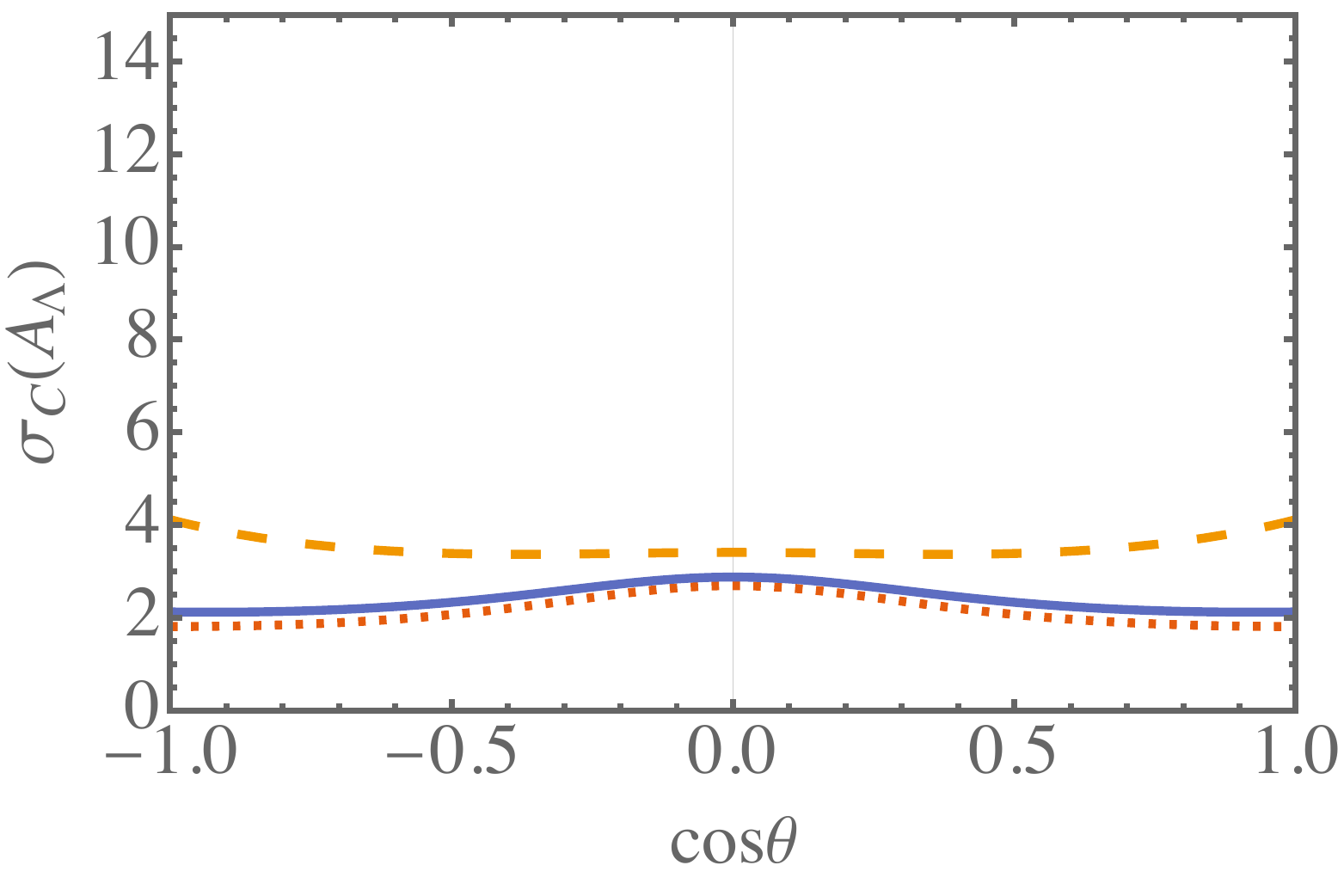}
\put(-180,120){\large(b)}
\caption{Uncertainties $\sigma_C(A_{\Lambda})$ in the DT measurement in 
    (a) $e^+e^-\to J/\psi\to\Lambda\overline\Lambda$ and (b) $e^+e^-\to J/\psi\to\Xi\overline\Xi$ processes as a function of the production angle $\cos\theta$ where dashed line shows $P_e=0$, solid line is for $P_e=0.8$ and dotted line representes $P_e=1$. }
    \label{fig:ALth}
\end{figure}

\paragraph{Large-angle collision scheme} The SCTF will use 
crab-waist collision scheme meaning larger crossing angle than at BEPCII (22 mrad). The presently considered crossing angle is 60 mrad~\cite{Levichev:2018cvd,Luo:2018njj}. However, in Ref.~\cite{Telnov:2020rxp} much larger crossing angles, up to 500 mrad, are considered in conjunction with a novel c.m. energy monochromatization scheme. The monochromatization could increase the number of the $J/\psi$ events and therefore it is worthwhile to discuss some of the consequences of such collision arrangement for the acceptance in the hyperon CP-violation tests.

\begin{figure}
    \centering
\includegraphics[width=0.49\columnwidth]{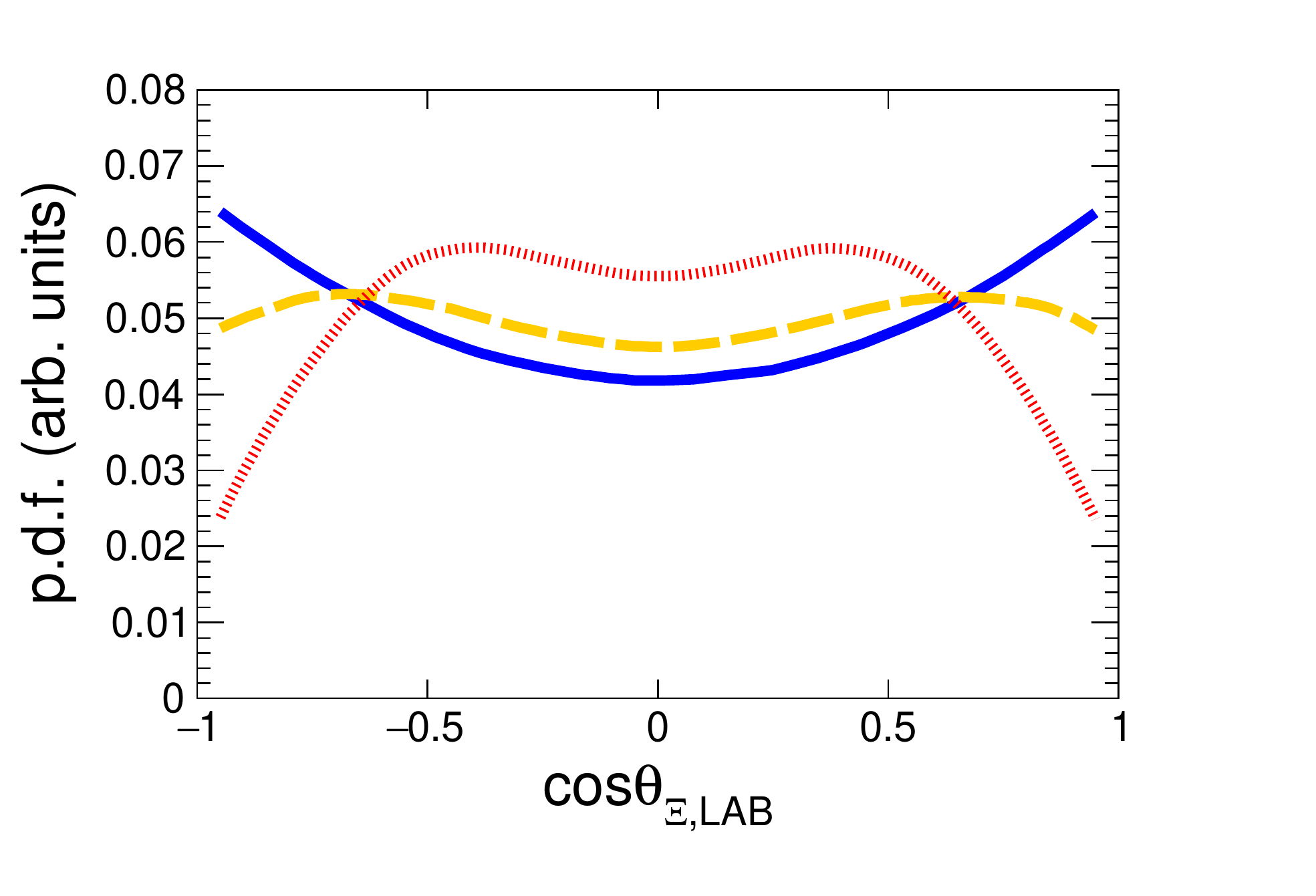}
\put(-180,120){\large(a)}    \includegraphics[width=0.49\columnwidth]{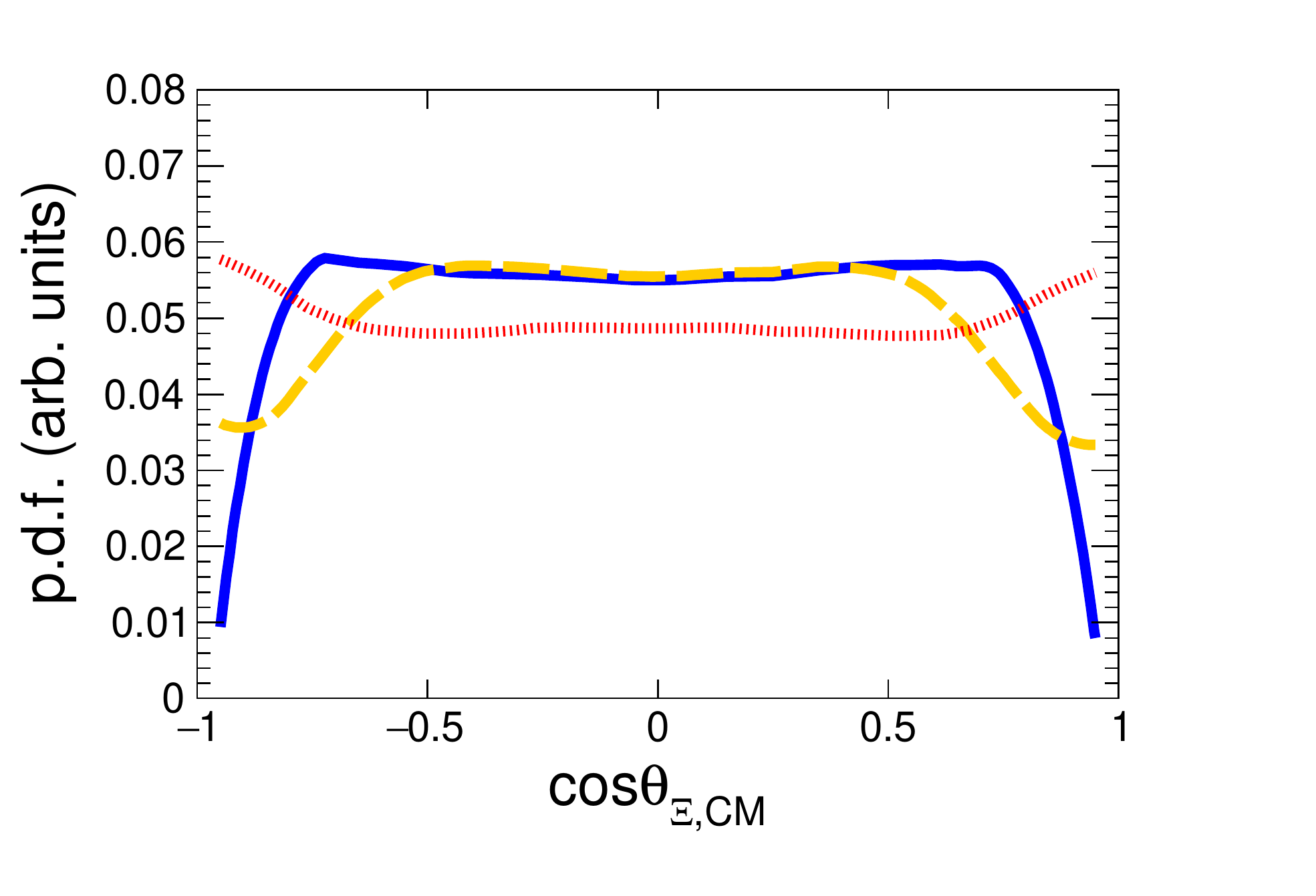}
\put(-180,120){\large(b)}    
    \caption{(color online) Production angle distribution for beam-crossing angles 0 rad (blue solid), 0.3 rad (orange dashed) and 0.5 rad (red dotted). (a) The $\Xi$ production angle in the detector frame. (b) The $\Xi$ production angle in the electron--positron c.m. frame for the events where all six charged tracks are accepted in the detector $|\cos\theta_{\rm LAB}|<0.93$.}
    \label{fig:xangle1}
\end{figure}

\begin{figure}
    \centering
    \includegraphics[width=0.49\columnwidth]{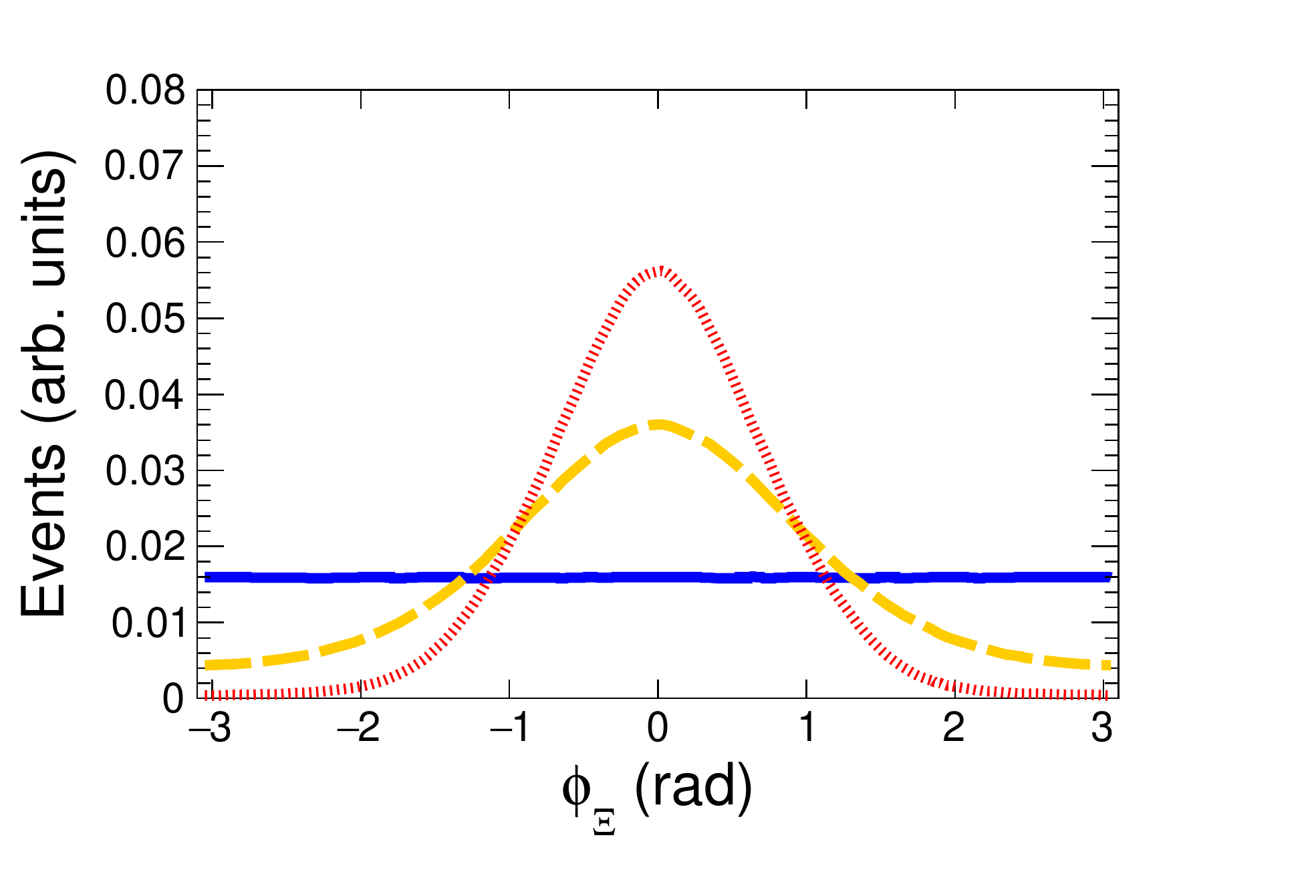}
\put(-180,120){\large(a)}    \includegraphics[width=0.49\columnwidth]{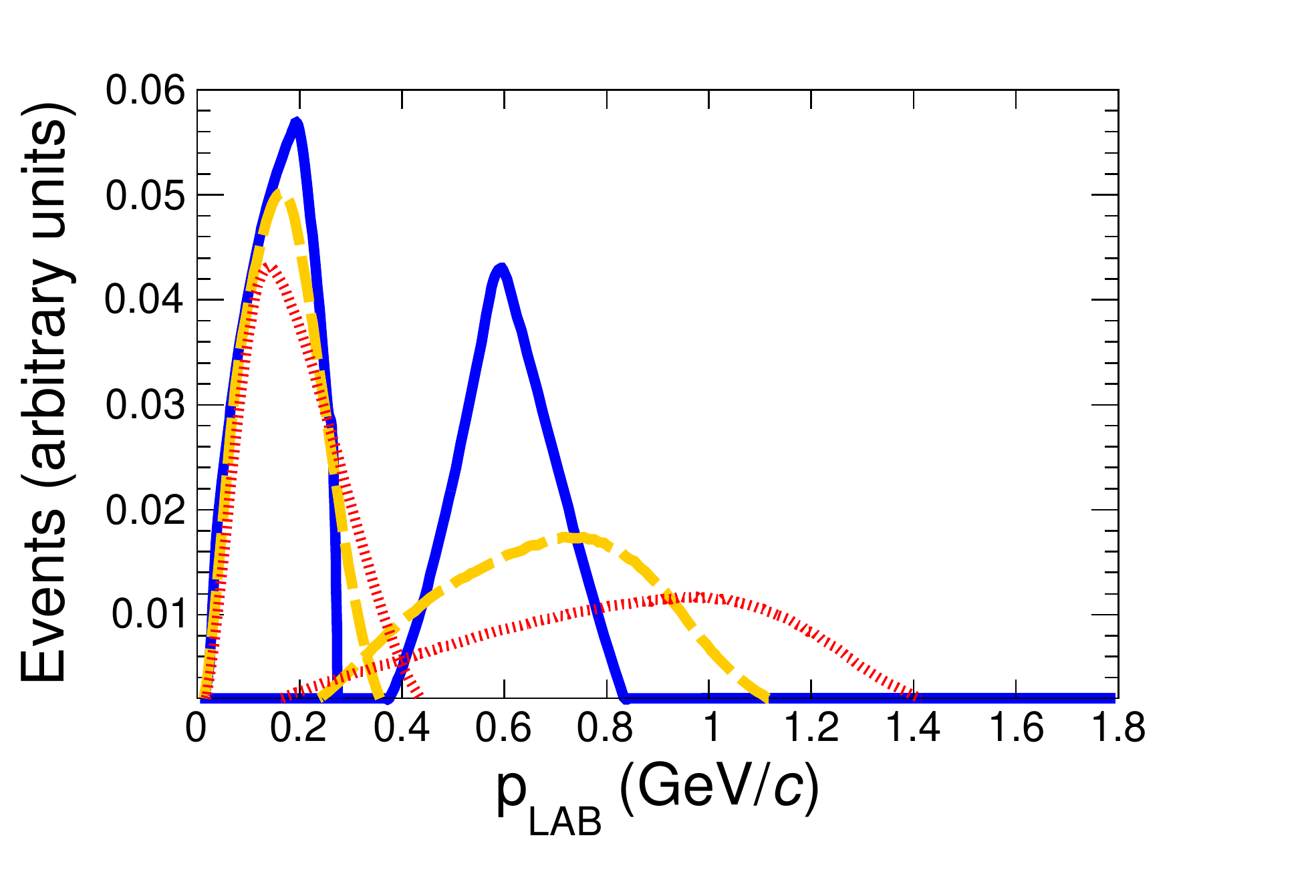}
\put(-160,120){\large(b)}    
\caption{(color online) (a) Azimuthal distribution of $\Xi_{\rm LAB}$ and (b) momentum distributions for all final state particles for beam scattering angles 0 rad (blue solid), 0.3 rad (orange dashed) and 0.5 rad (red dotted).} 
    \label{fig:xangle2}
\end{figure}

In such collision scheme the detector reference frame is significantly different from the electron--positron c.m. system.
This has impact on both angular acceptance and the detection efficiency as a function of the measured-particles momenta
and it has to be considered in the detector design. For example, the polar-angle, $\theta_{\rm LAB}$, distribution of $\Xi^-$ in the detector rest frame  is given in Fig.~\ref{fig:xangle1}(a) for the 0.0, 0.3 and 0.5 rad crossing angles. If the decay particles are measured only in the $|\cos\theta_{\rm LAB}|<0.93$ range as in the BESIII detector the observed $\Xi$ production-angle distribution in the electron--positron c.m. system is as in Fig.~\ref{fig:xangle1}(b).  A large beam-crossing angle will also significantly affect the azimuthal-angle distribution, as shown in  Fig.~\ref{fig:xangle2}(a), and the momenta distributions of the final-state protons and pions will overlap with each other, as shown in Fig.~\ref{fig:xangle2}(b).

\section{Outlook}
\label{sec:Conclusion}
We have advocated the importance of  CPV studies in  hyperon decays as a complementary tool to the studies in kaon decays.  Using recent experimental results, we have revised and updated the amplitudes of the $\Lambda$ and $\Xi$ hadronic two-body decays.

The main part of this report discusses the implications of the polarized-electron beams for CPV tests in the non-leptonic hyperon decays at SCTF, using entangled baryon--antibaryon pairs from $J/\psi$ decays  with data sets of $10^{12}$ $J/\psi$ events. The use of the polarization, together with additional improvements of the analysis techniques, shows the potential to reach a precision compatible with the size of the predicted SM signal.

Using an analytical approximation for the Fisher information matrices of the CPV observables, we can understand how the precision of such measurements depends on the polarization and spin-correlation terms in the production processes. Some of the obtained analytical results can be directly extended to charm baryon CPV studies. At SCTF, they can be studied in close-to-threshold $e^+e^-\to B\overline B$ processes. For such processes, the analytic results of Sec.~\ref{sec:SSdecay} and Sec.~\ref{sec:DSdecay} can be taken as a starting point.
 The main difference in the strategy for charmed baryons is due to the fact that the branching fractions for two-body nonleptonic decays are small, and the DT analysis likely will not be feasible.

In addition to the $e^+e^-\to B\overline B$ processes, the HyperCP-type experiments can be an interesting option for CP tests and decay parameter determination, provided that sources of (anti)baryons with large initial polarization are available. Possible candidate processes are semileptonic decays of charmed baryons $\Xi^0_c\to\Xi^-\ell^+\nu_\ell$ or two-body hadronic decays like $\Xi^0_c\to\Xi^-\pi^+$ with large value of the decay parameter $\alpha=0.6(4)$ and relatively large branching fraction $1.2$~\%
\cite{ParticleDataGroup:2020ssz}. For such studies, unpolarized charmed baryons that are abundantly produced at the LHC in $pp$ collisions can be used. Again, our analytic formulas can be used to provide a first estimate of the statistical uncertainties for such experiments.

We have left out a potentially interesting discussion of the uncertainties of the decay parameters $\alpha_D$ and $\phi_D$. The  $\alpha_D$ parameter is correlated with production parameters and 
extraction of uncertainties and correlation coefficients requires inverting information matrices with larger dimensions and the analytical results might be difficult to interpret. The same is valid for the production parameters $\alpha_\psi$ and $\Delta\Phi$ that are relevant for the experiments where the goal is to study the properties of the production process. Usually such experiments have a limited number of the collected events and analysis is done assuming the decay parameters are known.  

\begin{acknowledgments}
This  work was supported in part by National Natural Science Foundation of China (NSFC) under Contract No. 
  11935018 and {Polish National Science Centre through the Grant 2019/35/O/ST2/02907}. P.A. is grateful for the support from Olle Engkvist Foundation and Lundström-Åman Foundation. 
   V.B. contribution is supported by the CAS President’s International Fellowship Initiative (PIFI) (Grant No. 2021PM0014).  In addition a support from the Swedish Research Council grant No. 2021-04567 is acknowledged.

\end{acknowledgments}
\appendix

\section{Isospin decomposition}
\label{sec:Isospin}

For the analysis presented here, we {\it assume} isospin symmetry for the elementary weak decay process 
but want to account for the different phase space caused 
by isospin-violating mass splittings. The basic parameters (see also Appendix \ref{sec:AppEFT}) in the Feynman matrix element of a weak decay process of a spin-1/2 
baryon to another spin-1/2 baryon and a pseudoscalar meson are related to the partial-wave amplitudes via 
(see mini-review 79.2.\ ``Hyperon nonleptonic decays'' of \cite{ParticleDataGroup:2020ssz})
\begin{eqnarray}
  g_S=S \,, \qquad  g_P = P\ \frac{E+M}{\vert {\bf q}\vert}  \,.  \label{eq:gSgPetc}  
\end{eqnarray}
We want to account for isospin breaking caused by the P-wave kinematical factor $\vert {\bf q}\vert/(E+M)$. 
Here $\vert \bf q\vert$, $E$ and $M$ denote the momentum, energy and mass, respectively, of the final baryon 
in the rest frame of the initial baryon. 

Suppose we consider two processes, labelled by 1 and 2, connected by isospin symmetry. 
We start with our isospin symmetry assumption for the basic parameters:
\begin{eqnarray}
  g_{S1} \stackrel{!}{=} g_{S2} \,, \qquad  g_{P1} \stackrel{!}{=} g_{P2} \,.
  \label{eq:ab-isos}
\end{eqnarray}
Corrections $\Delta_1$ and $\Delta_2$ to the kinematic terms due to different masses in processes 1 and 2 can be written as
\begin{eqnarray}
  S_1 = S_2 \,, \qquad  P_1(1+\Delta_1) =: P_1\frac{E_1+M_1}{E+M} \frac{\vert {\bf q}\vert}{\vert {\bf q}_1\vert}=  P_2(1+\Delta_2)\ ,
  \label{eq:SP-isos}
\end{eqnarray}
where $S_{1,2}$ and $P_{1,2}$ are the $S$ and $P$ amplitudes, respectively,  for the two processes, and $M$ and $\vert {\bf q}\vert$ are arbitrary parameters such that $|M-M_{1,2}|\le|M_1-M_2|$ and $\left|\vert {\bf q}\vert-\vert {\bf q}_{1,2}\vert\right|\le|\vert {\bf q}_1\vert-\vert {\bf q}_2||$, ensuring that $\Delta_{1,2} \ll 1$. The kinematic-correction  factors are related as
\begin{eqnarray}
  k_{12}:=\frac{1+\Delta_1}{1+\Delta_2} = \frac{\vert {\bf q}_2\vert}{\vert {\bf q}_1\vert} \frac{E_1+M_1}{E_2+M_2}\approx 1+\Delta_1-\Delta_2 \,.
  \label{eq:defratio}
\end{eqnarray}
For the studied reactions we have:
\begin{eqnarray}
  k_\Lambda = \frac{\vert {\bf q}_p\vert}{\vert {\bf q}_n\vert} \frac{E_n+M_n}{E_p+M_p}\myeq  1-0.031158(8)
  \label{eq:concrLam}
\end{eqnarray}
and
\begin{eqnarray}
  k_\Xi = \frac{\vert {\bf q}_{\Xi^-} \vert}{\vert {\bf q}_{\Xi^0}\vert} \frac{E_{\Xi^0}+M_{\Xi^0}}{E_{\Xi^-}+M_{\Xi^-}} \myeq1+0.0319(19)\,.
  \label{eq:concrXi}  
\end{eqnarray}
In \eqref{eq:concrXi} the labels refer to the state that decays, while
in \eqref{eq:concrLam} to the final baryon.

The isospin decomposition,  using notation similar to Ref.~\cite{Overseth:1970qv},  for the $L=S,P$ amplitudes of $\Lambda\to p\pi^-$ $([\Lambda p])$ and  $\Lambda\to n\pi^0$  $([\Lambda n])$ reads:
\begin{align}
    L_{[\Lambda p]}&=-\sqrt{\frac{2}{3}}L_{1,1}\exp\!{(i\xi^{L}_{1,1}+i\delta_{1}^L)}\!+\!
    \sqrt{\frac{1}{3}}L_{3,3}\exp\!{(i\xi^{L}_{3,3}+i\delta_{3}^L)},\\ \nonumber
    L_{[\Lambda n]}&=\phantom{-}\sqrt{\frac{1}{3}}L_{1,1}\exp\!{(i\xi^{L}_{1,1}+i\delta_{1}^L)}\!+\!
    \sqrt{\frac{2}{3}}L_{3,3}\exp\!{(i\xi^{L}_{3,3}+i\delta_{3}^L)} \ . \nonumber
\end{align}
The equations imply that the $P$ amplitudes are multiplied by the kinematic correction factors:
$P_{[\Lambda p]}\to(1+\Delta_{[\Lambda p]})P_{[\Lambda p]}$ and $P_{[\Lambda n]}\to(1+\Delta_{[\Lambda n]})P_{[\Lambda n]}$.
For $\Xi^-\to \Lambda\pi^-$ $([\Xi -])$ and $\Xi^0\to \Lambda\pi^0$ $([\Xi 0])$, one has: 
\begin{align}
    L_{[\Xi -]}&=L_{1,2}\exp\!{(i\xi^{L}_{1,2}+i\delta_{2}^L)}\!+\!\frac{1}{2}L_{3,2}\exp\!{(i\xi^{L}_{3,2}+i\delta_{2}^L)},\\
    L_{[\Xi 0]}&=\sqrt{\frac{1}{2}}L_{1,2}\exp\!{(i\xi^{L}_{1,2}+i\delta_{2}^L)}\!-\!\sqrt{\frac{1}{2}}L_{3,2}\exp\!{(i\xi^{L}_{3,2}+i\delta_{2}^L)}\ \label{eq:Lisospin}.
\end{align}
Similarly, $P_{[\Xi -]}\to(1+\Delta_{[\Xi -]})P_{[\Xi -]}$ and $P_{[\Xi 0]}\to(1+\Delta_{[\Xi 0]})P_{[\Xi 0]}$.
In practice, we want to separate the kinematic isospin breaking effects in the overall normalisation 
and in the $P$ amplitude. Therefore, we choose a relation between $\Delta_1$ and $\Delta_2$ such as these corrections cancel in the lowest order for the overall normalisation. For $\Lambda$ decays, this is achieved by setting $\Delta_{\Lambda}:=\Delta_{[\Lambda p]}=(k_\Lambda-1)/3$ and $\Delta_{[\Lambda n]}=-2(k_\Lambda-1)/3=-2\Delta_{\Lambda}$ where
$k_\Lambda$ is defined in \eqref{eq:concrLam}. In the same way, for $\Xi$ decays this obtained through $\Delta_{\Xi}:=\Delta_{[\Xi -]}=(k_\Xi-1)/3$ and $\Delta_{[\Xi 0]}=-2(k_\Xi-1)/3=-2\Delta_{\Xi}$, where
$k_\Xi$ is defined in \eqref{eq:concrXi}.

The $\Delta I=1/2$ transitions dominate what one sees in the comparison of the decay widths of the related decays. 
The decay widths are related to  the amplitude squared by the kinematic pre-factor given in Eq.~\eqref{eq:Gamma}.
The decays of the two cascades in the  $\Delta I=1/2$ limit are:
\begin{align}
 |{\cal A}_{[\Xi -]}|^2  &=  S_{1,2}^2 + P_{1,2}^2(1 + 2\Delta_{\Xi}) \  \text{and}\ \
  2|{\cal A}_{[\Xi 0]}|^2  =S_{1,2}^2 + P_{1,2}^2(1 - 4\Delta_{\Xi}) \ .
\end{align}
This relation agrees well with the measured ratio of the cascade life times  of 1.77(6) derived using the life-time values $\tau(\Xi^0)=2.90(9)\times10^{-10}$\,s and  $\tau(\Xi^-)=1.639(15)\times10^{-10}$\,s~\cite{ParticleDataGroup:2020ssz}. 
For $\Lambda$ decays, the relation between the amplitudes squared in the $\Delta I =1/2$ limit reads:
\begin{align}
 |{\cal A}_{[\Lambda p]}|^2 &= \frac{2}{3}\left[S_{1,1}^2 + P_{1,1}^2(1 + 2\Delta_{\Lambda})\right]   \  \text{and}\ \
  2|{\cal A}_{[\Lambda n]}|^2  =  \frac{2}{3}\left[S_{1,1}^2 + P_{1,1}^2(1 -4\Delta_{\Lambda})\right]  \ .
\end{align}
In this case, the relation between the amplitudes is supported by the ratio of the branching fractions of the two decays of
1.785(29). The phase space volume ratio $-/0$ is $r_\Lambda=0.9658(1)$, however, since the decay channel $\Lambda\to p\pi^-$ includes two charged particles, significant e.m. corrections are expected~\cite{Jarlskog:1967xsa,Belavin:1968wtg}.

The leading-order (LO) corrections due to $\Delta I=3/2$  contributions can be deduced from the partial widths and values of the decay parameters of the isospin modes. For the cascades the sum of the amplitudes squared, $|{\cal A}_{[\Xi -]}|^2+|{\cal A}_{[\Xi 0]}|^2$, does not receive any correction and is 
\begin{align}
 |{\cal A}_{[\Xi -]}|^2+|{\cal A}_{[\Xi 0]}|^2&=
\frac{3}{2}\left(S_{1,2}^2+P_{1,2}^2\right)\label{eq:GXi2}\ .
\end{align}
The sum can be treated as the overall normalisation of the amplitudes by setting ${S_{1,2}^2+P_{1,2}^2}$ to one. Instead, the following linear combination of the amplitudes squared changes and receives the leading-order correction expressed as: 
\begin{align}
 |{\cal A}_{[\Xi -]}|^2-2|{\cal A}_{[\Xi 0]}|^2&=
3\left(S_{1,2}S_{3,2}+P_{1,2}P_{3,2} +2 P_{1,2}^2\Delta_{\Xi} \right) \label{eq:GXi1}\ .
\end{align}
To translate these relations to the life-time measurements, they have to be corrected by  the ratio $r_\Xi$ of the  kinematic pre-factors for the charged and neutral cascades from Eq.~\eqref{eq:Gamma}.
Using the PDG values for the masses~\cite{ParticleDataGroup:2020ssz} the numerical value of the ratio is $r_\Xi=1.027(2)$.  Dividing Eq.~\eqref{eq:GXi1} by Eq.~\eqref{eq:GXi2} and converting the  the amplitudes squared to the life times, one derives the  experimental constraint: 
\begin{align}
{\frac{1}{2}\frac{\tau(\Xi^0) -2\tau(\Xi^-)r_\Xi}{ \ \tau(\Xi^0) +\phantom{2}\tau(\Xi^-)r_\Xi}} &=s^2s_{3}+(1-s^2)(p_{3} + 2\Delta_{\Xi}) \label{eq:Xi1} \\
&\myeq -0.051(11)\nonumber \ ,
\end{align}
where  $s_3$ and $p_3$ denote the ratios $S_{3,2}/S_{1,2}$ and $P_{3,2}/P_{1,2}$, respectively, and $s:=S_{1,2}$.
The values of the decay parameters $\alpha$ for the two cascades can be used to derive two additional constraints. The isospin average of the decay parameters\footnote{Here we use simplified notation $\alpha_{[\Xi0]}\to\alpha_0$ and $\alpha_{[\Xi-]}\to\alpha_-$. }
\begin{align}
\alpha_{\Xi}:=\frac{2{\alpha_-}+ {\alpha_0}}{3}&={2S_{1,2}P_{1,2}}
\cos(\delta_{2}^P-\delta_{2}^S)\label{eq:Xi2}\\
&\approx 2s\sqrt{1-s^2}\label{eq:Xi2a}\\
&\myeq -0.368(4)\nonumber
\end{align}
remains unchanged and represents the decay parameter in the $\Delta I =1/2$ limit, the same value for the two decay modes. To obtain the approximate form  in Eq.~\eqref{eq:Xi2a} we use the fact that the experimental value  $\cos(\delta_{2}^P-\delta_{2}^S)=0.999(1)$, {\textit i.e.} it is consistent with one~\cite{Ablikim:2021qkn}.  Solving for $s$ one gets two solutions,
\begin{equation}
   s^2=\frac{1}{2}\left(1\pm\sqrt{1-\alpha_{\Xi}^2}\right)\ .
\end{equation}
The solution with $|S|^2>|P|^2$ is selected since  Eq.~\eqref{eq:PDGdef} and the experimental value of the $\phi_\Xi$ parameter imply that $\sgn(|S|^2-|P|^2)=\sgn(\cos\phi_\Xi)>0$.
The measurement of the difference $\alpha_--\alpha_0$ provides the second relation to determine $s_3$ and $p_3$:
\begin{align}
  {\alpha_-}- {\alpha_0}&={3}\left({S_{1,2}^2-P_{1,2}^2}\right) \left(S_{1,2}P_{3,2}-S_{3,2}P_{1,2}+2P_{1,2}S_{1,2}\Delta_\Xi\right)\cos(\delta_{2}^P-\delta_{2}^S)\label{eq:Xi3},\\
 \frac{{\alpha_-}- {\alpha_0}}{\alpha_{\Xi}}&=\frac{3}{2}(2s^2-1) \left(p_{3}-s_{3}+2\Delta_{\Xi} \right)\label{eq:Xi3a}\\
  &\myeq0.092(25)\ .
\end{align}
The determined amplitudes are given in Table~\ref{tab:SPamp}.  The size of the  $S_{3,2}$ amplitude is 5\%  of $S_{1,2}$, while the   $P_{3,2}$ amplitude is consistent with zero (less than 3\% of $P_{1,2}$ within one s.d.).
The strong-phase difference $\delta_{2}^P-\delta_{2}^S$ can be determined from the measurement of the $\beta$ decay parameters:
\begin{align}
    \frac{\beta_-}{\alpha_-}=\frac{\beta_0}{\alpha_0}=
    \tan(\delta_{2}^P-\delta_{2}^S) \ .\label{eq:Xi4}
\end{align}
Using the measured $\phi_-$ and $\phi_0$ parameters for the cascades listed in Table~\ref{tab:decayproperties} the average experimental value of the strong phase $\delta_{2}^P-\delta_{2}^S$ of $1(4)^\circ$ is consistent with zero. The size of the error indicates considerably smaller value than the pion--nucleon phase shifts relevant for the $\Lambda$ and $\Sigma$ decays (given in Table.~\ref{tab:Sphases}).
\begin{table}[ht]
    \caption{Amplitudes for the $\Delta I=1/2$ and $\Delta I=3/2$ transitions in the $\Lambda$- and $\Xi$-hyperons non-leptonic decays.\label{tab:SPamp}}
    \centering
\begin{ruledtabular} 
\begin{tabular}{lrrrrrr}
         &\multicolumn{2}{c}{$\Delta I=1/2$}&\multicolumn{2}{c}{$\Delta I=3/2$}&
         \multicolumn{2}{c}{$(\Delta I=3/2 )/(\Delta I=1/2)$}\\
         &\multicolumn{1}{c}{$S$}&\multicolumn{1}{c}{$P$} &\multicolumn{1}{c}{$S$}&\multicolumn{1}{c}{$P$}& \multicolumn{1}{c}{$s_3$}& \multicolumn{1}{c}{$p_3$}  \\
   $\Xi\to\Lambda\pi$      & $-0.9823(04)$ &$0.187(02)$    &  $0.052(11)$ &$-0.002(4)$ & $-0.053(11)$  & $-0.008(20)$  \\
 $\Lambda\to N\pi$        &  $0.9145(18)$ &$0.405(04)$&  $0.011(06)$& $0.016(06)$ & $0.012(06)$ &$0.038(15)$  \\
\end{tabular}
\end{ruledtabular}    
\end{table}

The corresponding relations for the $\Lambda$ decays are more complicated since there are four strong phases, however, the required strong phases for the pion--nucleon final state interactions are well known and given in Table~\ref{tab:Sphases}. Including the leading-order $\Delta I=3/2$ corrections the relations between the decay widths can be expressed as
\begin{align}
|{\cal A}_{[\Lambda p]}|^2+\phantom{2}|{\cal A}_{[\Lambda n]}|^2&=
S_{1,1}^2+P_{1,1}^2\label{eq:GLa2}
\end{align}
and 
\begin{align}
 |{\cal A}_{[\Lambda p]}|^2-2|{\cal A}_{[\Lambda n]}|^2&=
-\sqrt{8}\left(P_{1,1}P_{3,3}\cos(\delta_{1}^P-\delta_{3}^P)+
S_{1,1}S_{3,3}\cos(\delta_{1}^S-\delta_{3}^S)\right)+4P_{1,1}^2\Delta_{\Lambda} \label{eq:GLa1}\ .
\end{align}
We fix the overall normalisation of   the amplitudes by setting ${S_{1,1}^2+P_{1,1}^2}$ to one and set the notation for the amplitudes: $s:=S_{1,1}$, $s_3:=S_{3,3}/S_{1,1}$ and $p_3:=P_{3,3}/P_{1,1}$, where the reuse
of the same symbols as for the cascades should not lead to a confusion. By dividing Eq.~\eqref{eq:GLa1} by  Eq.~\eqref{eq:GLa2} and expressing the amplitudes squared by the branching fractions $\cal B$, one gets
\begin{align}
 -\frac{1}{\sqrt{8}}\frac{{\cal B}({[\Lambda p]})-2{\cal B}({[\Lambda n]})r_\Lambda}{{\cal B}({[\Lambda p]})+\phantom{2}{\cal B}({[\Lambda n]})r_\Lambda}&=
(1-s^2)(p_{3}\cos(\delta_{1}^P-\delta_{3}^P)-\sqrt{2} \Delta_{\Lambda})+
s^2s_{3}\cos(\delta_{1}^S-\delta_{3}^S) \\
  &\myeq 0.019(4) \nonumber\ ,
\end{align}
where $r_\Lambda$ is the ratio of the kinematic factors in Eq.~\eqref{eq:Gamma} for the $\Lambda\to p\pi^-$ and $\Lambda\to n\pi^0$ decay modes.  The results for the $\alpha$ parameters are\footnote{Here, the simplified notation reads $\alpha_{[\Lambda n]}\to\alpha_0$ and $\alpha_{[\Lambda p]}\to\alpha_-$. }
\begin{align}
\alpha_\Lambda:=\frac{2{\alpha_-}+ {\alpha_0}}{3}&={2S_{1,1}P_{1,1}}\cos(\delta_{1}^P-\delta_{1}^S)\\
&=2s\sqrt{1-s^2}\cos(\delta_{1}^P-\delta_{1}^S)\\
  &\myeq 0.734(6)\nonumber,\\
  \frac{{\alpha_-}-\alpha_0}{\alpha_\Lambda}&=\frac{3}{\sqrt{2}}\frac{\Delta\alpha_{3/2}}{\cos(\delta^P_1-\delta^S_1)} +3(2s^2-1)\Delta_{\Lambda}\label{eq:LamDal}\\
  &\myeq 0.086(24)\nonumber\ , 
\end{align}
where the average in the first row is the same as the values in the $|\Delta I|=1/2$ limit. The first order correction is given as:
\begin{equation}
    \begin{split}
        \Delta\alpha_{3/2}=& p_3\left[(1-s^2)\cos(2\delta^P_1-\delta^S_1-\delta^P_3)-
          s^2\cos(\delta^S_1-\delta^P_3)\right]
      \\
+&s_3\left[s^2 \cos(2\delta^S_1-\delta^P_1-\delta^S_3)-
 (1-s^2)\cos(\delta^P_1-\delta^S_3)\right] \ .
    \end{split} 
\end{equation}
The three relations and the $|S|^2>|P|^2$ condition allow one to determine the $s$, $s_3$ and $p_3$
amplitudes for the $\Lambda$ decays. The results are given in Table~\ref{tab:SPamp}. The size of the $\Delta I=3/2$ amplitudes is 3\% of the $\Delta I=1/2$ ones. 
Finally, the value of the $\beta_-$ and $\beta_0$
can be calculated 
\begin{align}
    \frac{\beta_\Lambda}{\alpha_\Lambda}:=\frac{1}{3}\left(\frac{2\beta_-}{\alpha_-}+\frac{\beta_0}{\alpha_0}\right)&=
    \tan(\delta_{1}^P-\delta_{1}^S) \ .\label{eq:La4}\\
    \frac{\beta_-}{\alpha_-}-\frac{\beta_0}{\alpha_0}&=3\frac{ {p_3} \sin (\delta_{1}^P-\delta_{3}^P)-{s_3} \sin (\delta_{1}^S-\delta_{3}^S)}{\sqrt{2}\cos^2(\delta_{1}^P-\delta_{1}^S)}\ ,
\end{align}
where $\beta_\Lambda/\alpha_\Lambda$ is the value in the $\Delta I=1/2$ limit. Using the strong phases from Table~\ref{tab:Sphases} $\beta_\Lambda/\alpha_\Lambda=-0.128(2)$
what translates to $\phi_\Lambda$ of $-0.139(3)$. The LO $\Delta I=3/2$ correction gives $\beta_-/\alpha_-=-0.130(3)$ and $\phi_-$ of $-0.148(4)$, {\textit i.e.} the result with order of magnitude better precision than the direct measurement of $-0.113(61)$ (Table~\ref{tab:decayproperties}).

\section{Effective Lagrangian and parameterization of amplitudes}
\label{sec:AppEFT}
A  hermitian  effective Lagrangian for the initial decay $B \to b \pi$ where all baryons have spin-$1/2$ is given by 
\begin{eqnarray}
  \label{eq:lagreff}
  {\cal  L}= 
  \aaa \, i \overline b B \pi - \aaa^* \, i\overline B b \, \overline\pi - \bb \, \overline b i \gamma_5 B \pi - \bb^* \, \overline B i \gamma_5 b \, \overline\pi  \,.
\end{eqnarray}
The $\aaa$ terms lead to s-waves for the decay products, the $\bb$ terms lead to p-waves and  the $\aaa$ terms break parity (P) symmetry while the $\bb$ terms do not.
If $\aaa$ is real, then the $\aaa$ terms break P and C, but conserve CP symmetry. 
If $\bb$ is real, then the $\bb$ terms conserve C symmetry and therefore also CP. 
One can make $\bb$ real and positive by moving its phase into a redefinition of the $B$-baryon field (and a redefinition of the 
discrete transformations by an additional phase). The CP symmetry is then conserved, if the parameter $\aaa$ is real. 

Except for an irrelevant overall phase, one might write the decay matrix elements as 
\begin{eqnarray}
  {\cal M}_{B\to b\pi} \sim \overline u_b \left( \aaa-\bb\gamma_5 \right) u_B \,, \quad
  {\cal M}_{\overline B\to \overline b \overline\pi} \sim \overline v_B \left( -\aaa^*-\bb^* \gamma_5 \right) v_b \,.
  \label{eq:matr-match-PDG}  
\end{eqnarray}
This fits to the conventions of the Particle Data Group. 
Then one reads off: $S_{\rm ini} \sim \aaa$, $\overline S_{\rm ini} \sim -\aaa^*$, $P_{\rm ini} \sim \bb$, $\overline P_{\rm ini} \sim \bb^*$ 
where the p-waves pick up an additional 
phase space factor that we have not displayed explicitly. The relations between partial-wave amplitudes and parameters from 
the Lagrangian suggest writing for the initial amplitudes
\begin{alignat}{2}
  S_{\rm ini}& = \phantom{-}\vert S \vert \, e^{i\xi_S} \,, \quad   P_{\rm ini}&& = \vert P \vert \, e^{i\xi_P} \,, \nonumber \\
  \overline S_{\rm ini}& = -\vert S \vert \, e^{-i\xi_S} \,, \quad   \overline P_{\rm ini}&& = \vert P \vert \, e^{-i\xi_P} \,.
  \label{eq:dec-ini}  
\end{alignat}
Strictly speaking, $\xi_P$ is not needed. What matters is the relative phase between $S$ and $P$, which can be expressed via 
$\xi_S-\xi_P$ but equally well via $\xi_S$ if one puts $\xi_P=0$. In principle, phases can vary between 0 and $2\pi$ 
or $-\pi$ and $\pi$. However, an overall minus sign for all amplitudes would not lead to an observable consequence. Therefore, 
it is sufficient to consider $\xi_S-\xi_P \in [0,\pi)$ or $\in [-\pi/2,+\pi/2)$. If \eqref{eq:dec-ini} were the complete amplitudes, then one would always find $\overline\alpha = -\alpha$ and $\overline\beta = \beta$, irrespective 
of CP violation or conservation. 
For the case of CP symmetry, one would find $\beta =0$. 

This whole analysis leaves out final-state interactions. Rescattering is a non-local phenomenon that cannot be treated 
by a tree-level calculation using a local, hermitian Lagrangian. 
Instead, one can use explicit loop calculations if one has a microscopic picture of the 
reaction, or one can use an Omn\`es-function matrix that parameterizes the final-state interactions. This is discussed in more detail in Appendix 
\ref{sec:FSI-app}.

\section{Treatment of final-state interactions}
\label{sec:FSI-app}

In the following, we discuss in some detail the treatment of final-state interactions for the main decays of the $\Lambda$ and
$\overline \Lambda$ baryons. The case of $\Xi^{0,-}$ decays is just simpler.

The relevant decay channels of $\Lambda$ are $(p \pi^-)_S$, $(p \pi^-)_P$, $(n \pi^0)_S$, $(n \pi^0)_P$ 
where the subscript denotes the partial wave. It is more convenient to build linear combinations with respect to the isospin 
of the final states. Then the four decay channels are $(N \pi)_{S,I=1/2}$, $(N \pi)_{S,I=3/2}$, $(N \pi)_{P,I=1/2}$, $(N \pi)_{P,I=3/2}$.
Following the conventions of the main text, we denote the corresponding decay amplitudes by $L_{1,1}$ for $I=1/2$
and by $L_{3,3}$ for $I=3/2$; here $L=S,P$. The initial decay amplitudes that emerge from the weak 
process are denoted by $L_{\dots}^{\rm ini}$. 
For the corresponding antiparticle decays, we use $\overline L_{\dots}$.
We assume baryon number conservation. Then there are no oscillations
between $\Lambda$ and its antiparticle $\overline \Lambda$. 
But the final-state interactions (FSI) might allow for transitions
between the 4 final states $(N \pi)_{S,I=1/2}$, $(N \pi)_{S,I=3/2}$, $(N \pi)_{P,I=1/2}$, and $(N \pi)_{P,I=3/2}$. This defines a 
coupled-channel problem.  
We assume that the weak process is of short-distance nature such that no structure is resolved. Therefore, the discontinuity 
of a decay amplitude is solely given by the FSI. This defines an Omn\`es problem \cite{Omnes:1958hv}; 
for an analogous situation, see \textit{\textit{e.g.}}\ \cite{Daub:2015xja}. 

In general, one has a $4 \times 4$ Omn\`es-function matrix $\Omega$ that parameterizes the FSI. This matrix maps the 
``bare'' amplitudes of the initial decay onto the ``full'' amplitudes that contain the FSI:
\begin{eqnarray}
  \left(
    \begin{array}{c}
      S_{1,1} \\ S_{3,3} \\ P_{1,1} \\ P_{3,3} 
    \end{array}
  \right) = \Omega   \left(
    \begin{array}{c}
      S^{\rm ini}_{1,1} \\ S^{\rm ini}_{3,3} \\ P^{\rm ini}_{1,1} \\ P^{\rm ini}_{3,3} 
    \end{array}
  \right) \,.
  \label{eq:full4x4}
\end{eqnarray}
The corresponding equation for the antiparticle sector reads
\begin{eqnarray}
  \left(
    \begin{array}{c}
      \overline S_{1,1} \\ \overline S_{3,3} \\ \overline P_{1,1} \\ \overline P_{3,3} 
    \end{array}
  \right) = \overline \Omega   \left(
    \begin{array}{c}
      \overline S^{\rm ini}_{1,1} \\ \overline S^{\rm ini}_{3,3} \\ \overline P^{\rm ini}_{1,1} \\ \overline P^{\rm ini}_{3,3} 
    \end{array}
  \right) \,.
  \label{eq:full4x4-anti}
\end{eqnarray}

Next, we assume that parity and charge conjugation are both conserved by the FSI. This is true for strong and electromagnetic 
FSI. In this case, the FSI are the same in the particle and antiparticle sector, {\textit i.e.}\ $\Omega = \overline\Omega$, and there is no 
cross talk between the parity-even p-waves and the parity-odd s-waves: 
\begin{eqnarray}
  \Omega = \overline \Omega = \left(
    \begin{array}{cc}
      \Omega_S & 0 \\ 0 & \Omega_P
    \end{array}
  \right)
  \label{eq:only2x2}  
\end{eqnarray}
with $2 \times 2$ matrices $\Omega_S$ and $\Omega_P$. 

Finally, we assume isospin symmetry. Then the $2 \times 2$ matrices become diagonal. Watson's theorem \cite{Watson:1954uc} 
ensures that the phase of the pertinent Omn\`es function agrees with the scattering phase shift $\delta_{L,2I}$ 
of the corresponding $N$-$\pi$ scattering:
\begin{eqnarray}
  \label{eq:omnes-diag}
  \Omega_L = 
  {\rm diag}\left(\vert \Omega_{L,2I=1} \vert \, e^{i\delta_{L,1}}, \vert \Omega_{L,2I=3} \vert \, e^{i\delta_{L,3}} \right) 
\end{eqnarray}
with $L=S,P$. 
Here it is of advantage that we changed from the particle basis ($p\pi^-$ and $n\pi^0$) to the isospin basis 
($I=1/2$ and $I=3/2$). In the particle basis, the $2 \times 2$ Omn\`es matrices would not be diagonal.

Of course, if required by precision, the assumptions that lead from more general $4 \times 4$ matrices
in \eqref{eq:full4x4} and \eqref{eq:full4x4-anti} to \eqref{eq:only2x2} and \eqref{eq:omnes-diag} can be relaxed one by one.

\section{Average polarization and spin-correlation terms}
\label{app:avpol}
\begin{figure}
    \centering
    \includegraphics[width=0.7\textwidth]{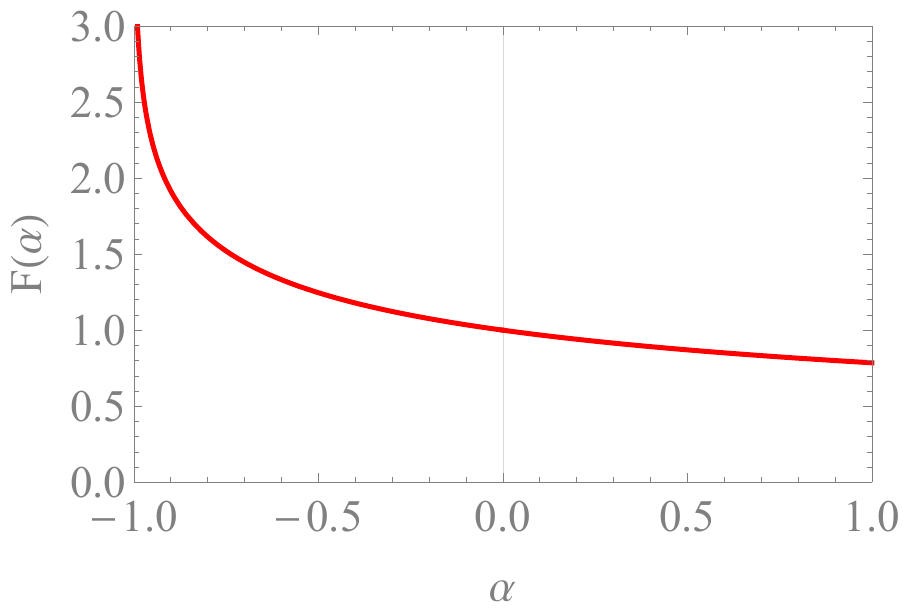}
    \caption{Function $F(\alpha)$ defined in Eq.~\eqref{eq:funcF}. The function is divergent for $\alpha\to -1$.}
    \label{fig:Function}
\end{figure}

Expression for average polarization squared
\begin{equation}
\begin{split}
   \braket{{\bf P}_B^2} =& \int{\bf P}_B^2 \left(\frac{1}{\sigma}\frac{\dd\sigma}{\dd\Omega_B}\right)\dd\Omega_B = \frac{3}{2}\int{\bf P}_B^2\frac{ 1+\alpha_\psi\cos^2\!\theta}{3+\alpha_\psi}\ \dd\!\cos\theta\ ,
\end{split}
\end{equation}
where ${\bf {P}}_B$ is given by Eq.~\eqref{eq:PolB}. The integral can be calculated exactly, and the result is expressed as
\begin{equation}
   \braket{{\bf P}_B^2} =\mathbb{p}_0+\mathbb{p}_2P_e^2\ , 
\end{equation}
where 
\begin{align}
    \mathbb{p}_0&=\frac{\left(1-\alpha_\psi ^2\right) \sin ^2(\Delta \Phi )}{\alpha_\psi ^2 (3+\alpha_\psi)} \left\{3+2 \alpha_\psi -3 (1+\alpha_\psi) \text{F}(\alpha_\psi )\right\},\\
    \mathbb{p}_2&=\frac{3 (1+\alpha_\psi)^2}{ \alpha_\psi  (3+\alpha_\psi)}
    \left\{1-\frac{1-\alpha_\psi }{1+\alpha_\psi} \cos^2 (\Delta \Phi ) -  \left(1-(1\!-\!\alpha_\psi ) \cos^2 ( \Delta \Phi )\right)\text{F}(\alpha_\psi )\right\}\ .
\end{align}
The function $F(\alpha)$ is
\begin{equation}
    F(\alpha):=\int_{0}^1\frac{\dd x}{1+\alpha x^2}=\left\{\begin{array}{cc}
    \frac{\arctan\sqrt{\alpha}}{\sqrt{\alpha}}     & 0<\alpha\le 1 \\
        1 & \alpha=0\\
    \frac{{\rm arctanh}\sqrt{|\alpha|}}{\sqrt{|\alpha|}}        & -1<\alpha<0
    \end{array}\right.
    \ .\label{eq:funcF}
\end{equation}
 Properties of the function  $F(1)=\frac{\pi}{4}$ and $\lim_{\alpha\to-1} F(\alpha)=\infty$. The function is drawn in Fig.~\ref{fig:Function}.
For $\alpha_\psi=1$ the coefficients are 
\begin{align}
     \mathbb{p}_0=0\ \ {\rm and}\ \
     \mathbb{p}_2=\frac{3(4-\pi)}{4}\approx0.6438  \ 
\end{align}
and for $\alpha_\psi=0$
\begin{align}
     \mathbb{p}_0=\frac{2}{15}\sin ^2(\Delta \Phi )\ \ {\rm and}\ \
   \mathbb{p}_2=\frac{2+\cos (2 \Delta \Phi )}{3} \ .\label{eq:PolBa0}
\end{align}
One derives similar expressions for the sum of the squares of the spin-correlation terms. The result can be expressed as
\begin{equation}
   \braket{{\mathbb{S}}_{B\overline B}^2} =\mathbb{s}_0+\mathbb{s}_2P_e^2\ , 
\end{equation}
where 
\begin{align}
    \mathbb{s}_0&=\frac{1 }{\alpha_\psi ^2 (3+\alpha_\psi)} \left\{
    \left(1-\alpha_\psi ^2\right) (2 \alpha_\psi +3) \cos (2 \Delta \Phi )-7 \alpha_\psi ^3-2 \alpha_\psi -3\right.\nonumber\\
    &\left.-3 (\alpha_\psi +1)^2 {F}(\alpha_\psi ) \left[(1-\alpha_\psi) \cos (2 \Delta \Phi )+\alpha_\psi-2 \alpha_\psi^2-1\right]
    \right\},
    \\
    \mathbb{s}_2&=\frac{6 (1-\alpha_\psi^2)\sin ^2(\Delta \Phi )}{ \alpha_\psi  (3+\alpha_\psi)}
    \left\{(1+\alpha_\psi) {F}(\alpha_\psi)-1 \right\}
    \ .
\end{align}

For the $e^+e^-\to B\overline B$ process specified by the parameters $\alpha_\psi$, $\beta_\psi$ and  $\gamma_\psi$ the polarization and spin-correlation terms as a function of the $B$-baryon production angle $\theta$ are
\begin{align}
    \mathbb{P}^2_B(\cos\theta)&=2\frac{(\alpha_\psi +1)^2 P_e^2 \cos ^2\theta +\sin ^2\theta  \left(\beta _{\psi }^2 \cos ^2\theta +P_e^2 \gamma _{\psi }^2\right)}{\left(1+\alpha_\psi  \cos ^2\theta\right)^2},\\
    \mathbb{S}^2_{B\overline B}(\cos\theta)&=\frac{\left(\alpha_\psi ^2+1\right) \sin ^4\theta +\left(\alpha_\psi +\cos ^2\theta \right)^2+2 \sin ^2\theta  \left(\gamma _{\psi }^2 \cos ^2\theta +P_e^2 \beta _{\psi }^2\right)}{\left(1+\alpha_\psi  \cos ^2\theta\right)^2}\ .
\end{align}

\bibliographystyle{apsrev4-2} \bibliography{refBB}

\end{document}